\let\accentvec\vec 

\documentclass{style/technical_report}

\usepackage{graphicx}
\usepackage{xcolor}
\usepackage{latexsym,amssymb,textcomp}
\let\vec\accentvec 
\usepackage[fleqn, cmex10]{amsmath}
\usepackage{stmaryrd}
\usepackage{listings}

\usepackage{ifthen}
\usepackage{comment}
\usepackage{mathptmx}
\usepackage{cite}
\usepackage{array}

\newenvironment{fortechnicalreport}
	{}
	{}
\newenvironment{forjournal}
	{}
	{}

\includecomment{fortechnicalreport}
\excludecomment{forjournal}

\newcommand{\sectionreference}[1]{Section~\ref{#1}}
\newcommand{\externalsectionreference}[1]{Section~#1}
\newcommand{\figurereference}[1]{Figure~\ref{#1}}
\newcommand{\listingreference}[1]{Listing~\ref{#1}}
\newcommand{\mathematicaldefinition}{\buildrel def\over =}
\newcommand{\mathematicalelementdefinition}{\buildrel def\over \in}
\newcommand{\journaltitle}{A Formal Reference for {SCOOP}}
\newcommand{\technicalreporttitle}{A comprehensive operational semantics of the {SCOOP} programming model}
\begin{fortechnicalreport}
\newcommand{\elementfontsize}{\normalsize}
\end{fortechnicalreport}
\begin{forjournal}
\newcommand{\elementfontsize}{\footnotesize}
\end{forjournal}
\newcommand{\elementverticalspacingoffset}{-1pt}
\newcommand{\elementverticalmarginoffsetfront}{-2ex}
\newcommand{\elementverticalmarginoffsetback}{-1.5ex}

\newcommand{\ecreate}{\mbox{\lstinline[language=SCOOP]!create!}\ }
\newcommand{\eif}[3]{\mbox{\lstinline[language=SCOOP]!if!}\ #1\ \mbox{\lstinline[language=SCOOP]!then!}\ #2\ \mbox{\lstinline[language=SCOOP]!else!}\ #3\ \mbox{\lstinline[language=SCOOP]!end!}}

\newcommand{\euntil}[2]{\mbox{\lstinline[language=SCOOP]!until!}\ #1\ \mbox{\lstinline[language=SCOOP]!loop!}\ #2\ \mbox{\lstinline[language=SCOOP]!end!}}
\newcommand{\eassignment}{\thinspace \mathit{:=} \thinspace}

\newcommand{\voidliteral}{\mbox{\lstinline[language=SCOOP]!Void!}}
\newcommand{\currententityname}{\mbox{\lstinline[language=SCOOP]!Current!}}
\newcommand{\resultentityname}{\mbox{\lstinline[language=SCOOP]!Result!}}

\newcommand{\newidentifierfeature}{\mathit{new\_id}}

\newcommand{\freshchanneldefinition}[1]{#1 \ is \ fresh}
\newcommand{\datafeature}{\mathit{data}}

\definecolor{operationcolor}{rgb}{0.12, 0.41, 0.43}
\newcommand{\OPERATION}[1]{{\color{operationcolor}\mathtt{#1}}}
\newcommand{\evaluateoperation}{\OPERATION{eval}}
\newcommand{\writevalueoperation}{\OPERATION{write}}
\newcommand{\readvalueoperation}{\OPERATION{read}}
\newcommand{\waitoperation}{\OPERATION{wait}}
\newcommand{\resultoperation}{\OPERATION{result}}
\newcommand{\singlelineconditionaloperation}[3]{\conditionaloperationname\ #1\ \OPERATION{then}\ #2\ \OPERATION{else}\ #3\ \OPERATION{end}}
\newcommand{\multilineconditionaloperation}[3]{& \conditionaloperationname\ #1\ \OPERATION{then} \\ & \indentation #2 \\ & \OPERATION{else} \\ & \indentation #3 \\ & \OPERATION{end}}
\newcommand{\conditionaloperationname}{\OPERATION{provided}}
\newcommand{\nooperation}{\OPERATION{nop}}
\newcommand{\issueoperation}{\OPERATION{issue}}
\newcommand{\executedelegatedoperation}{\OPERATION{execute\_delegated}}
\newcommand{\leavedelegatedexecutionoperation}{\OPERATION{leave\_delegated}}
\newcommand{\applyoperation}{\OPERATION{apply}}
\newcommand{\setonceroutinenotfreshoperation}{\OPERATION{set\_not\_fresh}}
\newcommand{\setonceroutinenotfreshwithresultoperation}{\OPERATION{set\_not\_fresh\_with\_result}}
\newcommand{\checkpreconditionandlockrequestqueuesoperation}{\OPERATION{check\_pre\_and\_lock\_rqs}}
\newcommand{\popobtainedrequestqueuelocksoperation}{\OPERATION{pop\_obtained\_rq\_locks}}
\newcommand{\checkpostconditionandunlockrequestqueuesoperation}{\OPERATION{check\_post\_and\_unlock\_rqs}}
\newcommand{\returnoperation}{\OPERATION{return}}
\newcommand{\lockrequestqueuesoperation}{\OPERATION{lock}}
\newcommand{\unlockrequestqueueoperation}{\OPERATION{unlock}}
\newcommand{\calloperation}{\OPERATION{call}}
\newcommand{\notifyoperation}{\OPERATION{notify}}
\newcommand{\statementseparator}{;}

\newcommand{\statetype}{\mathbf{STATE}}
\newcommand{\state}{\sigma}
\newcommand{\storefeature}{\mathit{store}}
\newcommand{\processorregionsfeature}{\mathit{regions}}
\newcommand{\heapfeature}{\mathit{heap}}
\newcommand{\setallfeature}{\mathit{set}}
\newcommand{\setvaluefeature}{\mathit{set\_val}}
\newcommand{\setenvironmentvaluefeature}{\mathit{set\_env\_val}}
\newcommand{\deepimportfeature}{\mathit{deep\_import}}
\newcommand{\deepimportrecursivewithmapfeature}{\mathit{deep\_import\_rec\_with\_map}}
\newcommand{\deepimportrecursivewithoutmapfeature}{\mathit{deep\_import\_rec\_without\_map}}
\newcommand{\lastimportedreferencefeature}{\mathit{last\_imported\_ref}}
\newcommand{\newprocessorfeature}{\mathit{new\_proc}}
\newcommand{\newobjectfeature}{\mathit{new\_obj}}

\newcommand{\pushenvironmentwithfeaturefeature}{\mathit{push\_env\_with\_feature}}
\newcommand{\environmentvaluefeature}{\mathit{env\_val}}
\newcommand{\makefeature}{\mathit{make}} 

\newcommand{\storetype}{\mathbf{STORE}}
\newcommand{\store}{s}
\newcommand{\environmentsfeature}{\mathit{envs}}
\newcommand{\pushenvironmentfeature}{\mathit{push\_env}}
\newcommand{\popenvironmentfeature}{\mathit{pop\_env}}

\newcommand{\environmenttype}{\mathbf{ENV}}
\newcommand{\environment}{e}
\newcommand{\namesfeature}{\mathit{names}}
\newcommand{\updatefeature}{\mathit{update}}
\newcommand{\valuefeature}{\mathit{val}} 

\newcommand{\processorregionstype}{\mathbf{REG}}
\newcommand{\processorregions}{k}
\newcommand{\processorsfeature}{\mathit{procs}}
\newcommand{\handledobjectsfeature}{\mathit{handled\_objs}}
\newcommand{\lastaddedprocessorfeature}{\mathit{last\_added\_proc}}
\newcommand{\isrequestqueuelockedfeature}{\mathit{rq\_locked}}
\newcommand{\iscallstacklockedfeature}{\mathit{cs\_locked}}
\newcommand{\obtainedrequestqueuelocksfeature}{\mathit{obtained\_rq\_locks}}
\newcommand{\obtainedcallstacklockfeature}{\mathit{obtained\_cs\_lock}}
\newcommand{\retrievedrequestqueuelocksfeature}{\mathit{retrieved\_rq\_locks}}
\newcommand{\retrievedcallstacklocksfeature}{\mathit{retrieved\_cs\_locks}}
\newcommand{\arelockspassedfeature}{\mathit{locks\_passed}}
\newcommand{\addprocessorfeature}{\mathit{add\_proc}}
\newcommand{\removeobjectfeature}{\mathit{remove\_obj}}
\newcommand{\lockrequestqueuesfeature}{\mathit{lock\_rqs}}
\newcommand{\delegateobtainedrequestqueuelocksfeature}{\mathit{delegate\_obtained\_rq\_locks}}
\newcommand{\unlockrequestqueuefeature}{\mathit{unlock\_rq}}
\newcommand{\popobtainedrequestqueuelocksfeature}{\mathit{pop\_obtained\_rq\_locks}}
\newcommand{\passlocksfeature}{\mathit{pass\_locks}}
\newcommand{\revokelocksfeature}{\mathit{revoke\_locks}}
\newcommand{\handlerfeature}{\mathit{handler}}
\newcommand{\requestqueuelocksfeature}{\mathit{rq\_locks}}
\newcommand{\callstacklocksfeature}{\mathit{cs\_locks}}
\newcommand{\addobjectfeature}{\mathit{add\_obj}} 

\newcommand{\heaptype}{\mathbf{HEAP}}
\newcommand{\heap}{h}
\newcommand{\objectsfeature}{\mathit{objs}}
\newcommand{\referencesfeature}{\mathit{refs}}
\newcommand{\referencedobjectfeature}{\mathit{ref\_obj}}
\newcommand{\lastaddedobjectfeature}{\mathit{last\_added\_obj}}
\newcommand{\isonceroutinefreshfeature}{\mathit{is\_fresh}}
\newcommand{\oncefunctionresultfeature}{\mathit{once\_result}}
\newcommand{\updatereferencefeature}{\mathit{update\_ref}}
\newcommand{\setoncefunctionnotfreshfeature}{\mathit{set\_once\_func\_not\_fresh}}
\newcommand{\setonceprocedurenotfreshfeature}{\mathit{set\_once\_proc\_not\_fresh}}
\newcommand{\referencefeature}{\mathit{ref}}

\newcommand{\objecttype}{\mathbf{OBJ}}
\newcommand{\object}{o}
\newcommand{\identifierfeature}{\mathit{id}} 
\newcommand{\classtypefeature}{\mathit{class\_type}} 
\newcommand{\attributevaluefeature}{\mathit{att\_val}}
\newcommand{\setattributevaluefeature}{\mathit{set\_att\_val}}
\newcommand{\copyfeature}{\mathit{copy}}

\newcommand{\referencetype}{\mathbf{REF}}

\newcommand{\voidvalue}{\mathit{void}}

\newcommand{\processortype}{\mathbf{PROC}}

\newcommand{\classtypetype}{\mathbf{CLASS\_TYPE}}
\newcommand{\namefeature}{\mathit{name}} 
\newcommand{\isexpandedclasstypefeature}{\mathit{is\_exp}}
\newcommand{\isreferenceclasstypefeature}{\mathit{is\_ref}}
\newcommand{\attributesfeature}{\mathit{attributes}}
\newcommand{\functionsfeature}{\mathit{functions}}
\newcommand{\proceduresfeature}{\mathit{procedures}}
\newcommand{\invariantexistsfeature}{\mathit{inv\_exists}}
\newcommand{\invariantfeature}{\mathit{inv}}
\newcommand{\booleanclasstype}{BOOLEAN}
\newcommand{\booleanclasstypeitemattributename}{\mathit{item}}
\newcommand{\featurebynamefeature}{\mathit{feature\_by\_name}}

\newcommand{\featuretype}{\mathbf{FEATURE}}
\newcommand{\formalargumentsfeature}{\mathit{formals}}
\newcommand{\isonceroutinefeature}{\mathit{is\_once}}
\newcommand{\preconditionexistsfeature}{\mathit{pre\_exists}}
\newcommand{\preconditionfeature}{\mathit{pre}}
\newcommand{\postconditionexistsfeature}{\mathit{post\_exists}}
\newcommand{\postconditionfeature}{\mathit{post}}
\newcommand{\localsfeature}{\mathit{locals}}
\newcommand{\bodyfeature}{\mathit{body}}
\newcommand{\isexportedfeature}{\mathit{exported}}

\newcommand{\routinetype}{\mathbf{ROUTINE}}

\newcommand{\functiontype}{\mathbf{FUNCTION}}

\newcommand{\proceduretype}{\mathbf{PROCEDURE}}

\newcommand{\attributetype}{\mathbf{ATTRIBUTE}}

\newcommand{\entitytype}{\mathbf{ENTITY}}
\newcommand{\contextfeaturefeature}{\mathit{context\_feature}}
\newcommand{\currententity}{\mathit{current}}
\newcommand{\resultentity}{\mathit{result}}

\newcommand{\expressiontype}{\mathbf{EXPRESSION}}

\newcommand{\instructiontype}{\mathbf{INSTRUCTION}}

\newcommand{\nametype}{\mathbf{NAME}}

\newcommand{\literaltype}{\mathbf{LITERAL}}
\newcommand{\objectfeature}{\mathit{obj}}

\newcommand{\identifiertype}{\mathbf{ID}}

\newcommand{\settype}[1]{\mathbf{SET[#1]}}
\newcommand{\isemptyfeature}{\mathit{is\_empty}} 
\newcommand{\containsfeature}{\mathit{has}} 
\newcommand{\countfeature}{\mathit{count}} 
\newcommand{\addfeature}{\mathit{add}} 
\newcommand{\set}[1]{\{#1\}}
\newcommand{\setinference}[2]{\protect\begin{split}\{#1 \thickspace \vert \thickspace #2\}\protect\end{split}}

\newcommand{\tupletype}[1]{
	\mathbf{
		TUPLE \ifthenelse{\equal{#1}{}}{}{[#1]}
	}
}
\newcommand{\tuple}[1]{(#1)}

\newcommand{\stacktype}[1]{\mathbf{STACK[#1]}}
\newcommand{\pushfeature}{\mathit{push}}
\newcommand{\popfeature}{\mathit{pop}}
\newcommand{\topfeature}{\mathit{top}}
\newcommand{\flattenedfeature}{\mathit{flat}}

\newcommand{\maptype}[2]{\mathbf{MAP[#1, #2]}}
\newcommand{\keysfeature}{\mathit{keys}}

\newcommand{\booleantype}{\mathbf{BOOLEAN}}
\newcommand{\truevalue}{\mathit{true}}
\newcommand{\falsevalue}{\mathit{false}}

\newcommand{\creation}[2]{\mathit{new} \thickspace #1.#2}

\newcommand{\condition}[1]{\protect\begin{split}\mathit{if} #1\protect\end{split}}
\newcommand{\multilinecondition}[1]{\protect\begin{split}& \mathit{if} \protect\\ & \smallindentation \protect\begin{split}#1\protect\end{split} \protect\\ & \mathit{then}\protect\end{split}}
\newcommand{\otherwisecondition}{\mathit{otherwise}}

\newcommand{\typingenvironment}{\Gamma}
\newcommand{\iscontrolledfeature}{\mathit{controlled}}
\newcommand{\controllingentityfeature}{\mathit{controlling\_entity}}
\newcommand{\typingenvironmentderivation}[1]{\typingenvironment \vdash #1}
\newcommand{\typefromtypingenvironment}{\mathit{type\_of}}

\newcommand{\featurekeyword}[1]{\mathbf{#1}}
\newcommand{\feature}[4]{
	\vspace{\elementverticalmarginoffsetfront}
	\setlength{\jot}{\elementverticalspacingoffset}
	\elementfontsize
	\begin{gather*}
		\begin{split}
			& \begin{split}
				#1
			\end{split}
			\ifthenelse
				{\equal{#3}{}}
				{}
				{
						\\
						& \indentation \begin{split} 
							& #2 \thickspace \featurekeyword{require} \\
							& \indentation \begin{split}
								#3
							\end{split}
						\end{split}
				}
			\ifthenelse
				{\equal{#4}{}}
				{}
				{
					\\
					& \indentation \begin{split}
						& \featurekeyword{axioms} \\
						& \indentation \begin{split}
							#4
						\end{split}
					\end{split}
				}
		\end{split}
	\end{gather*}
	\normalsize
	\setlength{\jot}{0pt}
	\vspace{\elementverticalmarginoffsetback}
}
\newcommand{\where}[2]{
	\protect\begin{split}
		& \protect\begin{split}
			#1
		\protect\end{split} \protect\\
		& \smallindentation \protect\begin{split}
			& \featurekeyword{where} \protect\\
			& \smallindentation \protect\begin{split}
				#2
			\protect\end{split}
		\protect\end{split}
	\protect\end{split}
}

\newcommand{\function}[1]{
	\vspace{\elementverticalmarginoffsetfront}
	\setlength{\jot}{\elementverticalspacingoffset}
	\elementfontsize
	\begin{gather*}
		\begin{split}
			#1
		\end{split}
	\end{gather*}
	\normalsize
	\setlength{\jot}{0pt}
	\vspace{\elementverticalmarginoffsetback}
}
\newcommand{\isolateddefinition}[1]{%
	\vspace{\elementverticalmarginoffsetfront}
	\setlength{\jot}{\elementverticalspacingoffset}%
	\elementfontsize%
	\begin{gather*}%
		\begin{split}%
			#1%
		\end{split}%
	\end{gather*}%
	\normalsize%
	\setlength{\jot}{0pt}%
	\vspace{\elementverticalmarginoffsetback}
}

\newcommand{\inferencerule}[4]{
	\vspace{\elementverticalmarginoffsetfront}
	\setlength{\jot}{\elementverticalspacingoffset}
	\elementfontsize
	\begin{gather*}
		\begin{split}
			& \mbox{\rm\bf{#1}}\\[1ex]
			& \frac
				{
					\textstyle\rule[-1.0ex]{0cm}{3ex}
					\begin{array}{l}
						#2
					\end{array}
				}
	    	{
	    		\textstyle\rule[-.5ex]{0cm}{3ex}
	    		\transition
						{#3}
						{#4}
	    	}
	  \end{split}
 	\end{gather*}
 	\normalsize
 	\setlength{\jot}{0pt}
 	\vspace{\elementverticalmarginoffsetback}
}
\newcommand{\singlelineinferencerule}[4]{
	\vspace{\elementverticalmarginoffsetfront}
	\setlength{\jot}{\elementverticalspacingoffset}
	\elementfontsize
	\begin{gather*}
		\begin{split}
			& \mbox{\rm\bf{#1}}\\[1ex]
			& \frac
				{
					\textstyle\rule[-1.0ex]{0cm}{3ex}
					\begin{array}{l}
						#2
					\end{array}
				}
	    	{
	    		\textstyle\rule[-.5ex]{0cm}{3ex}
	    		\singlelinetransition
						{#3}
						{#4}
	    	}
		\end{split}
 	\end{gather*}
 	\normalsize
 	\setlength{\jot}{0pt}
	\vspace{\elementverticalmarginoffsetback}
}
\newcommand{\configuration}[2]{
	\begin{split}
		\langle #1, #2 \rangle
	\end{split} 
}
\newcommand{\singlelineconfiguration}[2]{
	\langle #1, #2 \rangle
}
\newcommand{\isolatedconfiguration}[2]{
	\vspace{\elementverticalmarginoffsetfront}
	\setlength{\jot}{\elementverticalspacingoffset}
	\elementfontsize
	\begin{gather*}
		\begin{split}
			& \langle \\
			& \indentation \begin{split}
				#1
			\end{split} \\
			& , \\ 
			& \indentation \begin{split}
				#2
			\end{split} \\
			& \rangle		
		\end{split}
	\end{gather*}
	\normalsize
	\setlength{\jot}{0pt}
	\vspace{\elementverticalmarginoffsetback}
}
\newcommand{\isolatedsinglelineconfiguration}[2]{
	\vspace{\elementverticalmarginoffsetfront}
	\setlength{\jot}{\elementverticalspacingoffset}
	\elementfontsize
	\begin{gather*}
		\begin{split}
			\langle #1, #2 \rangle
		\end{split}
	\end{gather*}
	\normalsize
	\setlength{\jot}{0pt}
	\vspace{\elementverticalmarginoffsetback}
}
\newcommand{\processorseparator}{\thickspace \vert \thickspace}
\newcommand{\substitution}[2]{[ #1 / #2 ]}

\newcommand{\transition}[2]{
	\begin{split}
		\typingenvironment \vdash & #1 \rightarrow \\
		& #2
	\end{split}
}
\newcommand{\singlelinetransition}[2]{
	\typingenvironment \vdash #1 \rightarrow #2
}

\newcommand{\isolatedsinglelinetransition}[2]{
	\vspace{\elementverticalmarginoffsetfront}
	\setlength{\jot}{\elementverticalspacingoffset}
	\elementfontsize
	\begin{gather*}
		\begin{split}
			\typingenvironment \vdash #1 \rightarrow #2
		\end{split}
	\end{gather*}
	\normalsize
	\setlength{\jot}{0pt}
	\vspace{\elementverticalmarginoffsetback}
}

\newcommand{\simplifiedstate}[4]{
	& \begin{split}
		& \mbox{locks} \colon \\
		& \indentation \begin{split}
			#1
		\end{split}
	\end{split} \\
	& \begin{split}
		& \mbox{objects} \colon \\
		& \indentation \begin{split}
			#2
		\end{split}
	\end{split} \\
	& \begin{split}
		& \mbox{once status} \colon
		\ifthenelse
			{\equal{#3}{}}
			{}
			{
				\\
				& \indentation \begin{split}
					#3
				\end{split}
			}
	\end{split} \\
	& \begin{split}
		& \mbox{environments} \colon \\
		& \indentation \begin{split}
			#4
		\end{split}
	\end{split}
}
\newcommand{\isolatedsimplifiedstate}[4]{
	\vspace{\elementverticalmarginoffsetfront}
	\setlength{\jot}{\elementverticalspacingoffset}
	\elementfontsize
	\begin{gather*}
		\begin{split}
			\simplifiedstate{#1}{#2}{#3}{#4}
		\end{split}
	\end{gather*}
	\normalsize
	\setlength{\jot}{0pt}
	\vspace{\elementverticalmarginoffsetback}
}
\newcommand{\simplifiedstatelocksentry}[6]{
	#1 :: \ \mbox{orq} \colon (#2) \thickspace \thickspace \mbox{rrq} \colon (#3) \thickspace \thickspace \mbox{rcs} \colon (#4) \thickspace \thickspace #5 \thickspace \thickspace #6
}
\newcommand{\simplifiedstateunlockedindicator}{\mbox{unlocked}}
\newcommand{\simplifiedstatelockedindicator}{\mbox{locked}}
\newcommand{\simplifiedstatepassedlocksindicator}{\mbox{passed}}
\newcommand{\simplifiedstatenopassedlocksindicator}{}

\newcommand{\simplifiedstateobjectsentry}[2]{
	\begin{split}#1 :: \ #2\end{split}
}
\newcommand{\simplifiedstatereferencedobject}[2]{
	#1 \rightarrow #2
}
\newcommand{\simplifiedstateoncestatusentry}[2]{
	#1 :: \ #2
}
\newcommand{\simplifiedstateallprocessorsindicator}{\mbox{all}}
\newcommand{\simplifiedstateoncefunctionstatus}[3]{
	\{\mbox{\lstinline[language=SCOOP]!#1!}\}.\mbox{\lstinline[language=SCOOP]!#2!} \rightarrow #3
}
\newcommand{\simplifiedstateonceprocedurestatus}[2]{
	\{\mbox{\lstinline[language=SCOOP]!#1!}\}.\mbox{\lstinline[language=SCOOP]!#2!}
}
\newcommand{\simplifiedstateenvironmentsentry}[2]{
	\begin{split}#1 :: \ #2\end{split}
}
\newcommand{\simplifiedstateenvironmentsentryseparator}{\thickspace / \thickspace}
\newcommand{\simplifiedstateentityvalue}[2]{
	\mbox{\lstinline[language=SCOOP]!#1!} \rightarrow #2
}
\newcommand{\simplifiedstatecurrententityvalue}[1]{
	\currententityname \rightarrow #1
}
\newcommand{\simplifiedstateresultentityvalue}[1]{
	\resultentityname \rightarrow #1
}

\newcommand{\targetsfeature}{\mathit{targets}}
\newcommand{\actualargumentsfeature}{\mathit{args}}

\newcommand{\indentation}{\quad \thickspace}
\newcommand{\smallindentation}{\thickspace \thickspace}

\newcounter{clarification}
\newenvironment{clarification}[1][]
	{
		\stepcounter{clarification}
		\begin{trivlist}
			\item[\hskip \labelsep {\itshape Clarification \theclarification \ifthenelse{\equal{#1}{}}{}{\thinspace (#1)}.}]
	}
	{
		$\boxempty$ \end{trivlist}
	}

\begin{document}

\begin{forjournal}
\frontmatter          
\pagestyle{headings}  
\mainmatter           
\end{forjournal}

\begin{forjournal}
\title{\journaltitle}
\titlerunning{\journaltitle}
\end{forjournal}

\begin{fortechnicalreport}
\title{\technicalreporttitle}
\end{fortechnicalreport}

\begin{forjournal}
\author{Benjamin Morandi \and Sebastian Nanz \and Bertrand Meyer}
\authorrunning{Benjamin Morandi et al.}
\tocauthor{Benjamin Morandi, Sebastian Nanz, Bertrand Meyer}
\institute{Chair of Software Engineering, ETH Zurich, Switzerland, \\ \email{firstname.lastname@inf.ethz.ch}, \\ \texttt{http://se.inf.ethz.ch/}}
\end{forjournal}

\begin{fortechnicalreport}
\author{Benjamin Morandi, Sebastian Nanz, and Bertrand Meyer
\and
{\small
\begin{tabular}{c}
Chair of Software Engineering, ETH Zurich, Switzerland \\
\texttt{firstname.lastname@inf.ethz.ch} \\ 
\texttt{http://se.inf.ethz.ch/}
\end{tabular}}
}
\end{fortechnicalreport}

\maketitle 

\begin{abstract}
  Operational semantics is a flexible but rigorous means to describe
  the meaning of programming languages. Small semantics are often
  preferred, for example to facilitate model checking. However,
  omitting too many details in a semantics limits results to a core
  language only, leaving a wide gap towards real implementations.  In
  this paper we present a comprehensive semantics of the concurrent
  programming model SCOOP (Simple Concurrent Object-Oriented
  Programming). The semantics has been found detailed enough to guide
  an implementation of the SCOOP compiler and runtime system, and to
  detect and correct a variety of errors and ambiguities in the
  original informal specification and prototype implementation.  In
  our formal specification, we use abstract data types with
  preconditions and axioms to describe the state, and introduce a
  number of special operations to model the runtime system with our
  inference rules.  This approach makes our large formal specification
  manageable, providing a first step towards reference documents for
  specifying concurrent object-oriented languages based on operational
  semantics.
\end{abstract}

\setlength{\mathindent}{0mm}

\section{Introduction}
Concurrent programming has become an important part of mainstream
software development, caused by the widespread use of multicore
processors. The notorious difficulty of writing concurrent programs
correctly has on the other hand spawned work into novel language
abstractions to express concurrency and synchronization. One such
language is SCOOP \cite{meyer:1997:oosc,nienaltowski:2007:SCOOP}, an
object-oriented programming model for concurrency based on the idea of
contracts.

The main idea of SCOOP is to simplify the writing of correct
concurrent programs for developers, who can use familiar concepts from
object-oriented programming but are protected by the model from common
concurrency errors such as data races. This is achieved by a runtime
system that automatically takes care of operations such as obtaining
and releasing of necessary locks, without the need for explicit
program statements. While being based on conceptually simple ideas,
the semantics of the language concepts and runtime system turns out to
be very complex.

The question is therefore how the semantics can be properly
documented. The initial version of SCOOP has been defined
in~\cite{meyer:1997:oosc}, where all the main concepts are outlined
but implementation aspects are neglected for the most part. A first
prototype implementation was then introduced
in~\cite{nienaltowski:2007:SCOOP}, where the semantics was described only
informally, with the exception of a formalization of the type
system. In this paper we provide a full formalization of the
operational behavior of SCOOP, specified by a structural operational
semantics. The main contributions of the paper are:
\begin{itemize}
\item A formal specification of SCOOP that treats all important
  language elements.
\item Clarification and correction of the informal specification
  in~\cite{nienaltowski:2007:SCOOP}.
\end{itemize}
This work does not provide a formal completeness and soundness proof with respect to an axiomatic semantics. \sectionreference{sec:conclusion} discusses this possibility as part of future work. This work focuses on a formal reference for a concurrent programming language. We argue that this formal reference reflects and corrects the informal description by following a systematic approach.

\begin{forjournal}
This article is a condensed version of our technical report \cite{morandi-nanz-meyer:2011:operational_semantics_for_SCOOP} on the same subject.
\end{forjournal}
This paper is structured as follows. \sectionreference{sec:background} gives a brief overview of the main ideas of the SCOOP
model to provide a basic intuition for the main part of the paper.
\sectionreference{sec:related-work} gives an overview of related work.
\sectionreference{sec:language-overview} gives an overview of the considered language. The two following chapters contain the main parts of the
formalization: \sectionreference{sec:state-formalization} describes the state
formalization and \sectionreference{sec:execution-formalization} the formalization
of computations. \sectionreference{sec:conclusion} concludes and discusses future applications of the formalization.

\section{Background}\label{sec:background}
This section gives an overview of SCOOP. The starting idea is that every object is associated for its lifetime with a processor, called its \emph{handler}. A \emph{processor} is an autonomous thread of control capable of executing actions on objects. An object's class describes the possible actions as \emph{features}. A processor can be a CPU, but it can also be implemented in software, for example as a process or as a thread; any mechanism that can execute instructions sequentially is suitable as a processor. 

A variable \lstinline[language=SCOOP]!x! belonging to a processor can point to an object with the same handler (\emph{non-separate object}), or to an object on another processor (\emph{separate object}). In the first case, a \emph{feature call} \lstinline[language=SCOOP]!x.f! is \emph{non-separate}: the handler of \lstinline[language=SCOOP]!x! executes the feature synchronously. In this context, \lstinline[language=SCOOP]!x! is called the \emph{target} of the feature call. In the second case, the feature call is \emph{separate}: the handler of \lstinline[language=SCOOP]!x!, i.e., the \emph{supplier}, executes the call asynchronously on behalf of the requester, i.e., the \emph{client}. The possibility of asynchronous calls is the main source of concurrent execution. The asynchronous nature of separate feature calls implies a distinction between a feature call and a \emph{feature application}: the client logs the call with the supplier (feature call) and moves on; only at some later time will the supplier actually execute the body (feature application).

The producer-consumer problem serves as a simple illustration of these ideas. A root class defines the entities \lstinline[language=SCOOP]!producer!, \lstinline[language=SCOOP]!consumer!, and \lstinline[language=SCOOP]!buffer!.
\begin{lstlisting}[mathescape=true,language=SCOOP]
producer: separate PRODUCER
consumer: separate CONSUMER
buffer: separate BUFFER [INTEGER]
		-- The data structure for exchanging objects between the producer and the consumer.
\end{lstlisting}
The keyword \lstinline[language=SCOOP]!separate! specifies that the referenced objects may be handled by a processor different from the current one. A \emph{creation instruction} on a separate entity such as \lstinline[language=SCOOP]!producer! will create an object on another processor; by default the instruction also creates that processor.

Both the producer and the consumer access an unbounded buffer in feature calls such as \lstinline[language=SCOOP]!buffer.put (n)! and \lstinline[language=SCOOP]!buffer.item!. To ensure exclusive access, the consumer processor must lock the buffer processor before accessing the buffer. Such locking requirements of a feature must be expressed in the formal argument list: any target of separate type within the feature must occur as a formal argument; the arguments' handlers are locked for the duration of the feature execution, thus preventing data races. Such targets are called \emph{controlled}. For instance, in \lstinline[language=SCOOP]!consume!, \lstinline[language=SCOOP]!buffer! is a formal argument; the consumer processor has exclusive access to the buffer processor while executing \lstinline[language=SCOOP]!consume!.

Condition synchronization relies on preconditions (after the \lstinline[language=SCOOP]!require! keyword) to express wait conditions. Any precondition of the form \lstinline[language=SCOOP]!x.some_condition! makes the execution of the feature wait until the condition is true. For example, the precondition of \lstinline[language=SCOOP]!consume! delays the execution until the buffer is not empty. As the buffer is unbounded, the corresponding producer feature does not need a wait condition.
\begin{lstlisting}[language=SCOOP]
consume (buffer: separate BUFFER [INTEGER])
		-- Consume an item from the buffer.
	require
		not (buffer.count = 0)
	local
		consumed_item: INTEGER
	do
		consumed_item := buffer.item
	end
\end{lstlisting}
During a feature call, the consumer processor could pass its locks to the buffer processor if it has a lock that the buffer processor requires. This mechanism is known as \emph{lock passing}. In such a case, the consumer processor would have to wait for the passed locks to return. In \lstinline[language=SCOOP]!buffer.item!, the buffer processor does not require any locks from the consumer processor; hence, the consumer processor does not have to wait due to lock passing. However, the runtime system ensures that the result of the call \lstinline[language=SCOOP]!buffer.item! is properly assigned to the entity \lstinline[language=SCOOP]!consumed_item! using a mechanism called \emph{wait by necessity}: while the consumer processor usually does not have to wait for an asynchronous call to finish, it will do so if it needs the result.

The main part of the paper defines formally the implementation
that gives rise to the behavior outlined above. It also introduces
advanced concepts and additional language elements, which cannot be
covered in a brief overview, and shows how these give rise to a
complexity which can only be dealt with satisfactorily with a formal
specification.

\begin{fortechnicalreport}
\section{Related work}\label{sec:related-work}
\end{fortechnicalreport}
\begin{forjournal}
\section{Related Work}\label{sec:related-work}
\end{forjournal}
The discussion is divided into work on SCOOP and work on other languages.

\subsection{Approaches for SCOOP}
In his dissertation, Nienaltowski~\cite{nienaltowski:2007:SCOOP} worked out
the details of an implementation of SCOOP as suggested by
Meyer~\cite{meyer:1997:oosc}, and provided a prototype
implementation. The language semantics is described informally only,
with the exception of the type system which is defined using
an inference system. The informal description and the prototype
contain various ambiguities and omissions, which we are able to
clarify.

Torshizi et al.~\cite{torshizi-ostroff-paige-chechik:2009:JSCOOP} have defined and
implemented JSCOOP, a version of the SCOOP model for the Java
language. Only the most important language elements are considered,
and no attempt at formalization is made. In contrast, our
specification and implementation~\cite{eth:2011:SCOOP} on top of Eiffel
considers all language elements. We believe that our specification
could help to extend JSCOOP to a full treatment of the language
concepts.

Brooke, Paige and
Jacob~\cite{brooke-paige-jacob:2007:formal_semantics_for_SCOOP} have
used CSP~\cite{hoare:1985:CSP} to give a semantics to SCOOP as
described by Meyer~\cite{meyer:1997:oosc}. Their initial hope was to
use tools for analyzing CSP specifications, such as FDR, to
automatically check for deadlock in SCOOP programs, but found the size
of the specification prohibitive. A benefit of their approach is that
CSP provides the machinery needed to express concurrency and
synchronization, leading to a relatively concise model. Our goal is to
provide formal descriptions close to an actual implementation, and
therefore prefer to design an own operational semantics, rather than
going through the indirection of another process algebra.

Structural operational semantics, introduced by
Plotkin~\cite{plotkin:2004:structural_operational_semantics}, is a flavor of operational
semantics that has been used with great success to define various
concurrent systems. Our specification uses this style of semantics as
well. To model SCOOP we also make use of established modeling concepts
from process algebra, such as the notion of channels, which is present
in most calculi such as CSP~\cite{hoare:1985:CSP} or the
$\pi$-calculus~\cite{milner:1999:Pi_calculus}. We use the theory of abstract data types (ADT) 
\cite{liskov-zilles:1974:abstract_data_types} to model the elements of a program text and to model the
state of a SCOOP program.

Ostroff et
al.~\cite{ostroff-torshizi-huang-schoeller:2008:formal_semantics_for_SCOOP}
describe a structural operational semantics for SCOOP in the refined
version by Nienaltowski~\cite{nienaltowski:2007:SCOOP}. This
operational semantics inspired our work, and we have attempted to stay
close to their modeling ideas where possible, so that
\cite{ostroff-torshizi-huang-schoeller:2008:formal_semantics_for_SCOOP}
can be viewed as a reduced version of the semantics we describe in
this paper. While
\cite{ostroff-torshizi-huang-schoeller:2008:formal_semantics_for_SCOOP}
covers some of the most significant aspects of SCOOP, it falls short
of describing a number of other critical language concepts: in their
reduced model, a query routine handled by some processor $p$ must not
make calls to a processor other than $p$; lock passing, expanded
objects and the import mechanism, once routines, evaluation of
(asynchronous) postconditions and invariants, and explicit processor
tags are not considered. We clarify these aspects in this
paper. Furthermore,
\cite{ostroff-torshizi-huang-schoeller:2008:formal_semantics_for_SCOOP}
have pursued the goal to check temporal logic properties of SCOOP
programs using their semantics and the SPIN model checker, but were
limited to small programs by state space explosion. We have the
different goal of providing a reference document for SCOOP, and thus
don't have to sacrifice coverage of the language for keeping the
specification small.

\begin{fortechnicalreport}
\subsection{Approaches for other concurrent programming languages}
\end{fortechnicalreport}
\begin{forjournal}
\subsection{Approaches for Other Concurrent Programming Languages}
\end{forjournal}
Axum \cite{microsoft:2011:Axum} is a concurrent programming language based on the actor model. In Axum, actors are called agents. An agent is an isolated runtime component that executes in parallel with other agents. The agents communicate with each other by sending messages through channels. Each channel has input ports, output ports, and a protocol. The ports are queues of messages. The protocol is a state machine that defines how the channel behaves. Schemas define the structure of messages. Besides message passing, Axum also provides domains -- shared state between groups of actors. Erlang \cite{ericsson:2011:Erlang} and Scala \cite{odersky:2011:Scala} are further examples of actor-based programming languages.

C$\omega$ \cite{benton-cardelli-fournet:2004:C_omega} is an extension of C\# that integrates elements of the Join Calculus \cite{fournet-gonthier:1996:Cham}. C$\omega$ allows computations to be spawned off into different threads using asynchronous methods: while for synchronous methods the caller must wait until a routine completes, asynchronous methods return immediately while their body is scheduled for execution in another thread. C$\omega$ supports so-called chords, which associate the body of a routine with more than one method; the body is executed only if all methods have been called.

Another language is Cilk \cite{blumofe-joerg-kuszmaul-leiserson-randall-zhou:1995:Cilk}, which extends C with concurrency concepts. A method marked with the cilk keyword can be asynchronously spawned with the spawn keyword. The sync keyword requires the current method to wait for all previously spawned tasks to complete. An inlet function within a parent method receives the result of a spawned child method; the inlet functions of a parent method are guaranteed to execute atomically. Within an inlet function, the abort keyword tells the scheduler that any other child method spawned by the parent method can be aborted. Cilk also implements a work stealing mechanism to achieve high performance by dividing method executions efficiently among processors.

Ada \cite{iso-iec:1995:Ada} defines tasks -- units that can run in parallel. A task is declared within a procedure; it consists of a specification and an implementation. The task is activated when the procedure starts executing. The task specification can define a number of entry points with parameters; an entry point specifies an action the task can synchronize on. An accept statement within the task body indicates the point where the rendezvous can take place. Another task calls the entry point to take part in the rendezvous. With a select statement, one can wait for multiple entry points; alternatives may be guarded with boolean expressions. Ada defines protected objects -- a monitor-like construct with guards instead of conditional variables. A protected object is declared within a procedure; it has a specification and an implementation.

The occam programming language \cite{thomson:1995:occam} builds on the CSP process algebra \cite{hoare:1985:CSP}. A parallel construct defines a number of processes that execute concurrently; the parallel construct terminates when all spawned processes terminated. Processes communicate with each other through named channels. The alternation construct defines a number of processes, where only one of them gets executed; a guard defines when a process can be executed.

X10 \cite{charles-grothoff-saraswat-donawa-kielstra-ebcioglu-vonPraun-sarkar:2005:X10}, Fortress \cite{allen-chase-luchangco-maessen-steele:2004:Fortress}, and Chapel \cite{joyner-chamberlain-deitz:2006:Chapel} are based on the Partitioned Global Address Space (PGAS) model. PGAS uses a global shared memory. It defines portions on the global shared memory and associates them to specific processors to improve performance and scalability. X10 provides important abstractions such as places, asynchronous methods, future invocations, and barriers. However, it places a considerable burden on programmers. Fortress offers implicit parallelization of loops and operations on data structures. Chapel provides a higher-level multithreaded parallel programming model with abstractions for data parallelism, task parallelism, and nested parallelism.

Linda \cite{gelernter-carriero-chandran-chang:1985:Linda} is a coordination language to connect concurrent components; the components can be written in different programming languages. The coordination is based on a tuple space, which holds data tuples that can be stored and retrieved by the processes. Pattern matching is used to read and remove tuples; the operations block until a matching tuple is found. The eval construct creates a new process to evaluate an expression; the new process writes the evaluation result into the tuple space. Implementations of Linda can be found in several programming languages such as Java and C.

For the related languages mentioned above, we are not aware of rigorous behavioral specifications, with the exception of C$\omega$ and occam, which use the Join Calculus respectively CSP as the underlying model. For multi-threaded Java however, such formalizations have been attempted.

\'{A}brah\'{a}m, de Boer, de Roever, and Steffen \cite{abraham-deBoer-deRoever-Steffen:2003:formal_semantics_for_JavaMT} present an operational semantics for a subset of multi-threaded Java. They focus on the most important multi-threaded aspects, i.e., dynamic thread creation, thread termination, and re-entrant monitors. The semantics consists of two components: the semantics for isolated objects and the semantics for interacting objects. The authors want to use the semantics to develop a proof system that is based on an existing proof-system for isolated objects. A configuration is a set of instance configurations. An instance configuration contains the attribute values of one object. It also contains the local environment and the expression of each thread that is concurrently executing within the object. In modeling the state of a program, our semantics strictly separates the actions to be executed from the data. This makes it easier to derive implementations from the semantics because an implementation is likely to keep the program text and data separate. \'{A}brah\'{a}m et al. use transition labels to synchronize inference rules. The labels allow an external observer to follow the transitions. Our semantics is a pure reduction semantics without labels because we do not require observable transitions. 

Cenciarelli, Knapp, Reus, and Wirsing \cite{cenciarelli-knapp-reus-wirsing:1999:formal_semantics_for_Java_threads} also describe an operational semantics for a larger subset of multi-threaded Java. They cover a larger number of multi-threaded aspects than \cite{abraham-deBoer-deRoever-Steffen:2003:formal_semantics_for_JavaMT}. In particular they formalize Java's notification mechanism and the working memory. A configuration consists of a function that maps each thread to its expression and its local environments. The configuration also has a container with the objects and the static typing information. Lastly, the configuration consists of an event space. The event space is a partially ordered set of events that have been executed by the threads. The ordering reflects the order in which the events took place. An event space serves two purposes. First, it contains certain aspects of the state. For example, the lock and unlock actions tell us which thread owns which lock. Second, it records the history. A number of constraints state when an event space is valid. Hence, the event space indicates which further actions can take place. The authors use two different validity constraints for both Java's non-prescient semantics and its prescient semantics. Using this, they show that any prescient execution of a properly synchronized program can be simulated by a non-prescient execution. Compared to our semantics, there is no clean division between program text and the state and there is no clean division between the state and the typing information.

Lochbihler \cite{lochbihler:2008:formal_semantics_for_Java_threads} suggest a different operational semantics for a large subset of multi-threaded Java. Just like \cite{cenciarelli-knapp-reus-wirsing:1999:formal_semantics_for_Java_threads}, he covers the notification mechanism, but he does not formalize the working memory. He defines an instantiating semantics based on an extension of Jinja \cite{klein-nipkow:2006:Jinja}. Jinja is an operational semantics for a subset of single-threaded Java. The instantiating semantics is used for the sequential case. Lochbihler defines a generic formal framework to lift the instantiating semantics to the concurrent case. The configuration of the instantiating semantics consists of the expression, a container with the objects, and the local environments. The state of the framework semantics consists of the lock status, the thread information with the thread's expression along with the thread's local environments, a container with the objects, and the wait sets. Lochbihler formalizes the notion of deadlocks, where deadlocks are either based on locks or on wait sets. He then proves that every program that satisfies certain criteria either produces a final value, throws an exception, or deadlocks. He also shows that every such program preserves type safety.

\begin{fortechnicalreport}
\section{Language overview}\label{sec:language-overview}
\end{fortechnicalreport}
\begin{forjournal}
\section{Language Overview}\label{sec:language-overview}
\end{forjournal}
SCOOP is a programming language based on Eiffel, an object-oriented programming language, defined in the Eiffel ECMA standard \cite{ecma:2006:Eiffel}. SCOOP's concurrency model can be applied to other programming languages as well. For this reason, this work does not focus on SCOOP, but on its concurrency model. This section defines a subset of SCOOP, reduced to the parts that are relevant for the concurrency model. It presents the syntax of the subset and a list of simplifications. It then discusses the program representation that this formalization assumes.

\subsection{Syntax}
The following EBNF grammar defines the set of all considered programs:

\begin{lstlisting}[language=SCOOP_grammar, escapechar=\%]
program = {class_declaration} root_procedure_declaration ;
root_procedure_declaration = "{" class_name "}" "." feature_name ;
class_declaration =
  ["expanded"] "class" class_name
    ["create" feature_name {"," feature_name}]
    "feature" ["{" class_name {"," class_name} "}"] {feature_declaration}
    ["invariant" expression]
  "end" ;

feature_declaration = routine_declaration | attribute_declaration ;
routine_declaration =
  feature_name ["(" entity_declaration {";" entity_declaration} ")"] [":" type]
    ["require" expression]
    ["local" {entity_declaration}]
    ("do" | "once")
      {instruction [";"]}
    ["ensure" expression]
    "end" ;
attribute_declaration = entity_declaration ;
entity_declaration = entity_name ":" type ;

instruction =
  entity_name ":=" expression |
  expression "." feature_name ["(" expression {"," expression} ")"] |
  "create" entity_name "." feature_name ["(" expression {"," expression} ")"] |
  "if" expression "then" {instruction [";"]} "else" {instruction [";"]} "end" |
  "until" expression "loop" {instruction [";"]} "end" ;
	
expression =
  literal |
  entity_name |
  expression "." feature_name ["(" expression {"," expression} ")"] ;
literal = boolean_literal | integer_literal | character_literal | void_literal ;
boolean_literal = "True" | "False" ;
integer_literal = ["-"]("0" | %$\ldots$% | "9") {"0" | %$\ldots$% | "9"} ;
character_literal = "% ' %" "a" | %$\ldots$% | "z" | "A" | %$\ldots$% | "Z" | "0" | %$\ldots$% | "9" "% ' %" ;
void_literal = "Void" ;

type =
  [detachable_tag]
  ["separate"] [explicit_processor_specification]
  class_name [actual_generics] ;
detachable_tag =
  "attached" | "detachable" ;
explicit_processor_specification =
  qualified_explicit_processor_specification |
  unqualified_explicit_processor_specification ;
qualified_explicit_processor_specification =
  "<" entity_name "." "handler" ">" ;
unqualified_explicit_processor_specification =
  "<" entity_name ">" ;
	
class_name = name ;
feature_name = name ;
name = ("a" | %$\ldots$% | "z" | "A" | %$\ldots$% | "Z") {"a" | %$\ldots$% | "z" | "A" | %$\ldots$% | "Z"};
entity_name = feature_name | "Result" | "Current" ;
\end{lstlisting}

A \emph{class} consists of a number of features. A \emph{feature} is either a \emph{routine} -- a sequence of instructions -- or an \emph{attribute} -- a data storage. If a routine returns a result, then it is called a \emph{function}; otherwise, it is called a \emph{procedure}. If a routine is marked as a once routine (\lstinline[language=SCOOP]!once! keyword), then the routine gets executed only once in a given context. Functions and attributes are also called \emph{queries}; routines are also called \emph{commands}. A routine can define a \emph{precondition} (\lstinline[language=SCOOP]!require! keyword) and a \emph{postcondition} (\lstinline[language=SCOOP]!ensure! keyword). The enclosing class can define an \emph{invariant} (\lstinline[language=SCOOP]!invariant! keyword). Each feature can be exported to a list of classes, so that only these classes can use the feature. A number of procedures are dedicated \emph{creation procedures}. These procedures can be used in creation expression (\lstinline[language=SCOOP]!create! keyword) to create new objects. A class can be marked as an \emph{expanded class} (\lstinline[language=SCOOP]!expanded! keyword). Objects of expanded classes get copied when they get passed around; objects of non-expanded classes get aliased.

Formally, a type $t$ is a triple $(d, p, c)$. The component $d$ is the \emph{detachable tag}, $p$ is the \emph{processor tag}, and $c$ is the \emph{class type}. The detachable tag $d$ captures detachability. An entity of attached type (\lstinline[language=SCOOP]!attached! keyword), i.e., $d =\ !$, is statically guaranteed to store a value, i.e., to be non-void. An entity of detachable type (\lstinline[language=SCOOP]!detachable! keyword), i.e., $d =\ ?$, can be void. As discussed later, the detachable tag is also used for the selective locking mechanism to prevent a request queue from being locked. The processor tag $p$ captures the locality of objects accessed by an entity of the type $t$. The processor tag $p$ can be separate (\lstinline[language=SCOOP]!separate! keyword without explicit processor specification), i.e., $p = \top$. The object attached to the entity of the type $t$ is potentially handled by a different processor than the current processor. The processor tag $p$ can be explicit (\lstinline[language=SCOOP]!separate! keyword with explicit processor specification), i.e., $p = \alpha$. The object attached to the entity of the type $t$ is handled by the processor specified by $\alpha$. The processor tag $p$ can be non-separate (no \lstinline[language=SCOOP]!separate! keyword), i.e. $p = \bullet$. The object attached to the entity of the type $t$ is handled by the current processor. The processor tag $p$ can denote no processor, i.e., $p = \bot$. It is used in the type of the void reference. The explicit processor tag either has an unqualified or a qualified specification. An \emph{unqualified explicit processor specification}, i.e., $<p>$, is based on a processor attribute $p$. The processor attribute $p$ must have the type $(!, \bullet, \mathit{PROCESSOR})$ and it must be declared in the same class as the explicit processor specification or in one of the ancestors. The processor denoted by this explicit processor specification is the processor stored in $p$. A \emph{qualified explicit processor specification}, i.e., $<e.\handlerfeature>$, relies on an entity $e$ occurring in the same class as the explicit processor specification or in one of the ancestors. The entity $e$ must be a non-writable entity of attached type and the type of $e$ must not have a qualified explicit processor tag. The processor denoted by this explicit processor specification is the same processor as the one of the object referenced by $e$. Explicit processor tags support precise reasoning about object locality. Entities declared with the same processor tag represent objects handled by the same processor. The absence of both the keywords is treated as if there was an \lstinline[language=SCOOP]!attached! keyword.

\subsection{Simplifications}
This work makes the following simplifications:
\begin{itemize}
	\item It does not consider unqualified feature calls. It expects all feature calls to be in the qualified form. This includes accesses to attributes of the current object in expressions.
	\item It does not consider infix feature calls. It expects all feature calls in the non-infix form. For example, an expression \lstinline[language=SCOOP]!x > y! must be transformed into the equivalent form \lstinline[language=SCOOP]!x.is_greater(y)!.
	\item It simplifies the automatic initialization of entities. All entities, except for the current object entity, are initialized with the void reference.
	\item It neglects exception handling. The exception handling mechanism for SCOOP is still under development.
	\item It does not consider garbage collection because garbage collection is not refined in the SCOOP model.
	\item It does not consider agents. From this work's point of view, agents are normal objects.
\end{itemize}

\begin{figure}
  \centering
  \includegraphics[width=\textwidth]{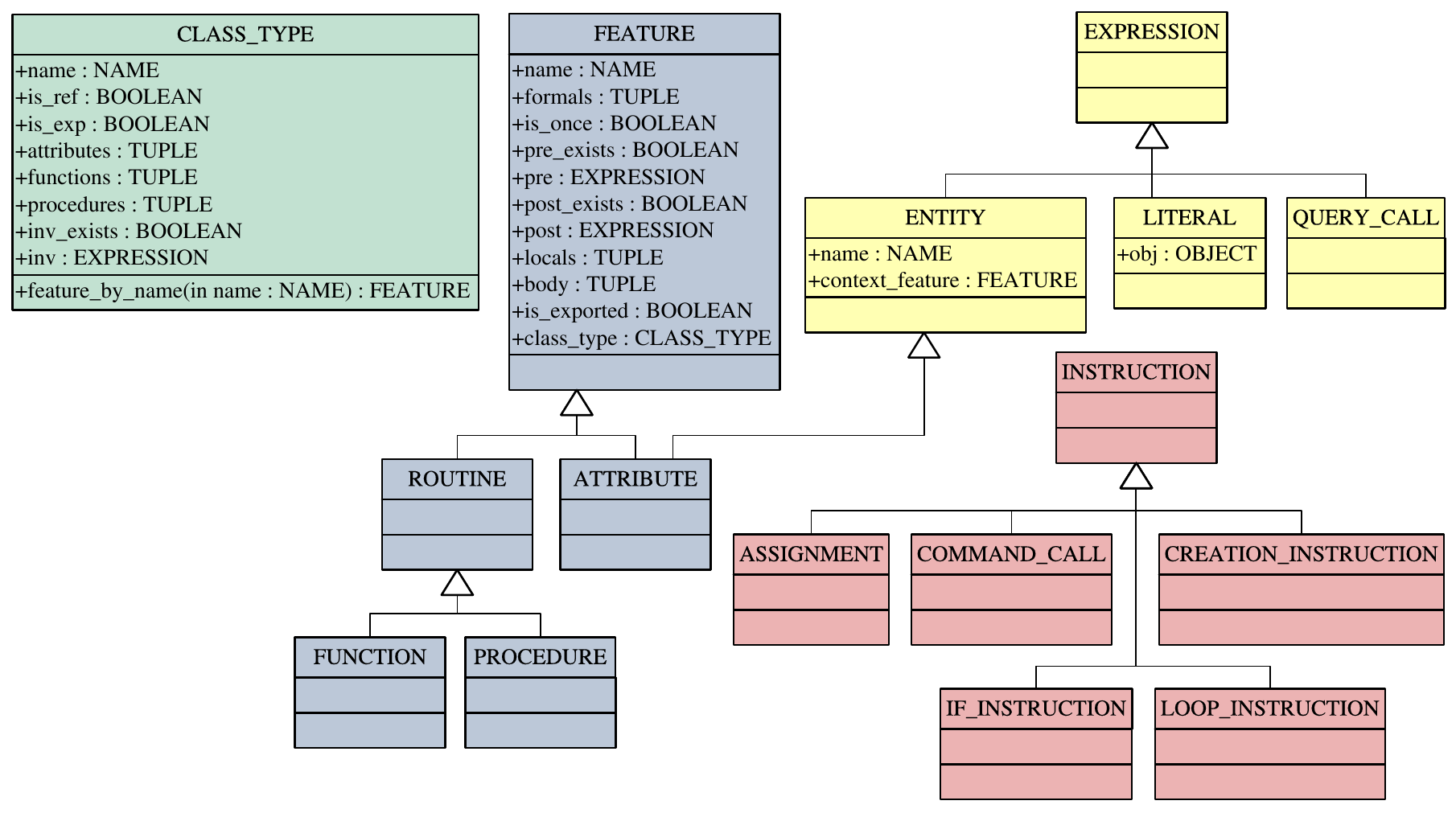}
  \caption{ADTs for the intermediate representation}
  \label{fig:intermediate representation ADTs}
\end{figure}

\begin{fortechnicalreport}
\subsection{Intermediate representation}
\end{fortechnicalreport}
\begin{forjournal}
\subsection{Intermediate Representation}
\end{forjournal}
For the purpose of the formalization, this work assumes that a program is given in an enriched intermediate representation, where the syntactical elements are replaced with instances of abstract data types. In particular, it assumes ADTs for class types, features, expressions, and instructions. \figurereference{fig:intermediate representation ADTs} summarizes these ADTs.
The instances of $\classtypetype$ are all possible class types, i.e., the types directly defined by all non-generic classes and all possible generic derivations of all possible generic classes. \sectionreference{sec:typing environment} discusses how to get these instances. The ADT $\classtypetype$ defines a name query $\namefeature$. Each class type can either be a reference class type or an expanded class type. The queries $\isreferenceclasstypefeature$ and $\isexpandedclasstypefeature$ provide this information. Each class type defines a number of features. These features can be divided into attributes, functions, and procedures. An attribute of an object stores a value. A function performs a computation and returns the result. This computation must not modify the state. A procedure performs a computation that modifies the state. Functions and procedures are also known as routines. For each of these categories, $\classtypetype$ defines a query that returns a tuple of features. The query $\attributesfeature$ returns a tuple of attributes, the query $\functionsfeature$ returns a tuple of functions, and the query $\proceduresfeature$ returns a tuple of procedures. If the name of a feature is known, then the query $\featurebynamefeature$ can be used to get the feature with that name. Each class type can have an invariant. The query $\invariantexistsfeature$ indicates whether such an invariant exists. In case an invariant exists, it can be accessed with the query $\invariantfeature$ as an expression. One of the instances of $\classtypetype$ is $\booleanclasstype$. This class type is expanded and it has an attribute with name $\booleanclasstypeitemattributename$. The value of this attribute is the represented boolean value, i.e., an instance of $\booleantype$.

In this formalization, a feature is an instance of $\featuretype$. The name of the feature can be retrieved with the query $\namefeature$ and the formal arguments are given by the query $\formalargumentsfeature$ that returns a tuple with the formal arguments as entities. Whether or not the feature is a once feature can be determined using the query $\isonceroutinefeature$. The queries $\preconditionfeature$ and $\postconditionfeature$ return an expression for the precondition respectively the postcondition, provided that the queries $\preconditionexistsfeature$ and $\postconditionexistsfeature$ indicate that the assertions exist. Next, there is the query $\localsfeature$ that gives the locals of the feature as entities. The query $\bodyfeature$ returns the body of the feature as a tuple of instructions. Each feature is either exported or not. A non-exported feature is only available in calls on the current object within the class that declared the feature. An exported feature can be called by other clients as well. The query $\isexportedfeature$ returns whether a feature is exported or not. Lastly, each feature has a link to the class it belongs to. This is given by the query $\classtypefeature$. This can be used for example to retrieve the invariant that must be preserved by a feature. For each feature category, there is an ADT that inherits from the $\featuretype$ ADT.

Expressions can either be entities, literals, or query calls. Every expression is an instance of $\expressiontype$. For each form of expression, there is one ADT that inherits from the $\expressiontype$ ADT. For entities there is an ADT with a query $\namefeature$ that returns the name of an entity. A query $\contextfeaturefeature$ links an entity to the feature in which the entity is declared. A literal is a character sequence that represents a constant value. As such, literals count as manifest expressions - programming constructs whose values can be deduced by the compiler statically. Literals are instances of an ADT $\literaltype$. This ADT has instances for booleans, integers, characters, and the void literal. Each literal except the void literal can be translated into an object with the query $\objectfeature$. This object matches the literal in both type and value. The following notation describes instances of $\expressiontype$:

\isolateddefinition{e \triangleq w \thickspace | \thickspace b \thickspace | \thickspace e.f(e, \ldots, e)}%

Here, $w$ is an element of $\literaltype$, $b$ is an instance of $\entitytype$, and $f$ is an instance of $\featuretype$. For instructions, there is an ADT $\instructiontype$ and an ADT for each kind of instruction. The following notation describes such instances:

\isolateddefinition%
	{%
		& h \triangleq \\
		& \indentation \begin{split}
			& b \eassignment e \thickspace | \\
			& e.f(e, \ldots, e) \thickspace | \\
			& \ecreate b.f(e, \ldots, e) \thickspace | \\
			& \eif{e}{[s \{; s\}*]}{[s \{; s\}*]} \thickspace | \\
			& \euntil{e}{[s \{; s\}*]}
		\end{split}
	}%

Here, $s$ stands either for an instance of $\instructiontype$ or an operation. Instructions are actions that occur in the intermediate representation (user syntax). Operations are actions that do not explicitly occur in the intermediate representation (run-time syntax).

This work builds on an existing type system formalization. It assumes the existence of a typing environment that can be queried for type information.

\begin{fortechnicalreport}
\subsection{Basic ADTs}\label{sec:basic ADTs}
This work assumes an ADT $\booleantype$ for the two boolean values $\truevalue$ and $\falsevalue$ along with the typical boolean operators. It assumes an ADT $\nametype$ for names and an ADT $\identifiertype$ for identifiers.

Based on these ADTs, there is a generic ADT $\settype{\mathbf{G}}$, whose instances are sets of elements with type $\mathbf{G}$. A set can be created with a call to the constructor $\makefeature$ and it can be populated with a call to the command $\addfeature$ that takes an element and returns a new set containing the element. However, we prefer to define an instance of this ADT with set inference or with a set expression $\set{x_{1}, \ldots, x_{n}}$. A set can be inspected with the query $\containsfeature$ that returns whether an element is part of the set or not. The $\in$ notation is an alias of the $\containsfeature$ query. A query $\isemptyfeature$ tells us whether the set is empty or not, and the query $\countfeature$ returns the number of elements in the set.

A tuple is an ordered list of elements. This work assumes a generic ADT $\tupletype{}$ that takes zero or more generic parameters for the types of each tuple element. Each generic parameter $\mathbf{G}$ restricts the set of possible tuples to the ones who have an element of type $\mathbf{G}$ in the respective position. For example, the instances of $\tupletype{}$ without any generic parameters are all possible tuples. The instances of $\tupletype{A}$ are all tuples whose first elements are of type $A$. Note that this also includes tuples with more than one element, as long as the first element is of type $A$. A tuple can be constructed with a call to the constructor $\makefeature$. Elements can be added to the end of the tuple with a call to the command $\addfeature$. We prefer to construct tuples with tuple expression $\tuple{x_{1}, \ldots, x_{n}}$. The query $\containsfeature$ checks whether an element is part of the tuple or not and the query $\isemptyfeature$ checks whether the tuple is empty or not. The query $\countfeature$ returns the number of elements in the tuple. 

Furthermore, this work assumes a generic ADT $\stacktype{\mathbf{G}}$ for stacks with elements of type $\mathbf{G}$. A stack gets constructed with a call to $\makefeature$. An element gets pushed with a call to $\pushfeature$ and is then available through the query $\topfeature$. The element can then be popped with a call to $\popfeature$. A query $\isemptyfeature$ returns whether the stack is empty or not. Another query $\flattenedfeature$ returns a set of type $\settype{\mathbf{G}}$. This set contains all the elements in the stack.

Next, this work assumes an ADT $\maptype{\mathbf{K}}{\mathbf{G}}$ for maps with keys of type $\mathbf{K}$ and values of type $\mathbf{G}$. As the name suggests, a map associates keys to values. A map gets constructed with a call to the constructor $\makefeature$. An association can be added with a call to $\addfeature$ where the first argument is the key and the second argument is the value. The set of all possible keys can be retrieved with a call to the query $\keysfeature$. The value for a specific key in this key set is returned by the query $\valuefeature$.
\end{fortechnicalreport}

\begin{fortechnicalreport}
\section{State formalization}\label{sec:state-formalization}
\end{fortechnicalreport}
\begin{forjournal}
\section{State Formalization}\label{sec:state-formalization}
\end{forjournal}
This section provides a formalization of the state of a SCOOP program. This is necessary to describe the effect of SCOOP constructs on the state. The discussion starts with the general approach and continues with the description of the state.

\begin{fortechnicalreport}
\subsection{General approach}
\end{fortechnicalreport}
\begin{forjournal}
\subsection{General Approach}
\end{forjournal}
This work considers the state of a SCOOP program to be a data structure that can be created, modified, and queried through features. For the specification of the state, this work uses Liskov's ADT theory \cite{liskov-zilles:1974:abstract_data_types}. The discussion begins with a justification and the consequences of this choice. The discussion finishes with an explanation on how to get types for elements in the intermediate representation.

\subsubsection{Abstract data types}
Meyer's work on a three-level approach to the description of data structure \cite{meyer:1981:data_structure_description} defines three levels on which a data structure can be described: functional, constructive, and physical. The functional specification is an algebraic approach that uses an implicit characterization of the data structure. The constructive specification provides a means to construct instances of the data structure. The instances constructed like this are mathematical entities. A physical description describes the layout of instances in memory. The constructive specification can be derived from the functional specification and the physical description can be derived from the constructive description.

This work models the state as an ADT instance, on the functional level in the hierarchy described above. This has several reasons. First of all, ADT theory allows us to describe the state on an abstract level without dealing with aspects of the implementation. The constructive and the physical level can be derived from the ADTs on the functional level. Second, ADT theory allows us to modularize the state. Different concerns of the state can be modeled as individual ADTs, while a single ADT can be used to consolidate the individual ADTs. This improves understandability and maintainability of the state description. Lastly, ADT theory is well established and suitable for the task at hand.

An ADT $t$ consists of queries, commands, and constructors. A query of $t$ provides information about an instance of $t$. The query takes as a first argument the target of type $t$, which is the instance to be queried. Next to the target, the query can take further arguments with types $t_{1}, \ldots, t_{n}$. Finally, the query returns a result of a type $t_{n + 1}$. The declaration of this query is written as $\mathit{query} \colon t \rightarrow t_{1} \rightarrow \ldots \rightarrow t_{n} \rightarrow t_{n + 1}$. For flexibility reasons, this work uses the curried form (as in Haskell) instead of the equivalent Cartesian form $\mathit{query} \colon t \times t_{1} \times \ldots \times t_{n} \rightarrow t_{n + 1}$. A command of $t$ returns an updated instance according to the command's semantics. The declaration of a command looks much like the one of a query. However, the result of the command is an instance of $t$. To simplify the discussions, the following terminology is used: an \emph{update of an ADT instance} is the act of calling a command on the instance; the \emph{updated instance} is the result of the command. A constructor of $t$ creates a new instance of $t$. In contrast to queries and commands, a constructor does not take the target as the first argument because its purpose is to create a new instance.

To describe an instance of an ADT, one can build an expression that starts with a constructor call. This expression can then be used as the first actual argument of a command call. The resulting expression can then be used as the first actual argument of the next command. This leads to a nested expression, in which the first feature call is in the root of the expression and the last feature call is on the outside of the expression. The instance described in such a way can then be queried. We find this functional notation hard to read. Therefore we use an equivalent object-oriented notation in which the first feature call is on the left and the last feature call is on the right. The main idea is not to write targets as arguments, but to write a target in front of the feature name and to use a dot to separate the two parts from each other. This leads to the following translation between the functional notation and the object-oriented notation:
\begin{itemize}
	\item The query expression $\mathit{query}(e_{0}, e_{1}, \ldots, e_{n})$ written in functional notation is equivalent to the expression $e_{0}.\mathit{query}(e_{1}, \ldots, e_{n})$ written in object-oriented notation.
	\item The command expression $\mathit{command}(e_{0}, e_{1}, \ldots, e_{n})$ written in functional notation is equivalent to the expression $e_{0}.\mathit{command}(e_{1}, \ldots, e_{n})$ written in object-oriented notation.
	\item The creation expression $\mathit{constructor}(e_{1}, \ldots, e_{n})$ for an instance of an ADT $t$ written in functional notation is equivalent to the expression $\creation{t}{\mathit{constructor}(e_{1}, \ldots, e_{n})}$ written in object-oriented notation.
\end{itemize}

The identity of an ADT instance is given by its query values. Hence, the following holds for all ADTs $t$: $\creation{t}{\mathit{constructor}(e_{1}, \ldots, e_{n})} = \creation{t}{\mathit{constructor}(e_{1}, \ldots, e_{n})}$.

\begin{example}[Functional notation versus object-oriented notation]
The expression in functional notation $\isemptyfeature(\popfeature(\pushfeature(\creation{\stacktype{\processortype}}{\makefeature}, \mathit{p})))$ can be written in object-oriented notation as $\creation{\stacktype{\processortype}}{\makefeature}.\pushfeature(\mathit{p}).\popfeature.\isemptyfeature$.
\end{example}

Each feature can have a precondition that must be satisfied before the feature gets called. A precondition is expressed as a number of assertions on the target and the arguments. A feature with a precondition is a partial feature. A partial feature is a feature whose domain is restricted. Such a partial feature is indicated with a crossed arrow $\nrightarrow$ after the type of each formal argument that got restricted by the feature's precondition. Non-restricted formal arguments are indicated with a normal arrow $\rightarrow$. The effect of an ADT command is described in a number of axioms. This work deviates from the practice of bundling all axioms for a specific ADT. Instead, all the axioms for a specific feature occur in the feature's declaration. Note that this work does not aspire a sufficiently complete ADT because this would lead to rule explosion. An ADT is sufficiently complete if its axioms make it possible to reduce any query expression to a form that does not involve an instance of the ADT. This requires that the axioms describe the effect of each command on each query. This work follows the practice to describe the effect of each command of an ADT on all the queries of the ADT that have been changed by the command. Unmentioned queries are unchanged.

\begin{example}[Command declaration]
The following declaration shows a command to set the value of an attribute $f$ of an object $\object$ to the value $v$. The value can either be a reference or a processor. The command takes the object as the target and returns an updated object whose attribute value is set.

\feature
	{\setattributevaluefeature \colon \objecttype \rightarrow \attributetype \nrightarrow \referencetype \cup \processortype \rightarrow \objecttype}
	{\object.\setattributevaluefeature(f, v)}
	{\object.\classtypefeature.\attributesfeature.\containsfeature(f)}
	{\object.\setattributevaluefeature(f, v).\attributevaluefeature(f) = v}

The command states in its precondition that the class type of the target object $\object$ must have an attribute $f$. This is expressed as an assertion after the require keyword. The part in front of the require keyword gives names to the target and the arguments. Note that the precondition makes the command partial. The updated object has the value of its attribute $f$ set to $v$. This is stated as an axiom after the ensure keyword.
\end{example}

So far the discussion covered queries, commands, and constructors for ADTs. This work extends the ADT theory with the notion of auxiliary features. Auxiliary features are convenience features that are not essential for the definition of the ADT, but nevertheless useful.

The remainder of this work declares various ADTs to model the state of a SCOOP program. Unless it would create confusion, it uses the same name for an instance of an ADT and the corresponding domain element. For example, the instance of the ADT $\objecttype$ is called an object.

\subsubsection{Identifier management}
This formalization models objects, references, and processors. All of these domain elements have an identity. These identities are automatically managed by the runtime system. The work by Khoshafian and Copeland \cite{khoshafian-copeland:1986:object_identity} on different levels of object identity provides good reasons for this decision. They introduce a scale that starts with identities given by the value, goes on with user-supplied identities, and ends with built-in identities. Built-in identities have the advantage that the identities are preserved in case of modifications. According to this hierarchy, our domain elements have a built-in identity. One straightforward way to reflect this, is to model each domain element as an instance of an ADT. However, this direct approach does not properly capture the identities of the domain elements because the identity of an ADT instance is not built-in, but based on the query values. This section describes a way to introduce built-in identities for ADT instances.

To model domain elements with built-in identities, one can define an ADT with an identifier query. A number of ADT instances represent a single domain element over time. Each of the ADT instances has the same value for the identifier query. A modification of the domain element can then be modeled as a new ADT instance where the value of the identity query is preserved and all other queries modulo the modification are preserved.

For this to work, the formalization ensures that no two ADT instances that model different domain elements have the same identity. This is ensured with a fresh identifier for each ADT instance that models a new domain element. For this purpose, the universal stateful query $\newidentifierfeature$ returns a fresh identifier. The formalization then preserves the identifier in every modification.

\subsubsection{Typing environment}\label{sec:typing environment}
Nienaltowski \cite{nienaltowski:2007:SCOOP} presents a formalization of the SCOOP type system for a core of SCOOP called SCOOP$_{\textrm{C}}$. The type system formalization is part of the base for this work. The \emph{typing environment} $\typingenvironment$ contains the class hierarchy of a SCOOP program along with all the type definitions of all features and entities. Type rules allow us to derive conclusions.

The notation $\typingenvironmentderivation{e: t}$ denotes that expression $e$ is of type $t$. Based on this derivation, the function $\typefromtypingenvironment(\typingenvironment, e)$ denotes the type of expression $e$ in the typing environment $\typingenvironment$. The type rules can be used to check whether an expression is controlled or not. In a SCOOP program, each processor $p$ that wants to apply a feature $f$ must make sure that all the processors $\tuple{q_{1}, \ldots, q_{n}}$ of all attached actual arguments of $f$ are exclusively available on behalf of processor $p$. This guarantees exclusive access on all objects handled by processors $\set{p, q_{1}, \ldots, q_{n}}$. Note that processor $p$ is in this set too because $p$ can exclusively access its objects during a feature execution. For safety, the type system only allows feature calls in $f$ on expressions, where the type system can derive that the value of the expression is a reference to an object and this object is handled by one of the processors $\set{p, q_{1}, \ldots, q_{n}}$. Such an expression is called \emph{controlled}. Whether or not an expression is controlled can be determined through the context in which the expression appears and the type of the expression. The context can either be the enclosing class, in case of expressions in invariants, or it can be the enclosing feature, in case of all other expressions. To be more precise, an expression $e$ of type $t = (d, p, c)$ is controlled if and only if $t$ is attached, i.e., $d =\ !$, and $t$ satisfies at least one of the following conditions:
\begin{itemize}
	\item The expression $e$ is non-separate, i.e., $p = \bullet$.
	\item The expression $e$ appears in a routine $f$ that has an attached formal argument $w$ with the same handler as $e$, i.e., $p = w.\mathit{handler}$.
\end{itemize}
The second condition is satisfied if and only if at least one of the following conditions is true:
\begin{itemize}
	\item The expression $e$ appears as an attached formal argument of $f$.
	\item The expression $e$ has a qualified explicit processor specification $w.\mathit{handler}$ and $w$ is an attached formal argument of $f$.
	\item The expression $e$ has an unqualified explicit processor specification $p$, and some attached formal argument of $f$ has $p$ as its unqualified explicit processor specification.
\end{itemize}
The notation $\typingenvironmentderivation{\iscontrolledfeature(t)}$ denotes that an expression $e$ of type $t$ is controlled. To establish the derivation $\typingenvironmentderivation{\iscontrolledfeature(t)}$ one has to find an attached formal argument $w$ in the enclosing routine such that the types suggest that $w$ and $e$ are handled by the same processor or one has the establish that the type $t$ is non-separate. One can therefore be sure that whenever an expression $e$ is controlled, either a matching formal argument exists or its type is non-separate. For the first case, the formal argument is the \emph{controlling entity} for $e$. For the second case, the current entity is the controlling entity. Although not present in Nienaltowski's formalization of the type system, this work introduces a new derivation $\typingenvironmentderivation{y = \controllingentityfeature(e)}$ that returns the controlling entity $y$ for an expression $e$ as an instance of $\entitytype$. This notion is essential to determine the handler of any controlled expression without evaluating the expression. One can simply determine the controlling entity and then determine the handler of the controlling entity.

\begin{fortechnicalreport}
\subsection{Components of the state}
\end{fortechnicalreport}
\begin{forjournal}
\subsection{Components of the State}
\end{forjournal}
The state is divided into three parts: the regions, the heap, and the store. The main purpose of the heap is to keep track of objects and to maintain the mapping of references to objects. It also maintains the once status of once routines, i.e., whether a once routine is fresh on a processor. The regions manage the association between objects and processors. Objects that are handled by the same processor form a region. The regions are also concerned with locking. The store is a map of names to references. It maps names of formal argument, names of local variables, the name of the current object entity, and the name of the result entity to references. A state ADT models the state with one query for each of the three parts.

\feature
	{\processorregionsfeature \colon \statetype \rightarrow \processorregionstype}
	{}
	{}
	{}
\feature
	{\heapfeature \colon \statetype \rightarrow \heaptype}
	{}
	{}
	{}
\feature
	{\storefeature \colon \statetype \rightarrow \storetype}
	{}
	{}
	{}

The next few sections introduce ADTs for each of the parts. A later section presents the state ADT.

\subsection{Heap ADT}
The \emph{heap} keeps track of the objects and the references associated to them. It also keeps track of the status of once routines. This section first defines an ADT for objects and references. Then it introduces an ADT for the heap.

\subsubsection{Objects and references}
There are two kinds of class types in the SCOOP type system: reference class types and expanded class types. The main difference lies in the semantics of using an instance of the types as the source of an attachment, such as assignment or argument passing. If an object of \emph{reference class type} is the source of an attachment, then the reference to the object gets copied over to the destination of the attachment. The object is then accessible both through the source of the attachment as well as through the destination of the attachment. If an object of \emph{expanded class type} is the source of an attachment, then a copy of the object gets attached to the destination of the attachment. The details can be found in \externalsectionreference{7.4} of the Eiffel ECMA standard \cite{ecma:2006:Eiffel}.

This formalization takes a unified view on objects and references that is compatible with the semantics described in the Eiffel ECMA standard. It does not consider objects of expanded class type as sub-objects in other objects or in an environment. Instead it locates expanded objects on the heap, just like objects of reference class type. For each object there is exactly one reference. Assigning references to objects of expanded type has one major advantage for the formalization. If an ADT instance $x$ that models an object gets updated, then one gets a new ADT instance $y$. If one would model expanded objects as sub-objects stored in other objects or in environments, then such an update might trigger a cascade of ADT instance updates: each ADT instance that has $x$ as a query value would have to be updated with $y$, and so on. A consequent usage of references avoids this issue. To do the update, one simply alters the reference to $x$ so that it points to $y$ from now on.

The ADT $\referencetype$ models references with an identity query $\identifierfeature$ and a constructor $\makefeature$. The constructor uses the query $\newidentifierfeature$ to create a fresh identifier for the newly created reference. The void reference $\voidvalue$ is an instance of this ADT.

\begin{fortechnicalreport}
\feature
	{\identifierfeature \colon \referencetype \rightarrow \identifiertype}
	{}
	{}
	{}

\feature
	{\makefeature \colon  \referencetype}
	{}
	{}
	{\makefeature.\identifierfeature = \newidentifierfeature}

\end{fortechnicalreport}

The ADT $\objecttype$ models objects. Each object has a query $\identifierfeature$ for its identifier, a query $\classtypefeature$ for its class type, and a query $\attributevaluefeature$ for its attribute values. An object can only have attribute values for attributes that are defined in its class type.

\begin{fortechnicalreport}
\feature
	{\identifierfeature \colon \objecttype \rightarrow \identifiertype}
	{}
	{}
	{} 

\feature
	{\classtypefeature \colon \objecttype \rightarrow \classtypetype}
	{}
	{}
	{}

\feature
	{\attributevaluefeature \colon \objecttype \rightarrow \attributetype \nrightarrow \referencetype \cup \processortype}
	{\object.\attributevaluefeature(f)}
	{\object.\classtypefeature.\attributesfeature.\containsfeature(f)}
	{}

\end{fortechnicalreport}

The attribute values of an object can be modified with the command $\setattributevaluefeature$. Only the attribute values for attributes that are defined in the class type can be modified. The result is an updated object where the attribute value of $f$ is set to $v$. Note that the value can either be a reference or a processor. Processor values are necessary to support processor attributes.

\feature
	{\setattributevaluefeature \colon \objecttype \rightarrow \attributetype \nrightarrow \referencetype \cup \processortype \rightarrow \objecttype}
	{\object.\setattributevaluefeature(f, v)}
	{\object.\classtypefeature.\attributesfeature.\containsfeature(f)}
	{\object.\setattributevaluefeature(f, v).\attributevaluefeature(f) = v}

The constructor $\makefeature$ can be used to create a new object. It creates a new object with the given class type. The new object has a new identifier that is given by the query $\newidentifierfeature$. The constructor initializes all the attribute values of the new object with the void reference.

\feature
	{\makefeature \colon \classtypetype \rightarrow \objecttype}
	{}
	{}
	{
		& \makefeature(c).\identifierfeature = \newidentifierfeature \protect\\
		& \makefeature(c).\classtypefeature = c \protect\\
		& \where
			{\forall {i \in \set{1, \ldots, n}} \colon {\makefeature(c).\attributevaluefeature(a_{i}) = \voidvalue}}
			{\set{a_{1}, \ldots, a_{n}} \mathematicaldefinition c.\attributesfeature}
	}

An object can also be copied with the auxiliary query $\copyfeature$. This is important for expanded objects with copy semantics. The copied object has the same class type and the same attribute values as the original object, but it has a new identity. The new identity comes from the call to the constructor $\makefeature$.

\feature
	{\copyfeature \colon \objecttype \rightarrow \objecttype}
	{}
	{}
	{
		\where
			{
				& \object.\copyfeature = \makefeature(\object.\classtypefeature) \protect\\
				& \indentation \protect\begin{split}
					& .\setattributevaluefeature(a_{1}, o.\attributevaluefeature(a_{1})) \protect\\
					& .\thickspace \ldots \protect\\
					& .\setattributevaluefeature(a_{n}, o.\attributevaluefeature(a_{n}))
				\protect\end{split}
			}
			{
				& n \mathematicaldefinition \object.\classtypefeature.\attributesfeature.\countfeature \protect\\
				& \set{a_{1}, \ldots, a_{n}} \mathematicaldefinition \object.\classtypefeature.\attributesfeature
			}
	}

\subsubsection{Mapping from references to objects}
The ADT $\heaptype$ makes use of $\objecttype$ and $\referencetype$ to model the mapping from references to objects. For this purpose, it declares the query $\objectsfeature$ to store all the objects on the heap and it declares the query $\referencesfeature$ to get all the references to these objects. The reference $\voidvalue$ is not part of the reference set. The query $\referencedobjectfeature$ defines the actual mapping. For each reference in $\referencesfeature$ an object in $\objectsfeature$ gets returned. The ADT also declares the query $\lastaddedobjectfeature$ to keep track of the last object that has been added to the heap. It uses this query to define the effect of adding an object to the heap.

\feature
	{\objectsfeature \colon \heaptype \rightarrow \settype{\objecttype}}
	{}
	{}
	{}

\feature
	{\referencesfeature \colon \heaptype \rightarrow \settype{\referencetype}}
	{}
	{}
	{}

\feature
	{\referencedobjectfeature \colon \heaptype \rightarrow \referencetype \nrightarrow \objecttype}
	{\heap.\referencedobjectfeature(r)}
	{\heap.\referencesfeature.\containsfeature(r)}
	{}

\feature
	{\lastaddedobjectfeature \colon \heaptype \rightarrow \objecttype}
	{\heap.\lastaddedobjectfeature}
	{\neg \heap.\objectsfeature.\isemptyfeature}
	{}

A number of commands are responsible for adding objects and for altering the mapping of references to objects. The command $\addobjectfeature$ takes an object $o$ and adds it to the heap. The result of the command is a new heap with the object $o$ and a new reference that points to $o$. The newly added object is indicated in the query $\lastaddedobjectfeature$. Note that this command does not create a new object. It simply adds an object that has been provided as an argument. The command requires that the object is not yet part of the heap.

\feature
	{\addobjectfeature \colon \heaptype \rightarrow \objecttype \nrightarrow \heaptype}
	{\heap.\addobjectfeature(o)}
	{
		& \forall u \in \heap.\objectsfeature \colon u.\identifierfeature \neq o.\identifierfeature \protect\\
		& \protect\begin{split}
			& \forall a \in o.\classtypefeature.\attributesfeature \colon \protect\\
			& \indentation o.\attributevaluefeature(a) \in \referencetype \rightarrow (o.\attributevaluefeature(a) = \voidvalue \vee \heap.\referencesfeature.\containsfeature(o.\attributevaluefeature(a)))
		\protect\end{split}
	}
	{
		& \where
			{
				& \heap.\addobjectfeature(o).\objectsfeature = \heap.\objectsfeature \cup \set{o} \protect\\
				& \heap.\addobjectfeature(o).\referencesfeature = \heap.\referencesfeature \cup \set{r} \protect\\
				& \heap.\addobjectfeature(o).\referencedobjectfeature(r) = o \protect\\
				& \heap.\addobjectfeature(o).\lastaddedobjectfeature = o
			}
			{
				& r \mathematicaldefinition \creation{\referencetype}{\makefeature}
			}
	}

If an object that is already part of the heap gets updated, then it is necessary to update the mapping from the reference to the object on the heap. This can be done with the command $\updatereferencefeature$ that takes a reference $r$ and an updated object $o$ and returns a heap where the reference $r$ points to $o$. The command requires that $r$ is a valid reference and that $o$ is an updated version of the original object. Because the remaining part of the state only deals with references rather than objects directly, a reference update does not require an update of these parts.

\feature
	{\updatereferencefeature \colon \heaptype \rightarrow \referencetype \nrightarrow \objecttype \nrightarrow \heaptype}
	{\heap.\updatereferencefeature(r, o)}
	{
		& \heap.\referencesfeature.\containsfeature(r) \protect\\
		& o.\identifierfeature = \heap.\referencedobjectfeature(r).\identifierfeature \protect\\
		& \protect\begin{split}
			& \forall a \in o.\classtypefeature.\attributesfeature \colon \protect\\
			& \indentation o.\attributevaluefeature(a) \in \referencetype \rightarrow (o.\attributevaluefeature(a) = \voidvalue \vee \heap.\referencesfeature.\containsfeature(o.\attributevaluefeature(a)))
		\protect\end{split}
	}
	{
		& \heap.\updatereferencefeature(r, o).\objectsfeature.\containsfeature(o) \protect\\
		& o \neq \heap.\referencedobjectfeature(r) \rightarrow \neg \heap.\updatereferencefeature(r, o).\objectsfeature.\containsfeature(\heap.\referencedobjectfeature(r)) \protect\\
		& \heap.\updatereferencefeature(r, o).\referencedobjectfeature(r) = o \protect\\
		& \heap.\lastaddedobjectfeature = \heap.\referencedobjectfeature(r) \rightarrow \heap.\updatereferencefeature(r, o).\lastaddedobjectfeature = o
	}

So far $\heaptype$ covers the mapping from references to objects. Occasionally it is necessary to have the inverse mapping. The commands $\addobjectfeature$ and $\updatereferencefeature$ ensure that there is exactly one reference for each object on the heap. Thus it is possible to define the inverse query $\referencefeature$ as an auxiliary query.

\feature
	{\referencefeature \colon \heaptype \rightarrow \objecttype \nrightarrow \referencetype}
	{\heap.\referencefeature(o)}
	{\heap.\objectsfeature.\containsfeature(o)}
	{
		\heap.\referencedobjectfeature(\heap.\referencefeature(o)) = o
	}

\subsubsection{Once routines}
A \emph{once routine} can either be a once function or a once procedure. A once routine gets executed at most once in a certain context. If a once routine has been executed in the context, then it is called \emph{non-fresh} in the context. Otherwise it is called \emph{fresh} in the context. The context is either the set of all processors in the system or a single processor. The heap remembers which once routines are fresh. For this purpose, $\heaptype$ declares the queries $\isonceroutinefreshfeature$ and $\oncefunctionresultfeature$. For any processor $p$ and any once routine $f$, the query $\isonceroutinefreshfeature$ states whether $f$ is fresh on $p$ or not. For a once function $f$ that is not fresh on a processor $p$, the query $\oncefunctionresultfeature$ returns the result of $f$ on $p$.

\begin{fortechnicalreport}
\feature
	{\isonceroutinefreshfeature \colon \heaptype \rightarrow \processortype \rightarrow \featuretype \nrightarrow \booleantype}
	{\heap.\isonceroutinefreshfeature(p, f)}
	{f.\isonceroutinefeature}
	{}

\feature
	{\oncefunctionresultfeature \colon \heaptype \rightarrow \processortype \rightarrow \featuretype \nrightarrow \referencetype}
	{\heap.\oncefunctionresultfeature(p, f)}
	{
		& f \in \functiontype \wedge f.\isonceroutinefeature \protect\\
		& \neg \heap.\isonceroutinefreshfeature(p, f)
	}
	{}

\end{fortechnicalreport}

Two commands change the once status of a fresh once routine to non-fresh. One version works for once functions and the other one for once procedures. Both commands take the once routine $f$ and the processor $p$. The version for once functions also takes a once result $r$. The two commands implement the semantics for once routines: a once routine has either a once per system or a once per processor semantics. Once functions declared as separate with or without an explicit processor specification have the once per system semantics. In this case, the command $\setoncefunctionnotfreshfeature$ defines $f$ as non-fresh on all processors. Once functions with a non-separate result type have the once per processor semantics. In this case, the command $\setoncefunctionnotfreshfeature$ sets $f$ as non-fresh on $p$ with the once result $r$. Once procedures have the once per processor semantics. In this case, the command $\setonceprocedurenotfreshfeature$ sets $f$ as non-fresh on $p$.

\feature
	{\setoncefunctionnotfreshfeature \colon & \heaptype \rightarrow \processortype \rightarrow \featuretype \nrightarrow \referencetype \nrightarrow \heaptype}
	{\heap.\setoncefunctionnotfreshfeature(p, f, r)}
	{
		& f \in \functiontype \wedge f.\isonceroutinefeature \protect\\
		& r \neq \voidvalue \rightarrow \heap.\referencesfeature.\containsfeature(r)
	}
	{
		& \protect\begin{split}
			& (\exists d, c \colon \typingenvironmentderivation{f: (d, \bullet, c)}) \rightarrow \protect\\
			& \indentation \protect\begin{split}
				& \neg \heap.\setoncefunctionnotfreshfeature(p, f, r).\isonceroutinefreshfeature(p, f) \wedge \protect\\
				& \heap.\setoncefunctionnotfreshfeature(p, f, r).\oncefunctionresultfeature(p, f) = r
			\protect\end{split}
		\protect\end{split} \protect\\
		& \protect\begin{split}
			& (\exists d, c \colon \typingenvironmentderivation{f: (d, p, c) \wedge p \neq \bullet}) \rightarrow \forall q \in \processortype \colon \protect\\
			& \indentation \protect\begin{split}
				& \neg \heap.\setoncefunctionnotfreshfeature(p, f, r).\isonceroutinefreshfeature(q, f) \wedge \protect\\
				& \heap.\setoncefunctionnotfreshfeature(p, f, r).\oncefunctionresultfeature(q, f) = r
			\protect\end{split}
		\protect\end{split}
	}

\feature
	{\setonceprocedurenotfreshfeature \colon \heaptype \rightarrow \processortype \rightarrow \featuretype \nrightarrow \heaptype}
	{\heap.\setonceprocedurenotfreshfeature(p, f)}
	{f \in \proceduretype \wedge f.\isonceroutinefeature}
	{\neg \heap.\setonceprocedurenotfreshfeature(p, f).\isonceroutinefreshfeature(p, f)}

\subsubsection{Creation}
A new heap can be created with the constructor $\makefeature$. A new heap has no objects and no references. All once routines are marked as fresh on all processors.

\feature
	{\makefeature \colon \heaptype}
	{}
	{}
	{
		& \makefeature.\objectsfeature.\isemptyfeature \protect\\
		& \makefeature.\referencesfeature.\isemptyfeature \protect\\
		& \forall p \in \processortype, f \in \featuretype \colon f.\isonceroutinefeature \rightarrow \makefeature.\isonceroutinefreshfeature(p, f)
	}

\subsection{Regions ADT}
The heap is partitioned into disjoint \emph{regions}, and each region is assigned to exactly one processor. This concept relates to the concept of a ken in Schmidt's work \cite{schmidt-chen:1995:concurrent_objects}. The processor of a region is the handler of all the objects in the region. Regions are also used to maintain locks. The following discussion first describes an ADT for processor and then describes an ADT for regions.

\subsubsection{Processors}
A \emph{processor} is an autonomous thread of control capable of executing features on objects. Each processor is responsible for a set of objects. As such a processor is called the \emph{handler} of its associated objects. Each object is assigned to exactly one processor that is the authority of feature executions on this object. If a processor $q$ wants to call a feature on an object handled by a different processor $p$, then $q$ needs to send a feature request to processor $p$. This is where the request queue of processor $p$ comes into place. The \emph{request queue} keeps track of features to be executed on behalf of other processors. Processor $q$ can add a request to this queue and processor $p$ will execute the request as soon as it executed all previous requests in the request queue. Processor $p$ uses its \emph{call stack} to execute the feature request at the beginning of the request queue. The call stack is responsible for the order of feature executions on the same processor. In a situation of a non-separate call, the call stack ensures that the calling feature execution resumes once the called feature execution terminated. The interaction between the call stack and the request queue is best described with the following loop through which each processor goes:
\begin{enumerate}
	\item Idle wait. If both the call stack and the request queue are empty, then wait for new requests to be enqueued.
	\item Request scheduling. If the call stack is empty but the request queue is not empty, then dequeue an item and push it onto the call stack.
	\item Request processing. If there is an item on the call stack, then pop the item from the call stack and process it. If the item is a feature request, then apply the feature. If the item is an operation, then execute the operation.
\end{enumerate}
For each processor there is a \emph{request queue lock} and a \emph{call stack lock}. A lock on the request queue grants permission to add a feature request to the end of the request queue. A lock on the call stack grants permission to add a feature request to the top of the call stack. Before processor $q$ can add a request to $p$'s request queue, it must have a lock on this request queue. Otherwise another processor could intervene. Once processor $q$ is done with the request queue of processor $p$ it can add an unlock operation to the end of the request queue. This makes sure that the request queue lock of $p$ will be released after all the previous feature requests have been executed. Similarly, processor $p$ must have a lock on its call stack to add features to its call stack. Initially, each processor has a lock on its own call stack and its request queue is not locked.

Processor $q$ could also make a synchronous call to $p$. However $q$ might be in possession of some locks that are necessary for the execution of the resulting feature request on $p$. In such a situation, $q$ is waiting for the synchronous call to terminate and $p$ is waiting for locks to be available. According to the conditions given by Coffman et al.~\cite{coffman-elphick-shoshani:1971:deadlocks} a deadlock occurred. This can be avoided if $q$ temporarily passes its locks to the $p$. This allows $p$ to finish the execution and hence $q$ can continue.

\begin{clarification}[Request queue locks and call stack locks]
The notion of request queue locks and call stack locks was not present in Nienaltowski's \cite{nienaltowski:2007:SCOOP} definition of SCOOP. He defines one lock for each processor. A lock on a processor means exclusive access to the whole processor. This lock model is not sufficient to describe SCOOP. In particular, this lock model creates a contradiction with respect to separate callbacks. A separate callback is a feature call in which processor $q$ made a direct or indirect call to processor $p$ and now $p$ is calling back processor $q$. The separate callback is only possible if $p$ has a lock on $q$. However, $p$ does not necessarily have this lock because the lock might be in possession of the processor that locked $q$ in the first place. Request queue locks and call stack locks allow us to clarify the situation. Thus we propose a new lock model with request queue locks and call stack locks.

The lock model used in Nienaltowski's work \cite{nienaltowski:2007:SCOOP} is an abstraction of the new lock model. The abstraction works under the assumption that no processor passes its locks. Under this assumption each processor keeps its call stack lock. In this abstraction, the request queue lock on a processor $p$ is called the lock on $p$. As long as the call stack lock on a processor $p$ is in possession of $p$, a request queue lock on $p$ in possession of a processor $q$ means that processor $p$ will be executing new feature requests in the request queue exclusively on behalf of $q$. This means that a request queue lock grants exclusive access to all the objects handled by $p$. Transferring this insight to the abstraction, a lock on processor $p$ denotes exclusive access to the objects handled by $p$.
\end{clarification}

The formalization defines the ADT $\processortype$ for processors. A processor has an identifier stored in the query $\identifierfeature$.

\begin{fortechnicalreport}
\feature
	{\identifierfeature \colon \processortype \rightarrow \identifiertype}
	{}
	{}
	{}

\end{fortechnicalreport}

The constructor $\makefeature$ returns a new processor with a fresh identifier. The fresh identifier is defined through the query $\newidentifierfeature$.

\feature
	{\makefeature \colon \processortype}
	{}
	{}
	{\makefeature.\identifierfeature = \newidentifierfeature}

The ADT $\processortype$ is very simple. It neither takes care of the mapping from processors to the handled objects nor does it take care of the locks. These aspects are taken care of by the ADT for regions.

\subsubsection{Mapping of processors to objects and locking}
This section introduces the ADT $\processorregionstype$ for regions. This ADT declares a query $\processorsfeature$ that keeps track of all the processors in the system. The query $\handledobjectsfeature$ defines a set of handled objects for each processor in $\processorsfeature$. Finally, the query $\lastaddedprocessorfeature$ denotes the last processor that has been added to $\processorsfeature$.

\feature
	{\processorsfeature \colon \processorregionstype \rightarrow \settype{\processortype}}
	{}
	{}
	{}

\feature
	{\handledobjectsfeature \colon \processorregionstype \rightarrow \processortype \nrightarrow \settype{\objecttype}}
	{\processorregions.\handledobjectsfeature(p)}
	{\processorregions.\processorsfeature.\containsfeature(p)}
	{}

\feature
	{\lastaddedprocessorfeature \colon \processorregionstype \nrightarrow \processortype}
	{\processorregions.\lastaddedprocessorfeature}
	{\neg \processorregions.\processorsfeature.\isemptyfeature}
	{}

Next to the queries that are concerned with the mapping from processors to objects, there are a number of queries that deal with locking. The feature $\isrequestqueuelockedfeature$ states whether the request queue of a processor in $\processorsfeature$ is locked or not. Similarly, the feature $\iscallstacklockedfeature$ states whether the call stack is locked.

The remaining queries specify the owners of the locks. For this, the formalization distinguishes between \emph{obtained} and \emph{retrieved} locks. Obtained locks are locks that got acquired by a processor. Retrieved locks are locks that got passed from another processor.

The query $\obtainedrequestqueuelocksfeature$ returns a stack of obtained processor sets for a processor. A stack of sets models the way processors obtain locks: they go through a nested series of feature applications and each feature application requires a set of locks before the feature can be executed. For each feature application the executing processor adds a new set on top of its stack. As soon as the feature application finished, the processor removes the top set from its stack. The query $\obtainedcallstacklockfeature$ returns the obtained call stack lock of a processor. Initially each processor starts with a lock on its own call stack and this call stack lock never changes. Thus this query is only declared for reasons of completeness. If a processor appears in a set of request queue locks, then the processor denotes its request queue lock. If a processor appears in a set of call stack locks, then the processor denotes its call stack lock.

A processor can pass its locks to another processor. There are several queries to formalize this aspect. The features $\retrievedrequestqueuelocksfeature$ and $\retrievedcallstacklocksfeature$ return the retrieved locks of a processor. Both of these queries return a stack of sets. The stack keeps track of the set of retrieved locks for each feature application. These two stacks grow and shrink in parallel to the stack $\obtainedrequestqueuelocksfeature$. Once a processor passed its locks, it cannot use them anymore until the locks are revoked. The query $\arelockspassedfeature$ returns whether a processor passed some or all of its locks or not.

\begin{fortechnicalreport}
\feature
	{\isrequestqueuelockedfeature \colon \processorregionstype \rightarrow \processortype \nrightarrow \booleantype}
	{\processorregions.\isrequestqueuelockedfeature(p)}
	{\processorregions.\processorsfeature.\containsfeature(p)}
	{}

\feature
	{\iscallstacklockedfeature \colon \processorregionstype \rightarrow \processortype \nrightarrow \booleantype}
	{\processorregions.\iscallstacklockedfeature(p)}
	{\processorregions.\processorsfeature.\containsfeature(p)}
	{}

\feature
	{\obtainedrequestqueuelocksfeature \colon \processorregionstype \rightarrow \processortype \nrightarrow \stacktype{\settype{\processortype}}}
	{\processorregions.\obtainedrequestqueuelocksfeature(p)}
	{\processorregions.\processorsfeature.\containsfeature(p)}
	{}

\feature
	{\obtainedcallstacklockfeature \colon \processorregionstype \rightarrow \processortype \nrightarrow \processortype}
	{\processorregions.\obtainedcallstacklockfeature(p)}
	{\processorregions.\processorsfeature.\containsfeature(p)}
	{}

\feature
	{\retrievedrequestqueuelocksfeature \colon \processorregionstype \rightarrow \processortype \nrightarrow \stacktype{\settype{\processortype}}}
	{\processorregions.\retrievedrequestqueuelocksfeature(p)}
	{\processorregions.\processorsfeature.\containsfeature(p)}
	{}

\feature
	{\retrievedcallstacklocksfeature \colon \processorregionstype \rightarrow \processortype \nrightarrow \stacktype{\settype{\processortype}}}
	{\processorregions.\retrievedcallstacklocksfeature(p)}
	{\processorregions.\processorsfeature.\containsfeature(p)}
	{}

\feature
	{\arelockspassedfeature \colon \processorregionstype \rightarrow \processortype \nrightarrow \booleantype}
	{\processorregions.\arelockspassedfeature(p)}
	{\processorregions.\processorsfeature.\containsfeature(p)}
	{}

\end{fortechnicalreport}

The following discussion first goes through the list of commands that add processors and commands that change the association of processors to objects. It then proceeds with the commands that handle locks. The command $\addprocessorfeature$ updates the regions with a new processor. Note that the processor must have been created beforehand. The axioms state that the new processor will be included in $\processorsfeature$ and that it will be stored in $\lastaddedprocessorfeature$. The axioms also state how the new processor is initialized. The new processor's request queue is unlocked and its call stack is locked. Apart from the initial lock on the call stack there are no obtained or retrieved locks and hence the processor did not pass its locks.

\feature
	{\addprocessorfeature \colon \processorregionstype \rightarrow \processortype \nrightarrow \processorregionstype}
	{\processorregions.\addprocessorfeature(p)}
	{\neg \processorregions.\processorsfeature.\containsfeature(p)}
	{
		& \processorregions.\addprocessorfeature(p).\processorsfeature.\containsfeature(p) \protect\\
		& \processorregions.\addprocessorfeature(p).\lastaddedprocessorfeature = p \protect\\
		& \processorregions.\addprocessorfeature(p).\handledobjectsfeature(p).\isemptyfeature \protect\\
		& \neg \processorregions.\addprocessorfeature(p).\isrequestqueuelockedfeature(p) \protect\\
		& \processorregions.\addprocessorfeature(p).\iscallstacklockedfeature(p) \protect\\
		& \processorregions.\addprocessorfeature(p).\obtainedrequestqueuelocksfeature(p).\isemptyfeature \protect\\
		&	\processorregions.\addprocessorfeature(p).\obtainedcallstacklockfeature(p) = p \protect\\
		& \processorregions.\addprocessorfeature(p).\retrievedrequestqueuelocksfeature(p).\isemptyfeature \protect\\
		& \processorregions.\addprocessorfeature(p).\retrievedcallstacklocksfeature(p).\isemptyfeature \protect\\
		& \neg \processorregions.\addprocessorfeature(p).\arelockspassedfeature(p)
	}

The command $\addobjectfeature$ takes a processor $p$ in $\processorsfeature$ and an object $o$ that is not handled by a processor in $\processorsfeature$ yet. It returns the updated regions in which $o$ is handled by $p$.

\feature
	{\addobjectfeature \colon \processorregionstype \rightarrow \processortype \nrightarrow \objecttype \nrightarrow \processorregionstype}
	{\processorregions.\addobjectfeature(p, o)}
	{
		& \processorregions.\processorsfeature.\containsfeature(p) \protect\\
		& \forall q \in \processorregions.\processorsfeature, u \in \processorregions.\handledobjectsfeature(q) \colon u.\identifierfeature \neq o.\identifierfeature
	}
	{\processorregions.\addobjectfeature(p, o).\handledobjectsfeature(p).\containsfeature(o)}

In the opposite direction, the command $\removeobjectfeature$ removes an object that is handled by a processor in $\processorsfeature$ from the regions.

\feature
	{\removeobjectfeature \colon \processorregionstype \rightarrow \objecttype \nrightarrow \processorregionstype}
	{\processorregions.\removeobjectfeature(o)}
	{\exists p \in \processorregions.\processorsfeature \colon \processorregions.\handledobjectsfeature(p).\containsfeature(o)}
	{
		\neg \exists p \in \processorregions.\processorsfeature \colon \processorregions.\removeobjectfeature(o).\handledobjectsfeature(p).\containsfeature(o)
	}

The following part discusses the commands that deal with the locking aspects of the regions. The command $\lockrequestqueuesfeature$ locks the request queues of a set of processors $\overline{q}$ on behalf of a processor $p$. None of these request queues must be locked beforehand.

\feature
	{\lockrequestqueuesfeature \colon \processorregionstype \rightarrow \processortype \nrightarrow \settype{\processortype} \nrightarrow \processorregionstype}
	{\processorregions.\lockrequestqueuesfeature(p, \overline{l})}
	{
		& \processorregions.\processorsfeature.\containsfeature(p) \protect\\
		& \forall {x \in \overline{l}} \colon {\processorregions.\processorsfeature.\containsfeature(x)} \protect\\
		& \forall {x \in \overline{l}} \colon {\neg \processorregions.\isrequestqueuelockedfeature(x)}
	}
	{
		& \processorregions.\lockrequestqueuesfeature(p, \overline{l}).\obtainedrequestqueuelocksfeature(p) = \processorregions.\obtainedrequestqueuelocksfeature(p).\pushfeature(\overline{l}) \protect\\
		& \forall{x \in \overline{l}} \colon {\processorregions.\lockrequestqueuesfeature(p, \overline{l}).\isrequestqueuelockedfeature(x)}
	}

At some point, processor $p$ will not require the obtained request queue locks anymore because $p$ made sure to enqueue all necessary features requests. Processor $p$ uses the command $\popobtainedrequestqueuelocksfeature$ to remove his claims on the obtained request queue locks. This requires that processor $p$ is in possession of these locks, i.e., that $p$ did not pass its locks.

\feature
	{\popobtainedrequestqueuelocksfeature \colon \processorregionstype \rightarrow \processortype \nrightarrow \processorregionstype}
	{\processorregions.\popobtainedrequestqueuelocksfeature(p)}
	{
		& \processorregions.\processorsfeature.\containsfeature(p) \protect\\
		& \neg \processorregions.\obtainedrequestqueuelocksfeature(p).\isemptyfeature \protect\\
		& \neg \processorregions.\arelockspassedfeature(p)
	}
	{
		& \processorregions.\popobtainedrequestqueuelocksfeature(p).\obtainedrequestqueuelocksfeature(p) = \processorregions.\obtainedrequestqueuelocksfeature(p).\popfeature
	}

Removing the locks from $p$'s obtained request queue locks stack does not mean that these request queues are unlocked. It just means that the request queue locks are not claimed by $p$ anymore and therefore $p$ will not enqueue further feature requests on the respective processors. The request queues remain locked until they get unlocked with a call to the command $\unlockrequestqueuefeature$. This happens after the processors whose request queues got locked by $p$ finished all the requested feature applications. The precondition of the command states that a request queue can only be unlocked if it is not claimed by any other processor. This precondition guarantees that the request queue can only be unlocked when it is not used as an obtained or retrieved lock by any other processor anymore. Note that there is no unlock command for call stack locks because the call stack never gets unlocked.

\feature
	{\unlockrequestqueuefeature \colon \processorregionstype \rightarrow \processortype \nrightarrow \processorregionstype}
	{\processorregions.\unlockrequestqueuefeature(p)}
	{
		& \processorregions.\processorsfeature.\containsfeature(p) \protect\\
		& \processorregions.\isrequestqueuelockedfeature(p) \protect\\
		& \forall q \in \processorregions.\processorsfeature \colon \neg \processorregions.\obtainedrequestqueuelocksfeature(q).\flattenedfeature.\containsfeature(p)
	}
	{\neg \processorregions.\unlockrequestqueuefeature(p).\isrequestqueuelockedfeature(p)}

The request queues remain locked until explicitly unlocked with a call to $\unlockrequestqueuefeature$. Between the call to $\popobtainedrequestqueuelocksfeature$ and the call to $\unlockrequestqueuefeature$, the owner of these locks is undefined. In some situations this is not satisfactory. A different solution must be found if another processor wants to claim the locks until they are unlocked. The command $\delegateobtainedrequestqueuelocksfeature$ serves this purpose. It takes a processor $p$ and a number of processors $\overline{l}$ and makes $p$ the owner of the request queue locks of all processors in $\overline{l}$ by adding these locks to the obtained request queue locks stack of $p$. This can only work if there is no current owner and the request queues are indeed locked.

\feature
	{\delegateobtainedrequestqueuelocksfeature \colon \processorregionstype \rightarrow \processortype \nrightarrow \settype{\processortype} \nrightarrow \processorregionstype}
	{\processorregions.\delegateobtainedrequestqueuelocksfeature(p, \overline{l})}
	{
		& \processorregions.\processorsfeature.\containsfeature(p) \protect\\
		& \forall {x \in \overline{l}} \colon {\processorregions.\processorsfeature.\containsfeature(x)} \protect\\
		& \forall {x \in \overline{l}} \colon {\neg \exists y \in \processorregions.\processorsfeature \colon \processorregions.\obtainedrequestqueuelocksfeature(y).\flattenedfeature.\containsfeature(x)} \protect\\
		& \forall {x \in \overline{l}} \colon {\processorregions.\isrequestqueuelockedfeature(x)}
	}
	{
		& \processorregions.\delegateobtainedrequestqueuelocksfeature(p, \overline{l}).\obtainedrequestqueuelocksfeature(p) = \processorregions.\obtainedrequestqueuelocksfeature(p).\pushfeature(\overline{l})
	}

Delegation is different from lock passing: \emph{delegation} is the permanent transfer of ownership and \emph{lock passing} is the temporary transfer of the right to use the locks. The following discussion looks at the commands to pass and revoke locks. The command $\passlocksfeature$ takes a processor $p$ and a processor $q$ as well as a set of request queue locks $\overline{l_{r}}$ along with a set of call stack locks $\overline{l_{c}}$. The result is an updated instance of $\processorregionstype$ in which $\overline{l_{r}}$ and $\overline{l_{c}}$ have been passed from $p$ to $q$. As a precondition for this task, processor $p$ must be in possession of all these locks. This means that all the locks in $\overline{l_{r}}$ and $\overline{l_{c}}$ must be obtained or retrieved locks of $p$ and the locks must not be passed. The updated result must reflect that some or all of $p$'s locks have been passed. However, because the two sets of locks can potentially be empty, $p$'s locks must only be marked as passed if at least one of the two sets of locks is non-empty. Lastly, the command must take care of one special case of the lock passing operation. If a processor $q$ different from processor $p$ passed its locks in a previous lock passing operation and now the command passes these locks back to $q$, then the command has to mark the locks of processor $q$ as not passed. This case is important to handle separate callbacks.

\feature
	{
		\passlocksfeature \colon & \processorregionstype \rightarrow \processortype \nrightarrow \processortype \nrightarrow \tupletype{\settype{\processortype}, \settype{\processortype}} \nrightarrow \processorregionstype
	}
	{\processorregions.\passlocksfeature(p, q, \tuple{\overline{l_{r}}, \overline{l_{c}}})}
	{
		& \processorregions.\processorsfeature.\containsfeature(p) \wedge \processorregions.\processorsfeature.\containsfeature(q) \protect\\
		& \forall {x \in \overline{l_{r}}} \colon  \processorregions.\processorsfeature.\containsfeature(x) \wedge \forall {x \in \overline{l_{c}}} \colon  \processorregions.\processorsfeature.\containsfeature(x) \protect\\
		& \forall x \in \overline{l_{r}} \colon \processorregions.\obtainedrequestqueuelocksfeature(p).\flattenedfeature.\containsfeature(x) \vee \processorregions.\retrievedrequestqueuelocksfeature(p).\flattenedfeature.\containsfeature(x) \protect\\
		& \forall x \in \overline{l_{c}} \colon x = \processorregions.\obtainedcallstacklockfeature(p) \vee \processorregions.\retrievedcallstacklocksfeature(p).\flattenedfeature.\containsfeature(x) \protect\\
		& \neg \processorregions.\arelockspassedfeature(p)
	}
	{
		& \processorregions.\passlocksfeature(p, q, \tuple{\overline{l_{r}}, \overline{l_{c}}}).\arelockspassedfeature(p) = 
			\left\{
				\protect\begin{array}{ll}
					\truevalue & \condition{\neg \overline{l_{r}}.\isemptyfeature \vee \neg \overline{l_{c}}.\isemptyfeature} \protect\\
					\falsevalue & \otherwisecondition
				\protect\end{array}
			\right. \protect\\
		& \processorregions.\passlocksfeature(p, q, \tuple{\overline{l_{r}}, \overline{l_{c}}}).\retrievedrequestqueuelocksfeature(q) = \processorregions.\retrievedrequestqueuelocksfeature(q).\pushfeature(\overline{l_{r}}) \protect\\
		& \processorregions.\passlocksfeature(p, q, \tuple{\overline{l_{r}}, \overline{l_{c}}}).\retrievedcallstacklocksfeature(q) = \processorregions.\retrievedcallstacklocksfeature(q).\pushfeature(\overline{l_{c}}) \protect\\
		& \left(
			\protect\begin{array}{l}
				p \neq q \wedge \protect\\
				\processorregions.\arelockspassedfeature(q) \wedge \protect\\
				\processorregions.\obtainedrequestqueuelocksfeature(q).\flattenedfeature \subseteq \overline{l_{r}} \wedge \protect\\
				\processorregions.\retrievedrequestqueuelocksfeature(q).\flattenedfeature \subseteq \overline{l_{r}} \wedge \protect\\
				\processorregions.\obtainedcallstacklockfeature(q) \in \overline{l_{c}} \wedge \protect\\
				\processorregions.\retrievedcallstacklocksfeature(q).\flattenedfeature \subseteq \overline{l_{c}}
			\protect\end{array}
		\right) \rightarrow \neg \processorregions.\passlocksfeature(p, q, \tuple{\overline{l_{r}}, \overline{l_{c}}}).\arelockspassedfeature(q)
	}

The command $\revokelocksfeature$ takes a processor $p$ and a processor $q$. It reverses the effect of a lock passing operation from a processor $p$ to $q$ and returns an updated instance of $\processorregionstype$. This is only allowed if processor $p$ passed locks to $q$ in a preceding lock passing operation. Note that the lock passing operation from $p$ to $q$ potentially marked the locks of $q$ as not passed. Revoking the locks from $q$ to $p$ requires the reverse action. If $p$ has retrieved locks in common with the locks of $q$, even after the retrieved locks from $p$ have been removed from $q$, then $q$'s locks must be marked as passed because they are now in possession of $p$.

\feature
	{\revokelocksfeature \colon \processorregionstype \rightarrow \processortype \nrightarrow \processortype \nrightarrow \processorregionstype}
	{\processorregions.\revokelocksfeature(p, q)}
	{
		& \processorregions.\processorsfeature.\containsfeature(p) \wedge \processorregions.\processorsfeature.\containsfeature(q) \protect\\
		& \neg \processorregions.\retrievedrequestqueuelocksfeature(q).\isemptyfeature \wedge \neg \processorregions.\retrievedcallstacklocksfeature(q).\isemptyfeature \protect\\
		& \processorregions.\retrievedrequestqueuelocksfeature(q).\topfeature \subseteq \processorregions.\obtainedrequestqueuelocksfeature(p).\flattenedfeature \cup \processorregions.\retrievedrequestqueuelocksfeature(p).\flattenedfeature \protect\\
		& \processorregions.\retrievedcallstacklocksfeature(q).\topfeature \subseteq \set{\processorregions.\obtainedcallstacklockfeature(p)} \cup \processorregions.\retrievedcallstacklocksfeature(p).\flattenedfeature \protect\\
		& \processorregions.\retrievedrequestqueuelocksfeature(q).\topfeature \cup \processorregions.\retrievedcallstacklocksfeature(q).\topfeature \neq \set{} \rightarrow \processorregions.\arelockspassedfeature(p) \protect\\
		& \neg \processorregions.\arelockspassedfeature(q)
	}
	{
		& \neg \processorregions.\revokelocksfeature(p, q).\arelockspassedfeature(p) \protect\\
		& \processorregions.\revokelocksfeature(p, q).\retrievedrequestqueuelocksfeature(q) = \processorregions.\retrievedrequestqueuelocksfeature(q).\popfeature \protect\\
		& \processorregions.\revokelocksfeature(p, q).\retrievedcallstacklocksfeature(q) = \processorregions.\retrievedcallstacklocksfeature(q).\popfeature \protect\\
		& \left(
			\protect\begin{array}{l}
				p \neq q \wedge \protect\\
				\left(
					\protect\begin{array}{l}
						\protect\begin{split}
							& \exists x \in \processorregions.\retrievedrequestqueuelocksfeature(p).\flattenedfeature \colon ( \protect\\
							& \indentation \protect\begin{split}
								& \processorregions.\obtainedrequestqueuelocksfeature(q).\flattenedfeature.\containsfeature(x) \vee \protect\\
						 		& \processorregions.\retrievedrequestqueuelocksfeature(q).\popfeature.\flattenedfeature.\containsfeature(x)
							\protect\end{split} \protect\\
							& ) \vee
						\protect\end{split} \protect\\
					  \protect\begin{split}
							& \exists x \in \processorregions.\retrievedcallstacklocksfeature(p).\flattenedfeature \colon ( \protect\\
							& \indentation \protect\begin{split}
								& x = \processorregions.\obtainedcallstacklockfeature(q) \vee \protect\\
						 		& \processorregions.\retrievedcallstacklocksfeature(q).\popfeature.\flattenedfeature.\containsfeature(x)
							\protect\end{split} \protect\\
							& )
						\protect\end{split}
					\protect\end{array}
				\right)
			\protect\end{array}
		\right) \protect\\ & \indentation \rightarrow	\processorregions.\revokelocksfeature(p, q).\arelockspassedfeature(q)
	}

These commands wrap up the mapping of processors to objects and the locking aspects. The discussion continues with a number of auxiliary queries to simplify access to the presented queries. The command $\addobjectfeature$ makes sure that a processor is assigned to each object that gets added. This mapping is available through the query $\handledobjectsfeature$. Thus it is possible to define an auxiliary query $\handlerfeature$ that is inverse to the query $\handledobjectsfeature$.

\feature
	{\handlerfeature \colon \processorregionstype \rightarrow \objecttype \nrightarrow \processortype}
	{\processorregions.\handlerfeature(o)}
	{\exists p \in \processorregions.\processorsfeature \colon \processorregions.\handledobjectsfeature(p).\containsfeature(o)}
	{
		\processorregions.\handledobjectsfeature(\processorregions.\handlerfeature(o)).\containsfeature(o)
	}

There are four different categories of locks that each processor can have. For both the request queue locks and the call stack locks, there are queries for obtained and retrieved locks. In some situations it is easier to just work with request queue locks and call stack locks without splitting them into obtained and retrieved locks. The auxiliary queries $\requestqueuelocksfeature$ and $\callstacklocksfeature$ serve this purpose. The auxiliary query $\requestqueuelocksfeature$ returns a set that contains all the obtained and the retrieved request queue locks of a processor $p$. Similarly, the auxiliary query $\callstacklocksfeature$ returns all the call stack locks of a processor $p$.

\feature
	{\requestqueuelocksfeature \colon \processorregionstype \rightarrow \processortype \nrightarrow \settype{\processortype}}
	{\processorregions.\requestqueuelocksfeature(p)}
	{\processorregions.\processorsfeature.\containsfeature(p)}
	{\processorregions.\requestqueuelocksfeature(p) = \processorregions.\obtainedrequestqueuelocksfeature(p).\flattenedfeature \cup \processorregions.\retrievedrequestqueuelocksfeature(p).\flattenedfeature}

\feature
	{\callstacklocksfeature \colon \processorregionstype \rightarrow \processortype \nrightarrow \settype{\processortype}}
	{\processorregions.\callstacklocksfeature(p)}
	{\processorregions.\processorsfeature.\containsfeature(p)}
	{\processorregions.\callstacklocksfeature(p) = \set{\processorregions.\obtainedcallstacklockfeature(p)} \cup \processorregions.\retrievedcallstacklocksfeature(p).\flattenedfeature}

\subsubsection{Creation}
The constructor $\makefeature$ creates a new instance of $\processorregionstype$. The new instance has no processors.

\feature
	{\makefeature \colon \processorregionstype}
	{}
	{}
	{\makefeature.\processorsfeature.\isemptyfeature}

\subsection{Store ADT}
Each processor in the system has a call stack to execute features. Every time a processor executes a feature, a new call stack frame gets created on top of the call stack. The new call stack frame stores the values of formal arguments, local variables, the current object entity, and the result entity for the current feature execution. The call stack is also responsible for the order of feature executions on the same processor. This formalization separates the two concerns of the call stack. The \emph{store} only models the values in each stack frame. A store has a stack of environments for each processor, where each \emph{environment} maps names to values. This section first presents an ADT for environments and then presents an ADT for the store.

\subsubsection{Environments}
The ADT $\environmenttype$ has a query $\namesfeature$ that stores all the defined names. The query $\valuefeature$ can then be used to get the value for each such name.

\feature
	{\namesfeature \colon \environmenttype \rightarrow \settype{\nametype}}
	{}
	{}
	{}

\feature
	{\valuefeature \colon \environmenttype \rightarrow \nametype \nrightarrow \referencetype \cup \processortype }
	{\environment.\valuefeature(n)}
	{\environment.\namesfeature.\containsfeature(n)}
	{}

The command $\updatefeature$ takes a name and a value and returns an updated environment. Note that it does not matter whether the name is already defined in the environment or not. In any case, the name will be defined in the updated environment and the name will be mapped to the value. The value can either be a reference or a processor. Environments with processor values are not strictly needed to describe SCOOP, however they make it possible to have a unified view on attribute values and environment values.

\feature
	{\updatefeature \colon \environmenttype \rightarrow \nametype \rightarrow \referencetype \cup \processortype \rightarrow \environmenttype}
	{}
	{}
	{
		& \environment.\updatefeature(n, v).\namesfeature = \environment.\namesfeature \cup \set{n} \protect\\
		& \environment.\updatefeature(n, v).\valuefeature(n) = v
	}

The constructor $\makefeature$ returns an empty environment.

\feature
	{\makefeature \colon \environmenttype}
	{}
	{}
	{\makefeature.\namesfeature.\isemptyfeature}

\subsubsection{Mapping from processors to environments}
The ADT $\storetype$ has a single query $\environmentsfeature$ that stores a stack of environments for each processor.

\feature
	{\environmentsfeature \colon \storetype \rightarrow \processortype \rightarrow \stacktype{\environmenttype}}
	{}
	{}
	{}

The command $\pushenvironmentfeature$ pushes a given environment on top a processor's stack of environments. The command $\popenvironmentfeature$ pops the top environment from a non-empty stack of environments.

\feature
	{\pushenvironmentfeature \colon \storetype \rightarrow \processortype \rightarrow \environmenttype \rightarrow \storetype}
	{}
	{}
	{\store.\pushenvironmentfeature(p, e).\environmentsfeature(p) = \store.\environmentsfeature(p).\pushfeature(e)}

\feature
	{\popenvironmentfeature \colon \storetype \rightarrow \processortype \nrightarrow \storetype}
	{\store.\popenvironmentfeature(p)}
	{\neg \store.\environmentsfeature(p).\isemptyfeature}
	{\store.\popenvironmentfeature(p).\environmentsfeature(p) = \store.\environmentsfeature(p).\popfeature}

The constructor $\makefeature$ creates an empty store.

\feature
	{\makefeature \colon \storetype}
	{}
	{}
	{\forall {p \in \processortype} \colon {\makefeature.\environmentsfeature(p).\isemptyfeature}}

\subsection{State ADT}\label{sec:state ADT}
The ADT $\statetype$ models the \emph{state} with three queries to retrieve the different parts of the ADT.

\feature
	{\processorregionsfeature \colon \statetype \rightarrow \processorregionstype}
	{}
	{}
	{}
\feature
	{\heapfeature \colon \statetype \rightarrow \heaptype}
	{}
	{}
	{}
\feature
	{\storefeature \colon \statetype \rightarrow \storetype}
	{}
	{}
	{}

The command $\setallfeature$ sets the regions, the heap, and the store at the same time. A precondition specifies consistency criteria between the parts of the state. The first two precondition clauses state that a processor can handle an object if and only if the object is on the heap. The third precondition clause states that if the heap declares a feature as non-fresh on a processor $p$, then the regions must know about this processor. The fourth precondition clause requires that all processors stored in attribute values are known by the regions. Note that $\heaptype$ already requires that the references stored in attribute values are known. The fifth precondition clause states that each non-empty environment in the store must belong to a processor that is known by the regions. The sixth precondition clause states that each value in the store must either be a known reference or a known processor.

\feature
	{\setallfeature \colon \statetype \rightarrow \processorregionstype \nrightarrow \heaptype \nrightarrow \storetype \nrightarrow \statetype}
	{\state.\setallfeature(k, h, s)}
	{
		& \forall p \in k.\processorsfeature, o \in k.\handledobjectsfeature(p) \colon h.\objectsfeature.\containsfeature(o) \protect\\
		& \forall o \in h.\objectsfeature \colon \exists p \in k.\processorsfeature \colon o \in k.\handledobjectsfeature(p) \protect\\
		& \forall p \in \processortype, f \in \featuretype \colon \neg h.\isonceroutinefreshfeature(p, f) \rightarrow k.\processorsfeature.\containsfeature(p) \protect\\
		& \protect\begin{split}
			& \forall o \in h.\objectsfeature, a \in o.\classtypefeature.\attributesfeature \colon \protect\\
			& \indentation o.\attributevaluefeature(a) \in \processortype \rightarrow k.\processorsfeature.\containsfeature(o.\attributevaluefeature(a))
		\protect\end{split} \protect\\
		& \forall p \in \processortype, e \in s.\environmentsfeature(p) \colon \neg e.\namesfeature.\isemptyfeature \rightarrow k.\processorsfeature.\containsfeature(p) \protect\\
		& \protect\begin{split}
			& \forall p \in k.\processorsfeature, e \in s.\environmentsfeature(p), x \in e.\namesfeature \colon \protect\\
			& \indentation \protect\begin{split}
				& (e.\valuefeature(x) \in \referencetype \rightarrow e.\valuefeature(x) = \voidvalue \vee h.\referencesfeature.\containsfeature(e.\valuefeature(x))) \wedge \protect\\
				& (e.\valuefeature(x) \in \processortype \rightarrow k.\processorsfeature.\containsfeature(e.\valuefeature(x)))
			\protect\end{split}
		\protect\end{split}
	}
	{
		& \state.\setallfeature(k, h, s).\processorregionsfeature = k \protect\\
		& \state.\setallfeature(k, h, s).\heapfeature = h \protect\\
		& \state.\setallfeature(k, h, s).\storefeature = s
	}

\subsubsection{Creation}
To create a state, one has to create the three parts of the state. This is done with the constructor $\makefeature$.

\feature
	{\makefeature \colon \statetype}
	{}
	{}
	{
		& \makefeature.\processorregionsfeature = \creation{\processorregionstype}{\makefeature} \protect\\
		& \makefeature.\heapfeature = \creation{\heaptype}{\makefeature} \protect\\
		& \makefeature.\storefeature = \creation{\storetype}{\makefeature}
	}

\subsubsection{Facade}
It is too cumbersome to work with $\statetype$ as it is. For example, the following expression defines a new state $\state'$ in which a new processor has been added to the state $\state$: $\state' \mathematicaldefinition \state.\setallfeature(\state.\processorregionsfeature.\addprocessorfeature(\creation{\processortype}{\makefeature}), \state.\heapfeature, \state.\storefeature)$. This expression is too long for this simple task, especially if the expression is used multiple times. It would be easier to have an auxiliary command that does this job for us. The \emph{facade} is an abstraction with auxiliary features that provide easy access to the state functionality. The facade is divided into different aspects. The following discussion dedicates one section to each aspect. It starts with the mapping of processors to objects and the mapping of references to objects. It continues with a section on how to set values, followed by a section on how to get values. It concludes with a section on locking.

\subsubsection{Mapping of processors to objects and mapping of references to objects}
The regions and the heap manage the references, the objects, the processors, and the mapping between them. The facade unifies all related features in one aspect. This section first defines a number of auxiliary queries for the mapping of processors to objects. Next, it defines auxiliary queries for the mapping of references to objects. It then defines auxiliary commands that work on both aspects.

The two auxiliary queries $\processorsfeature$ and $\lastaddedprocessorfeature$ give access to all the processors and the last added processor.

\begin{fortechnicalreport}
\feature
	{\processorsfeature \colon \statetype \rightarrow \settype{\processortype}}
	{}
	{}
	{\state.\processorsfeature = \state.\processorregionsfeature.\processorsfeature}

\feature
	{\lastaddedprocessorfeature \colon \statetype \nrightarrow \processortype}
	{\state.\lastaddedprocessorfeature}
	{\neg \state.\processorregionsfeature.\processorsfeature.\isemptyfeature}
	{\state.\lastaddedprocessorfeature = \state.\processorregionsfeature.\lastaddedprocessorfeature}

\end{fortechnicalreport}

The auxiliary query $\handlerfeature$ gives the handler of an object referenced by $r$. The auxiliary query uses the heap to get the referenced object and then gives this object to the regions to get the handler. In contrast to the corresponding auxiliary query in $\processorregionstype$, the version here takes a reference instead of an object. The version in $\processorregionstype$ deals directly with objects rather than references because it does not know about the heap and thus the mapping from references to objects is not available. The facade, however, has access to both the regions and the heap and thus it can use the preferred way of identifying objects: references.

\begin{fortechnicalreport}
\feature
	{\handlerfeature \colon \statetype \rightarrow \referencetype \nrightarrow \processortype}
	{\state.\handlerfeature(r)}
	{\state.\heapfeature.\referencesfeature.\containsfeature(r)}
	{\state.\handlerfeature(r) = \state.\processorregionsfeature.\handlerfeature(\state.\heapfeature.\referencedobjectfeature(r))}

\end{fortechnicalreport}

The auxiliary query $\newprocessorfeature$ is a shorthand for processor creation. The auxiliary query $\lastaddedobjectfeature$ returns the object that has been added last to the heap. The auxiliary query $\referencedobjectfeature$ returns the object that is associated to a given reference. In the other direction, the auxiliary query $\referencefeature$ returns the reference to a given object. The auxiliary query $\newobjectfeature$ is a shorthand for object creation; it returns a new object with a given class type.

\begin{fortechnicalreport}
\feature
	{\newprocessorfeature \colon \statetype \rightarrow \processortype}
	{}
	{}
	{\state.\newprocessorfeature = \creation{\processortype}{\makefeature}}

\feature
	{\lastaddedobjectfeature \colon \statetype \nrightarrow \objecttype}
	{\state.\lastaddedobjectfeature}
	{\neg \state.\heapfeature.\objectsfeature.\isemptyfeature}
	{\state.\lastaddedobjectfeature = \state.\heapfeature.\lastaddedobjectfeature}

\feature
	{\referencedobjectfeature \colon \statetype \rightarrow \referencetype \nrightarrow \objecttype}
	{\state.\referencedobjectfeature(r)}
	{\state.\heapfeature.\referencesfeature.\containsfeature(r)}
	{\state.\referencedobjectfeature(r) = \state.\heapfeature.\referencedobjectfeature(r)}

\feature
	{\referencefeature \colon \statetype \rightarrow \objecttype \nrightarrow \referencetype}
	{\state.\referencefeature(o)}
	{\state.\heapfeature.\objectsfeature.\containsfeature(o)}
	{\state.\referencefeature(o) = \state.\heapfeature.\referencefeature(o)}

\feature
	{\newobjectfeature \colon \statetype \rightarrow \classtypetype \rightarrow \objecttype}
	{}
	{}
	{\state.\newobjectfeature(c) = \creation{\objecttype}{\makefeature(c)}}

\end{fortechnicalreport}

The discussion continues with the auxiliary commands that modify the mapping of processors to objects and the mapping of references to objects. Before an object can be added to the set of handled objects of a processor, the processor must exist. If the processor does not exist yet, the command $\addprocessorfeature$ can be used to update a state with a new processor.

\feature
	{\addprocessorfeature \colon \statetype \rightarrow \processortype \nrightarrow \statetype}
	{\state.\addprocessorfeature(p)}
	{\neg \state.\processorregionsfeature.\processorsfeature.\containsfeature(p)}
	{\state.\addprocessorfeature(p) = \state.\setallfeature(\state.\processorregionsfeature.\addprocessorfeature(p), \state.\heapfeature, \state.\storefeature)}

The auxiliary command $\addobjectfeature$ can then be used to add an object to the processor and the heap. The auxiliary command takes a processor $p$ and an object $o$ and it returns a state in which object $o$ is part of the heap and handled by processor $p$.

\feature
	{\addobjectfeature \colon \statetype \rightarrow \processortype \nrightarrow \objecttype \nrightarrow \statetype}
	{\state.\addobjectfeature(p, o)}
	{
		& \state.\processorregionsfeature.\processorsfeature.\containsfeature(p) \protect\\
		& \forall u \in \state.\heapfeature.\objectsfeature \colon u.\identifierfeature \neq o.\identifierfeature \protect\\
		& \protect\begin{split}
			& \forall a \in o.\classtypefeature.\attributesfeature \colon \protect\\
			& \indentation \protect\begin{split}
				& (o.\attributevaluefeature(a) \in \referencetype \rightarrow o.\attributevaluefeature(a) = \voidvalue \vee \state.\heapfeature.\referencesfeature.\containsfeature(o.\attributevaluefeature(a))) \wedge \protect\\
				& (o.\attributevaluefeature(a) \in \processortype \rightarrow \state.\processorregionsfeature.\processorsfeature.\containsfeature(o.\attributevaluefeature(a)))
			\protect\end{split}
		\protect\end{split}
	}
	{\state.\addobjectfeature(p, o) = \state.\setallfeature(\state.\processorregionsfeature.\addobjectfeature(p, o), \state.\heapfeature.\addobjectfeature(o), \state.\storefeature)}

The auxiliary command $\updatereferencefeature$ updates a reference with an updated object. It takes a reference $r$ on the heap and an object $o$ and it returns a state in which $o$ replaced the object $u$ referenced by $r$ on the heap and in the regions. Note that $o$ must indeed be an updated version of the object referenced by $r$. The auxiliary command first removes $u$ from the set of handled objects and then adds $o$ to the set of handled objects of $u$'s handler. Then it updates the heap with the command $\updatereferencefeature$, which is declared in $\heaptype$.

\feature
	{\updatereferencefeature \colon \statetype \rightarrow \referencetype \nrightarrow \objecttype \nrightarrow \statetype}
	{\state.\updatereferencefeature(r, o)}
	{
		& \state.\heapfeature.\referencesfeature.\containsfeature(r) \protect\\
		& o.\identifierfeature = \state.\heapfeature.\referencedobjectfeature(r).\identifierfeature \protect\\
		& \protect\begin{split}
			& \forall a \in o.\classtypefeature.\attributesfeature \colon \protect\\
			& \indentation \protect\begin{split}
				& (o.\attributevaluefeature(a) \in \referencetype \rightarrow o.\attributevaluefeature(a) = \voidvalue \vee \state.\heapfeature.\referencesfeature.\containsfeature(o.\attributevaluefeature(a))) \wedge \protect\\
				& (o.\attributevaluefeature(a) \in \processortype \rightarrow \state.\processorregionsfeature.\processorsfeature.\containsfeature(o.\attributevaluefeature(a)))
			\protect\end{split}
		\protect\end{split}
	}
	{
		\where
			{\state.\updatereferencefeature(r, o) = \state.\setallfeature(k, h, s)}
			{
				& u \mathematicaldefinition \state.\heapfeature.\referencedobjectfeature(r) \protect\\
				& k \mathematicaldefinition \state.\processorregionsfeature.\removeobjectfeature(u).\addobjectfeature(\state.\processorregionsfeature.\handlerfeature(u), o) \protect\\
				& h \mathematicaldefinition \state.\heapfeature.\updatereferencefeature(r, o) \protect\\
				& s \mathematicaldefinition \state.\storefeature
			}
	}

\subsubsection{Setting values}\label{sec:setting values}
This section takes a look at how to set values. To start, it looks at a prerequisite for this task: the deep import operation. Setting values includes setting values of formal arguments, values of local variables, the value of the current object entity, the value of the result entity, and attribute values of the current object. All of these values can be written and read without a feature call. This section concludes with auxiliary commands to set the status of once routines. The SCOOP validity rules exclude other types of value setting operations.

\paragraph{Deep import operation}
Expanded objects have a copy semantics: if an object $o$ of expanded class type is the source of an attachment, then a copy $u$ gets attached to the destination of the attachment. However, a shallow copy is not sufficient if $o$'s handler $p$ is different from $u$'s handler $q$. If $o$ has an attached non-separate entity, then $u$ now has a non-separate entity to which a separate object is attached. This would result in a \emph{traitor} -- a non-separate entity that points to a separate object. The SCOOP model, as defined by Nienaltowski \cite{nienaltowski:2007:SCOOP}, introduces the \emph{import operation} to solve this issue. Applied to $o$ the import operation creates a copied object structure that mirrors the original object structure in a way that $o$ and all the objects reachable from $o$ through non-separate references are replaced with copied objects that are handled by $q$. This data structure then gets attached to the destination of the attachment. The import operation computes the non-separate version of an object structure. 

\begin{clarification}[Deep import operation]
The import operation potentially results in a copied object structure that contains both copied and original objects. This can be an issue in case one of the copied objects has an invariant over the identities of objects, as shown in example \ref{ex:invariant violation as a result of the import operation}.

\begin{example}[Invariant violation as a result of the import operation]\label{ex:invariant violation as a result of the import operation}
\begin{figure}
  \centering
  \includegraphics[width=0.6\textwidth]{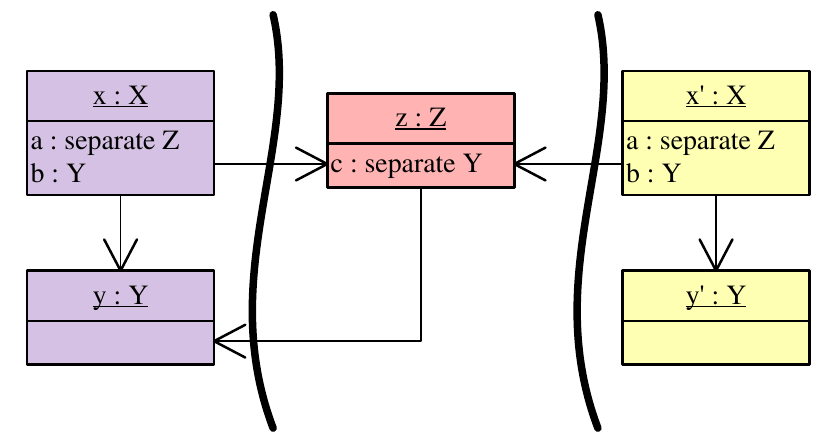}
  \caption{Invariant violation as a result of the import operation}
  \label{fig:import operation invariant violation example}
\end{figure}
Imagine two objects $x$ and $y$ handled by one processor and another object $z$ handled by another processor. Object $x$ has a separate entity $a$ that points to $z$ and a non-separate entity $b$ that points to $y$. Object $z$ has a separate entity $c$ that points to $y$. Object $x$ has an invariant with a query $a.c = b$. An import operation on $x$ executed by a third processor will result in two new objects $x'$ and $y'$ on the third processor. The reference $a$ of object $x'$ will point to the original $z$. The reference $b$ of object $x'$ will point to the new object $y'$. This situation is illustrated in \figurereference{fig:import operation invariant violation example}. Now object $x'$ is inconsistent, because $a.c$ and $b$ identify different objects, namely $y$ and $y'$.
\end{example}

The \emph{deep import operation} is a variant of the import operation that does not mix the copied and the original objects.
\end{clarification}

Instead of copying only the objects that are reachable through non-separate references, the deep import operation makes a full copy of the object structure. The deep importing processor handles all the copies of the objects that are non-separate with respect to the object to be imported. Each other separate object is handled by the processor of the respective original object. The deep import operation does not show the issue with invariants. The drawback of the deep import operation is that more objects must be copied. Nevertheless, we use the deep import operation in our formalization because we cannot tolerate violated invariants. Once routines complicate the deep import operation a bit. Consider a processor $p$ that wants to deep import an object $o$ handled by a different processor $q$. For each non-separate once function $f$ of each copied object the following must be done: if a non-separate once function $f$ is fresh on $p$ and non-fresh on $q$, then $f$ must be marked as non-fresh on $p$ and the value of $f$ on $q$ must be used as the value of $f$ on $p$. If a once procedure $f$ is fresh on $p$ and non-fresh on $q$, then $f$ must be marked as non-fresh on $p$. In all other cases, nothing must be done.

The auxiliary command $\deepimportfeature$ implements the deep import operation. The command takes an importing processor $p$ and a reference $r$ to be imported. The command returns a state in which the copied object structure exists on the heap and the objects are associated to the respective processors. The copied object structure is accessible through the auxiliary query $\lastimportedreferencefeature$.

\feature
	{\deepimportfeature \colon \statetype \rightarrow \processortype \nrightarrow \referencetype \nrightarrow \statetype}
	{\state.\deepimportfeature(p, r)}
	{
		& \state.\processorregionsfeature.\processorsfeature.\containsfeature(p) \protect\\
		& \state.\heapfeature.\referencesfeature.\containsfeature(r)
	}
	{
		\where
			{
				& \state.\deepimportfeature(p, r) = \state' \protect\\
				& \state.\deepimportfeature(p, r).\lastimportedreferencefeature = r'
			}
			{
				& w \mathematicaldefinition \creation{\maptype{\referencetype}{\referencetype}}{\makefeature} \protect\\
				& (r', w', \state') \mathematicaldefinition \deepimportrecursivewithmapfeature(p, \state.\handlerfeature(r), r, w, \state)
			}
	}

\begin{fortechnicalreport}
\feature
	{\lastimportedreferencefeature \colon \statetype \rightarrow \referencetype}
	{}
	{}
	{}

\end{fortechnicalreport}

The auxiliary command $\deepimportfeature$ is based on $\deepimportrecursivewithmapfeature$. This auxiliary function takes a tuple containing an importing processor $p$, a processor $q$ that handles the root of the object structure to be imported, a reference $r$ to be deep imported, and a state $\state$ to be modified. Note that the object referenced by $r$ is not necessarily handled by $q$ because this object might be on a different processor than the handler of the root of the object structure to be deep imported. The function returns another tuple with a reference $r''$ to the copied object structure and an updated state $\state''$.

\function
	{
		\where
			{\deepimportrecursivewithmapfeature(p, q, r, w, \state) = \tuple{r'', w'', \state''}}
			{
				& \tuple{r', w', \state'} \mathematicaldefinition \deepimportrecursivewithoutmapfeature(p, q, r, w, \state) \\
				& r'' \mathematicaldefinition 
						\left\{
							\begin{array}{ll}
								w.\valuefeature(r) & \condition{w.\keysfeature.\containsfeature(r)} \\
								r' & \condition{\neg w.\keysfeature.\containsfeature(r)}
							\end{array}
						\right. \\
				& w'' \mathematicaldefinition
					\left\{
						\begin{array}{ll}
							w & \condition{w.\keysfeature.\containsfeature(r)} \\
							w' & \condition{\neg w.\keysfeature.\containsfeature(r)}
						\end{array}
					\right. \\
				& \state'' \mathematicaldefinition
					\left\{
						\begin{array}{ll}
							\state & \condition{w.\keysfeature.\containsfeature(r)} \\
							\state' & \condition{\neg w.\keysfeature.\containsfeature(r)}
						\end{array}
					\right.
			}
	}

The auxiliary function $\deepimportrecursivewithmapfeature$ works hand in hand with the auxiliary function $\deepimportrecursivewithoutmapfeature$. They have the same signature and together they recursively traverse the object structure and make a deep copy of it. The functions must ensure that no object gets copied twice. For this purpose the functions take as an additional argument a map $w$ that maps references to objects in the input data structure to references in the copied data structure. A mapping from one reference $x$ to another reference $y$ means that the object referenced by $y$ is the copy of the object referenced by $x$. An updated map is returned as part of the result tuple. The auxiliary command $\deepimportfeature$ starts the recursion with an empty map. The auxiliary function $\deepimportrecursivewithmapfeature$ uses the map to determine whether the object referenced by $r$ has already been copied. In such a case, the result of the function comes from the map. Otherwise the auxiliary function $\deepimportrecursivewithmapfeature$ returns the result of the auxiliary function $\deepimportrecursivewithoutmapfeature$. The auxiliary function $\deepimportrecursivewithoutmapfeature$ creates a copy of the object referenced by $r$ and handles once routines. Finally, it returns a new reference $r'$, an updated map $w'$ in which $r$ is mapped to $r'$, and an updated state $\state'$.

\begin{forjournal}
\function
	{
		\where
			{\deepimportrecursivewithoutmapfeature(p, q, r, w, \state) = \tuple{r', w', \state'}}
			{
				& o \mathematicaldefinition \state.\referencedobjectfeature(r) \\
				& o'_{0} \mathematicaldefinition o.\copyfeature \\
				& \state'_{0} \mathematicaldefinition 
					\left\{
						\begin{array}{ll}
							\state.\addobjectfeature(p, o'_{0}) & \condition{\state.\handlerfeature(r) = q} \\
							\state.\addobjectfeature(\state.\handlerfeature(r), o'_{0}) & \otherwisecondition
						\end{array}
					\right. \\
				& w'_{0} \mathematicaldefinition 	w.\addfeature(r, \state'_{0}.\referencefeature(o'_{0})) \\
				& \set{a_{1}, \ldots, a_{n}} \mathematicaldefinition \setinference{a}{o.\attributevaluefeature(a) \in \referencetype \wedge o.\attributevaluefeature(a) \neq \voidvalue} \\
				& \forall {i \in \set{1, \ldots, n}} \colon \tuple{r'_{i}, w'_{i}, \state'_{i}} \mathematicaldefinition \deepimportrecursivewithmapfeature(p, q, o.\attributevaluefeature(a_{i}), w'_{i - 1}, \state'_{i - 1}) \\
				& \forall {i \in \set{1, \ldots, n}} \colon {o'_{i} \mathematicaldefinition o'_{i - 1}.\setattributevaluefeature(a_{i}, r'_{i})} \\
				& \state'_{x} \mathematicaldefinition \state'_{n}.\updatereferencefeature(\state'_{n}.\referencefeature(o'_{0}), o'_{n}) \\
				& \where
					{
						& \state'_{y} \mathematicaldefinition \state'_{x} \\
						& \indentation \begin{split}
							& .\setoncefunctionnotfreshfeature(\state'_{x}.\handlerfeature(\state'_{x}.\referencefeature(o'_{n})), f_{1}, \state'_{x}.\oncefunctionresultfeature(\state'_{x}.\handlerfeature(r), f_{1})) \\
							& .\thickspace \ldots \\
							& .\setoncefunctionnotfreshfeature(\state'_{x}.\handlerfeature(\state'_{x}.\referencefeature(o'_{n})), f_{w}, \state'_{x}.\oncefunctionresultfeature(\state'_{x}.\handlerfeature(r), f_{w})) \\
							& .\setonceprocedurenotfreshfeature(\state'_{x}.\handlerfeature(\state'_{x}.\referencefeature(o'_{n})), f_{w + 1}) \\
							& .\thickspace \ldots \\
							& .\setonceprocedurenotfreshfeature(\state'_{x}.\handlerfeature(\state'_{x}.\referencefeature(o'_{n})), f_{m})
						\end{split}
					}
					{
						& \set{f_{1}, \ldots, f_{w}} \mathematicaldefinition
							\setinference
								{x \in o.\classtypefeature.\functionsfeature}
								{
									& x.\isonceroutinefeature \wedge
									\exists c, d \colon \typingenvironmentderivation{x: (d, \bullet, c)} \wedge \\
									& \state'_{x}.\isonceroutinefreshfeature(\state'_{x}.\handlerfeature(\state'_{x}.\referencefeature(o'_{n})), x) \wedge \\
									& \neg 	\state'_{x}.\isonceroutinefreshfeature(\state'_{x}.\handlerfeature(r), x)
								} \\
						& \set{f_{w + 1}, \ldots, f_{m}} \mathematicaldefinition
							\setinference
								{x \in o.\classtypefeature.\proceduresfeature}
								{
									& x.\isonceroutinefeature \wedge \\
									& \state'_{x}.\isonceroutinefreshfeature(\state'_{x}.\handlerfeature(\state'_{x}.\referencefeature(o'_{n})), x) \wedge \\
									& \neg 	\state'_{x}.\isonceroutinefreshfeature(\state'_{x}.\handlerfeature(r), x)
								}
					} \\
				& r' \mathematicaldefinition \state'_{y}.\referencefeature(o'_{n}) \\
				& w' \mathematicaldefinition w'_{n} \\
				& \state' \mathematicaldefinition \state'_{y}
			}
	}

\end{forjournal}

The auxiliary function $\deepimportrecursivewithoutmapfeature$ is divided into several steps: a copy step, an attribute values update step, a clients update step, a once status update step, and a result generation step. Each of the steps has several definitions associated to it and each set of definitions depends on the definitions of the previous step. The following discussion goes through each of these steps in more details.

The copy step includes the definitions of $o$, $o'_{0}$, $\state'_{0}$ and $w'_{0}$. The definition $o$ is the object referenced by $r$, and the definition $o'_{0}$ makes a copy of $o$. In the next step, the function defines an updated state $\state'_{0}$ that includes the copy $o'_{0}$. There are two cases to be differentiated at this point. If $o$ is handled by $q$, then $o'_{0}$ must be handled by $p$. Otherwise $o'_{0}$ must be handled by the handler of $o$. The definition $w'_{0}$ is the updated map.

The attribute values update step recursively uses $\deepimportrecursivewithmapfeature$ to import all the non-void reference attribute values of $o$ using the updated map. This leads to an updated object with the deep imported values. This step includes the definition of $\set{a_{1}, \ldots, a_{n}}$, as well as the definitions of $\set{r'_{1}, \ldots, r'_{n}}$, $\set{w'_{1}, \ldots, w'_{n}}$, $\set{\state'_{1}, \ldots, \state'_{n}}$, and $\set{o'_{1}, \ldots, o'_{n}}$. The set $\set{a_{1}, \ldots, a_{n}}$ contains each attributes of $o$ whose value is a non-void reference. The function defines $\tuple{r'_{i}, w'_{i}, \state'_{i}}$ for $i = 1 \ldots n$ as a sequence of tuples. Each of the tuples is responsible for a single recursive deep import operation for one of the attributes in $\set{a_{1}, \ldots, a_{n}}$. Each such operation results in an updated map and an updated state that must be used in the next deep import operation. The result of this is an updated map $w'_{n}$, and updated state $\state'_{n}$, and references $r_{1}, \ldots, r_{n}$ to deep imported data structures. Finally, the function defines a sequence of updated objects $\set{o'_{1}, \ldots, o'_{n}}$ that ends with the updated object $o'_{n}$. The updated object has the values of the attributes $\set{a_{1}, \ldots, a_{n}}$ set to the deep imported data structures referenced by $r_{1}, \ldots, r_{n}$.

Until now, the function has an updated state $\state'_{n}$ that contains the initial copy $o'_{0}$. In the client update step, the function updates $\state'_{n}$ such that the reference to $o'_{0}$ points to the updated object $o'_{n}$. This is done in the clients update step. This step includes the definition $\state'_{x}$. Note that $\state'_{n}$ is derived from the state $\state'_{0}$, which includes the object $o'_{0}$.

\begin{fortechnicalreport}

\end{fortechnicalreport}

In a next step, the function takes care of the once routines of the imported object. For this, it defines a new state $\state'_{y}$ based on the state $\state'_{x}$. It defines $\set{f_{1}, \ldots, f_{w}}$ as the set of all non-separate once functions of $o$ that are fresh on the processor $\state'_{x}.\handlerfeature(\state'_{x}.\referencefeature(o'_{n}))$, which handles the copied object, but non-fresh on the processor $\state'_{x}.\handlerfeature(r)$, which handles the object referenced by $r$. Note that the two processor can be the same, in which case the set $\set{f_{1}, \ldots, f_{w}}$ is empty. Similarly, it defines the set $\set{f_{w + 1}, \ldots, f_{m}}$ for once procedures. For each once routine defined in this way, it updates the state $\state'_{x}$ such that the once status is taken over to the handler of the copied object. These definitions deal with the case where a once routine is fresh on the handler of the copied object, but non-fresh on the handler of the object referenced by $r$. Note that the remaining cases are implicitly taken care of because no change to the state is necessary. The result is the state $\state'_{y}$.

The last step defines the result of the function, based on the definitions of the preceding steps. The result generation step defines the resulting reference $r'$ to the imported object $o'_{n}$, the resulting map $w'$, and the resulting state $\state'$.

\paragraph{Setting values of formal arguments and the value of the current object entity}
The deep import operation is used in two ways. It is used when an expanded object handled by one processor gets used as an actual argument for a formal argument on another processor. The deep import operation also gets used when an expanded object handled by one processor gets returned to another processor. This section focuses on the argument passing aspect.

The auxiliary command $\pushenvironmentwithfeaturefeature$ defines a state in which a processor $p$ receives a new environment. The new environment is initialized for the execution of the feature $f$ with target reference $r_{0}$ and actual argument references $\tuple{r_{1}, \ldots, r_{n}}$. Actual arguments of expanded type must either be copied or they must be deep imported.

\feature
	{
		\pushenvironmentwithfeaturefeature \colon \statetype \rightarrow \processortype \nrightarrow \featuretype \rightarrow \referencetype \rightarrow \tupletype{} \nrightarrow \statetype
	}
	{\state.\pushenvironmentwithfeaturefeature(p, f, r_{0}, \tuple{r_{1}, \ldots, r_{n}})}
	{
		& \state.\processorregionsfeature.\processorsfeature.\containsfeature(p) \protect\\
		& f.\formalargumentsfeature.\countfeature = n \protect\\
		& \forall i \in \set{0, \ldots, n} \colon r_{i} \neq \voidvalue \rightarrow \state.\heapfeature.\referencesfeature.\containsfeature(r_{i})
	}
	{
		\where
			{
				\protect\begin{split}
					& \state.\pushenvironmentwithfeaturefeature(p, f, r_{0}, \tuple{r_{1}, \ldots, r_{n}}) = \protect\\
					& \indentation \state'_{n}.\setallfeature(\state'_{n}.\processorregionsfeature, \state'_{n}.\heapfeature, \state'_{n}.\storefeature.\pushenvironmentfeature(p, e))
				\protect\end{split}
			}	
			{
				& \state'_{0} \mathematicaldefinition \state \protect\\
				& \protect\begin{split}
					& \forall {i \in \set{1, \ldots, n}} \colon \tuple{\state'_{i}, r'_{i}} \mathematicaldefinition \protect\\ 
					& \indentation \left\{
						\protect\begin{split}
							& \condition{\exists d, q, c \colon \typingenvironmentderivation{f.\formalargumentsfeature(i): (d, q, c) \wedge c.\isexpandedclasstypefeature} \wedge r_{i} \neq \voidvalue \wedge \state'_{i - 1}.\handlerfeature(r_{i}) \neq p} \protect\\
							& \indentation \where
								{\tuple{\state_{x}, \state_{x}.\lastimportedreferencefeature}}
								{\state_{x} \mathematicaldefinition \state'_{i - 1}.\deepimportfeature(p, r_{i})} \protect\\
							& \condition{\exists d, q, c \colon \typingenvironmentderivation{f.\formalargumentsfeature(i): (d, q, c) \wedge c.\isexpandedclasstypefeature} \wedge r_{i} \neq \voidvalue \wedge \state'_{i - 1}.\handlerfeature(r_{i}) = p} \protect\\
							& \indentation \where
								{\tuple{\state_{x}, \state_{x}.\referencefeature(\state_{x}.\lastaddedobjectfeature)}}
								{\state_{x} \mathematicaldefinition \state'_{i - 1}.\addobjectfeature(p, \state'_{i - 1}.\heapfeature.\referencedobjectfeature(r_{i}).\copyfeature)} \protect\\
							& \otherwisecondition \protect\\
							& \indentation \tuple{\state'_{i - 1}, r_{i}}
						\protect\end{split}
					\right.
					\protect\end{split} \protect\\
				& w \mathematicaldefinition \creation{\environmenttype}{\makefeature} \protect\\
				& \indentation \protect\begin{split}
					& .\updatefeature(f.\formalargumentsfeature(1).\namefeature, r'_{1}) \thickspace \ldots \thickspace .\updatefeature(f.\formalargumentsfeature(n).\namefeature, r'_{n}) \protect\\
					& .\updatefeature(f.\localsfeature(1).\namefeature, \voidvalue) \thickspace \ldots \thickspace .\updatefeature(f.\localsfeature(f.\localsfeature.\countfeature).\namefeature, \voidvalue) \protect\\
					& .\updatefeature(\currententityname, r_{0})
				\protect\end{split} \protect\\
				& e \mathematicaldefinition
					\left\{
						\protect\begin{array}{ll}
							w & \condition{f \in \proceduretype} \protect\\
							w.\updatefeature(\resultentityname, \voidvalue) & \condition{f \in \functiontype}
						\protect\end{array}
					\right.
			}
		}

In a first step, the auxiliary command defines an updated state, in which $p$ gets a new initialized environment $e$. The updated state is based on an intermediate state $\state'_{n}$, which gets defined in a cascade of state updates with the goal of either copying or deep importing the actual arguments of expanded type. The cascade starts with the definition of a starting state $\state'_{0}$. For each formal argument, the cascade defines a tuple $\tuple{\state'_{i}, r'_{i}}$ with an updated state and a reference. If the corresponding actual argument is of reference class type, nothing needs to be done. If the actual argument is of expanded class type and the referenced object is not handled by $p$, then $p$ must deep import the object structure. This results in an updated state and a new reference to the deep imported object structure. If the actual argument is of expanded class type and the referenced object is handled by $p$, then the expanded object must be copied. This results in an updated state and a new reference to the copy. The resulting state $\state'_{n}$ contains all the deep imported and copied objects. The resulting references $r'_{1}, \ldots, r'_{n}$ will be used for values of the formal argument names.

In a next step, the command defines the environment $w$ as a new environment that gets updated to map formal argument names, local variable names, the current entity name, and the result entity name to the respective values. The names of the formal arguments get mapped to the references $r'_{1}, \ldots, r'_{n}$. Names of local variables are mapped to the void reference. The current entity name is mapped to the target reference.

The environment $w$ is the final environment $e$ in which the result name gets mapped to the void reference. This environment and the updated state $\state'_{n}$ define the result of the command. The auxiliary command $\pushenvironmentfeature$ pushes $e$ onto $p$'s stack of environments. The auxiliary command $\pushenvironmentfeature$ takes a processor $p$ and an environment $e$. It returns a state in which $e$ is pushed on top of $p$'s environment stack.

\begin{fortechnicalreport}
\feature
	{\pushenvironmentfeature \colon \statetype \rightarrow \processortype \nrightarrow \environmenttype \rightarrow \statetype}
	{\state.\pushenvironmentfeature(p, e)}
	{\state.\processorregionsfeature.\processorsfeature.\containsfeature(p)}
	{\state.\pushenvironmentfeature(p, e) = \state.\setallfeature(\state.\processorregionsfeature, \state.\heapfeature, \state.\storefeature.\pushenvironmentfeature(p, e))}

\end{fortechnicalreport}

The effect of a call to $\pushenvironmentwithfeaturefeature$ or a call to $\pushenvironmentfeature$ can be undone with a call to the auxiliary command $\popenvironmentfeature$. This auxiliary command takes a processor $p$ and removes the top environment from $p$'s stack of environments.

\begin{fortechnicalreport}
\feature
	{\popenvironmentfeature \colon \statetype \rightarrow \processortype \nrightarrow \statetype}
	{\state.\popenvironmentfeature(p)}
	{
		& \state.\processorregionsfeature.\processorsfeature.\containsfeature(p) \protect\\
		& \neg \state.\storefeature.\environmentsfeature(p).\isemptyfeature
	}
	{\state.\popenvironmentfeature(p) = \state.\setallfeature(\state.\processorregionsfeature, \state.\heapfeature, \state.\storefeature.\popenvironmentfeature(p))}

\end{fortechnicalreport}

\paragraph{Setting values of local variables and the value of the result entity}
The values of local variables and the value of the result entity are maintained in the store. The auxiliary command $\setenvironmentvaluefeature$ sets a value $v$ for the name $n$ in processor $p$'s top environment. For this, it defines an updated environment $e$ in which $n$ is set to $v$. It then defines an updated store $s$ by first removing the top environment and then adding the updated environment $e$. The updated store is then used to define an updated state. The updated state becomes the result of the auxiliary command.

\begin{fortechnicalreport}
\feature
	{\setenvironmentvaluefeature \colon \statetype \rightarrow \processortype \nrightarrow \nametype \nrightarrow \referencetype \cup \processortype \nrightarrow \statetype}
	{\state.\setenvironmentvaluefeature(p, n, v)}
	{
		& \state.\processorregionsfeature.\processorsfeature.\containsfeature(p) \protect\\
		& \neg \state.\storefeature.\environmentsfeature(p).\isemptyfeature \protect\\
		& v \in \referencetype \wedge v \neq \voidvalue \rightarrow \state.\heapfeature.\referencesfeature.\containsfeature(v) \protect\\
		& v \in \processortype \rightarrow \state.\processorregionsfeature.\processorsfeature.\containsfeature(v)
	}
	{
		\where
			{\state.\setenvironmentvaluefeature(p, n, v) = \state.\setallfeature(k, h, s)}
			{
				& e \mathematicaldefinition \state.\storefeature.\environmentsfeature(p).\topfeature.\updatefeature(n, v) \protect\\
				& k \mathematicaldefinition \state.\processorregionsfeature \protect\\
				& h \mathematicaldefinition \state.\heapfeature \protect\\
				& s \mathematicaldefinition \state.\storefeature.\popenvironmentfeature(p).\pushenvironmentfeature(p, e)
			}
	}

\end{fortechnicalreport}

\paragraph{Setting attribute values of the current object}
The auxiliary command $\setattributevaluefeature$ takes an object $o$, a name $n$, and a value $v$. It returns an updated state in which the attribute with name $n$ of object $o$ is set to the value $v$. In a first step, the auxiliary command defines an updated object with a call to $\setattributevaluefeature$. This updated object is then used to update the existing reference to $o$ in the state.

\begin{fortechnicalreport}
\feature
	{\setattributevaluefeature \colon & \statetype \rightarrow \objecttype \nrightarrow \nametype \nrightarrow \referencetype \cup \processortype \nrightarrow \statetype}
	{\state.\setattributevaluefeature(o, n, v)}
	{
		& \state.\heapfeature.\objectsfeature.\containsfeature(o) \protect\\
		& \exists a \in o.\classtypefeature.\attributesfeature \colon a.\namefeature = n \protect\\
		& v \in \referencetype \wedge v \neq \voidvalue \rightarrow \state.\heapfeature.\referencesfeature.\containsfeature(v) \protect\\
		& v \in \processortype \rightarrow \state.\processorregionsfeature.\processorsfeature.\containsfeature(v)
	}
	{
		\where
			{\state.\setattributevaluefeature(o, n, v) = \state.\updatereferencefeature(\state.\heapfeature.\referencefeature(o), o.\setattributevaluefeature(a, v))}
			{a \mathematicaldefinition o.\classtypefeature.\featurebynamefeature(n)}
	}

\end{fortechnicalreport}

\paragraph{Setting values of local variables, the value of the result entity, and attribute values of the current object in a unified way}
The auxiliary command $\setvaluefeature$ attaches a value $v$ to an entity with name $n$. The entity can either be a local variable or the result entity in the top environment of $p$. It can also be an attribute of the current object on $p$. In either case, the update affects an entity on $p$.

The definition of the resulting state is based on the auxiliary definitions $o$, $\state'$, and $v'$. The definition $o$ defines the current object, as defined by the top environment of processor $p$. The precondition makes sure that there is always such an environment on $p$ where the current object is defined. If $v$ is a reference and the referenced object is an object of reference class type, then $v$ can be attached directly to the entity with name $n$. If the object is an expanded object handled by processor $p$, then the referenced object must first be copied. Expanded objects handled by a processor different than $p$ must be deep imported. However, this is done right when the object gets returned from another processor to $p$. The definitions $\state'$ and $v'$ define a state and a value that are potentially updated according to these rules.

The state $\state'$ must be updated with the value $v'$. The update can either affect the current object on $p$ or it can affect the top environment of $p$. Attribute names of the current object, local variable names, and formal argument names are distinct. Therefore it is safe to first check whether the current object $o$ has an attribute with name $n$, in which case the current object gets updated with a $v'$. If the current object does not have such an attribute, then it is safe to assume that the top environment contains an entity with name $n$, in which case the top environment gets updated.

\feature
	{\setvaluefeature \colon \statetype \rightarrow \processortype \nrightarrow \nametype \nrightarrow \referencetype \cup \processortype \nrightarrow \statetype}
	{\state.\setvaluefeature(p, n, v)}
	{
		& \state.\processorregionsfeature.\processorsfeature.\containsfeature(p) \protect\\
		& \neg \state.\storefeature.\environmentsfeature(p).\isemptyfeature \wedge \state.\storefeature.\environmentsfeature(p).\topfeature.\namesfeature.\containsfeature(\currententityname) \protect\\
		& v \in \referencetype \wedge v \neq \voidvalue \rightarrow \state.\heapfeature.\referencesfeature.\containsfeature(v) \protect\\
		& v \in \processortype \rightarrow \state.\processorregionsfeature.\processorsfeature.\containsfeature(v)
	}
	{
		\where
			{
				& \state.\setvaluefeature(p, n, v) =
					\left\{
						\protect\begin{split}
							& \condition{\exists a \in o.\classtypefeature.\attributesfeature \colon a.\namefeature = n} \protect\\
							& \indentation \state'.\setattributevaluefeature(o, n, v') \protect\\
							& \otherwisecondition \protect\\
							& \indentation \state'.\setenvironmentvaluefeature(p, n, v')
						\protect\end{split}
					\right.
			}
			{
				& o \mathematicaldefinition \state.\heapfeature.\referencedobjectfeature(\state.\storefeature.\environmentsfeature(p).\topfeature.\valuefeature(\currententityname)) \protect\\
				& \protect\begin{split}
					& \tuple{\state', v'} \mathematicaldefinition \protect\\
					& \indentation \left\{
						\protect\begin{split}
							& \condition{v \in \referencetype \wedge v \neq \voidvalue \wedge \state.\heapfeature.\referencedobjectfeature(v).\classtypefeature.\isexpandedclasstypefeature \wedge \state.\handlerfeature(v) = p} \protect\\
							& \indentation \where
								{\tuple{\state_{x}, \state_{x}.\referencefeature(\state_{x}.\lastaddedobjectfeature)}}
								{\state_{x} \mathematicaldefinition \state.\addobjectfeature(p, \state.\heapfeature.\referencedobjectfeature(v).\copyfeature)} \protect\\
							& \otherwisecondition \protect\\
							& \indentation \tuple{\state, v}
						\protect\end{split}
					\right.
				\protect\end{split}
			}
	}

\paragraph{Setting values of once functions}
Values can also be stored in the status of once functions. A once function can be fresh or non-fresh. If the once function is non-fresh on a processor $p$, then there is a once result for the once function on $p$. A once function is set as non-fresh during the execution of the once function. The following discussion takes a look at how a processor can set the status of once routines in general, i.e., it considers both once functions and once procedures.

The auxiliary command $\setoncefunctionnotfreshfeature$ takes a processor $p$, a once function $f$, and a value $r$. It returns an updated state in which $f$ is set as non-fresh with the once result $r$. If $f$ is declared as non-separate, then $f$ is set as non-fresh on $p$ with the once result $r$. If $f$ is declared as separate with or without an explicit processor specification, then $f$ is set as non-fresh on all processors.

\begin{fortechnicalreport}
\feature
	{\setoncefunctionnotfreshfeature \colon & \statetype \rightarrow \processortype \nrightarrow \featuretype \nrightarrow \referencetype \nrightarrow \statetype}
	{\state.\setoncefunctionnotfreshfeature(p, f, r)}
	{
		& \state.\processorregionsfeature.\processorsfeature.\containsfeature(p) \protect\\
		& f \in \functiontype \wedge f.\isonceroutinefeature \protect\\
		& r \neq \voidvalue \rightarrow \state.\heapfeature.\referencesfeature.\containsfeature(r)
	}
	{
		\protect\begin{split}
			& \state.\setoncefunctionnotfreshfeature(p, f, r) = \protect\\
			& \indentation \state.\setallfeature(\state.\processorregionsfeature, \state.\heapfeature.\setoncefunctionnotfreshfeature(p, f, r), \state.\storefeature)
		\protect\end{split}
	}

\end{fortechnicalreport}

The auxiliary command $\setonceprocedurenotfreshfeature$ does the same for once procedures. It takes a processor $p$ and a once procedure $f$ and it returns a state in which $f$ is set as non-fresh on $p$.

\begin{fortechnicalreport}
\feature
	{\setonceprocedurenotfreshfeature \colon \statetype \rightarrow \processortype \rightarrow \featuretype \rightarrow \statetype}
	{\state.\setonceprocedurenotfreshfeature(p, f)}
	{
		& \state.\processorregionsfeature.\processorsfeature.\containsfeature(p) \protect\\
		& f \in \proceduretype \wedge f.\isonceroutinefeature
	}
	{
		\protect\begin{split}
			& \state.\setonceprocedurenotfreshfeature(p, f) = \protect\\
			& \indentation \state.\setallfeature(\state.\processorregionsfeature, \state.\heapfeature.\setonceprocedurenotfreshfeature(p, f), \state.\storefeature)
		\protect\end{split}
	}

\end{fortechnicalreport}

\subsubsection{Getting values}
This section takes a look at how a processor can read a value that got written with one of the mechanisms from \sectionreference{sec:setting values}.

\paragraph{Getting values of formal arguments, the value of the current object entity, values of local variables, and the value of the result entity}
The auxiliary query $\environmentsfeature$ takes a processor $p$ and returns the stack of environments for $p$. The auxiliary query $\environmentvaluefeature$ is more specialized. It takes a processor $p$ and a name $n$ and it returns the value stored under $n$ in the top environment of $p$.

\begin{fortechnicalreport}
\feature
	{\environmentsfeature \colon \statetype \rightarrow \processortype \nrightarrow \stacktype{\environmenttype}}
	{\state.\environmentsfeature(p)}
	{\state.\processorregionsfeature.\processorsfeature.\containsfeature(p)}
	{\state.\environmentsfeature(p) = \state.\storefeature.\environmentsfeature(p)}

\feature
	{\environmentvaluefeature \colon \statetype \rightarrow \processortype \nrightarrow \nametype \nrightarrow \referencetype \cup \processortype}
	{\state.\environmentvaluefeature(p, n)}
	{
		& \state.\processorregionsfeature.\processorsfeature.\containsfeature(p) \protect\\
		& \neg \state.\storefeature.\environmentsfeature(p).\isemptyfeature \wedge \state.\storefeature.\environmentsfeature(p).\topfeature.\namesfeature.\containsfeature(n)
	}
	{\state.\environmentvaluefeature(p, n) = \state.\storefeature.\environmentsfeature(p).\topfeature.\valuefeature(n)}

\end{fortechnicalreport}

\paragraph{Getting attribute values of the current object}
The auxiliary query $\attributevaluefeature$ takes an object $o$ and a name $n$ and returns the attribute value for the attribute with name $n$ of object $o$.

\begin{fortechnicalreport}
\feature
	{\attributevaluefeature \colon \statetype \rightarrow \objecttype \nrightarrow \nametype \nrightarrow \referencetype \cup \processortype}
	{\state.\attributevaluefeature(o, n)}
	{
		& \state.\heapfeature.\objectsfeature.\containsfeature(o) \protect\\
		& \exists a \in o.\classtypefeature.\attributesfeature \colon a.\namefeature = n
	}
	{
		\where
			{\state.\attributevaluefeature(o, n) = o.\attributevaluefeature(a)}
			{a \mathematicaldefinition o.\classtypefeature.\featurebynamefeature(n)}
	}

\end{fortechnicalreport}

\paragraph{Getting values of formal arguments, the value of the current object entity, values of local variables, the value of the result entity, and attribute values of the current object in a unified way}
The auxiliary queries $\environmentvaluefeature$ and $\attributevaluefeature$ define a new auxiliary query $\valuefeature$ that deals both with values in the top environment as well as with values stored in attributes of the current object. The auxiliary query $\valuefeature$ takes a processor $p$ and a name $n$ and it returns the value of $n$ in $p$'s current feature execution context. This context consists of the top environment and its reference to the current object. The auxiliary query requires that the execution context of processor $p$ is setup properly, i.e., there is a top environment with a reference to the current object. The precondition also states that either the top environment has the name $n$ registered or the current object has an attribute with name $n$. In any valid SCOOP program, any environment variable has a name that is distinct from the attribute names of the current object. This allows us to define the result of the auxiliary query in a simple way. If the name exists in the top environment, then the result is the value given by $\environmentvaluefeature$. Otherwise the name must be the name of an attribute of the current object, in which case the result is given by $\attributevaluefeature$. 

\feature
	{\valuefeature \colon \statetype \rightarrow \processortype \nrightarrow \nametype \nrightarrow \referencetype \cup \processortype}
	{\state.\valuefeature(p, n)}
	{
		& \where
			{
				& \state.\processorregionsfeature.\processorsfeature.\containsfeature(p) \protect\\
				& \neg \state.\storefeature.\environmentsfeature(p).\isemptyfeature \protect\\
				& e.\namesfeature.\containsfeature(\currententityname) \protect\\
				& e.\namesfeature.\containsfeature(n) \vee \exists a \in o.\classtypefeature.\attributesfeature \colon a.\namefeature = n
			}
			{
				& e \mathematicaldefinition \state.\storefeature.\environmentsfeature(p).\topfeature \protect\\
				& o \mathematicaldefinition \state.\heapfeature.\referencedobjectfeature(e.\valuefeature(\currententityname))
			}
	}
	{
		\state.\valuefeature(p, n) =
			\left\{
				\protect\begin{split}
					& \where
						{
							& \condition{e.\namesfeature.\containsfeature(n)} \protect\\
							& \indentation \state.\environmentvaluefeature(p, n)
						}
						{e \mathematicaldefinition \state.\storefeature.\environmentsfeature(p).\topfeature} \protect\\
					& \where
						{
							& \condition{\exists a \in o.\classtypefeature.\attributesfeature \colon a.\namefeature = n} \protect\\
							& \indentation \state.\attributevaluefeature(o, n)
						}
						{
							& e \mathematicaldefinition \state.\storefeature.\environmentsfeature(p).\topfeature \protect\\
							& o \mathematicaldefinition \state.\heapfeature.\referencedobjectfeature(e.\valuefeature(\currententityname))
						}
				\protect\end{split}
			\right.
	}

\paragraph{Getting values of once functions}
\begin{fortechnicalreport}
The following discussion takes a look at the auxiliary queries to access the status of once routines. It describes once routines in general, i.e., it also describe once procedures.
\end{fortechnicalreport}
The auxiliary query $\isonceroutinefreshfeature$ takes a processor $p$ and a once routine $f$. It returns whether $f$ is fresh on $p$ or not.

\begin{fortechnicalreport}
\feature
	{\isonceroutinefreshfeature \colon \statetype \rightarrow \processortype \nrightarrow \featuretype \nrightarrow \booleantype}
	{\state.\isonceroutinefreshfeature(p, f)}
	{
		& \state.\processorregionsfeature.\processorsfeature.\containsfeature(p) \protect\\
		& f.\isonceroutinefeature
	}
	{\state.\isonceroutinefreshfeature(p, f) = \state.\heapfeature.\isonceroutinefreshfeature(p, f)}

\end{fortechnicalreport}

For non-fresh once functions, the auxiliary query $\oncefunctionresultfeature$ returns the once result of $f$ on $p$.

\begin{fortechnicalreport}
\feature
	{\oncefunctionresultfeature \colon \statetype \rightarrow \processortype \nrightarrow \featuretype \nrightarrow \referencetype}
	{\state.\oncefunctionresultfeature(p, f)}
	{
		& \state.\processorregionsfeature.\processorsfeature.\containsfeature(p) \protect\\
		& f \in \functiontype \wedge f.\isonceroutinefeature \protect\\
		& \neg \state.\heapfeature.\isonceroutinefreshfeature(p, f)
	}
	{\state.\oncefunctionresultfeature(p, f) = \state.\heapfeature.\oncefunctionresultfeature(p, f)}

\end{fortechnicalreport}

\subsubsection{Locking}
This section explores the aspect of the facade that deals with locking. The auxiliary query $\isrequestqueuelockedfeature$ states whether a processor $p$'s request queue is locked or not. There are no auxiliary queries to distinguish between obtained and retrieved locks. Instead, the auxiliary queries $\requestqueuelocksfeature$ and $\callstacklocksfeature$ return the set of all request queue locks, respectively the set of all call stack locks of a processor $p$. These locks are only usable if they are not passed. This information can be retrieved with a call to the auxiliary query $\arelockspassedfeature$.

\begin{fortechnicalreport}
\feature
	{\isrequestqueuelockedfeature \colon \statetype \rightarrow \processortype \nrightarrow \booleantype}
	{\state.\isrequestqueuelockedfeature(p)}
	{\state.\processorregionsfeature.\processorsfeature.\containsfeature(p)}
	{\state.\isrequestqueuelockedfeature(p) = \state.\processorregionsfeature.\isrequestqueuelockedfeature(p)}

\feature
	{\requestqueuelocksfeature \colon \statetype \rightarrow \processortype \nrightarrow \settype{\processortype}}
	{\state.\requestqueuelocksfeature(p)}
	{\state.\processorregionsfeature.\processorsfeature.\containsfeature(p)}
	{\state.\requestqueuelocksfeature(p) = \state.\processorregionsfeature.\requestqueuelocksfeature(p)}

\feature
	{\callstacklocksfeature \colon \statetype \rightarrow \processortype \nrightarrow \settype{\processortype}}
	{\state.\callstacklocksfeature(p)}
	{\state.\processorregionsfeature.\processorsfeature.\containsfeature(p)}
	{\state.\callstacklocksfeature(p) = \state.\processorregionsfeature.\callstacklocksfeature(p)}

\feature
	{\arelockspassedfeature \colon \statetype \rightarrow \processortype \nrightarrow \booleantype}
	{\state.\arelockspassedfeature(p)}
	{\state.\processorregionsfeature.\processorsfeature.\containsfeature(p)}
	{\state.\arelockspassedfeature(p) = \state.\processorregionsfeature.\arelockspassedfeature(p)}

\end{fortechnicalreport}

The facade provides auxiliary commands for locking request queues, removing obtained request queue locks, unlocking request queues, delegating obtained request queue locks, passing locks, and revoking locks.

\begin{fortechnicalreport}
\feature
	{\lockrequestqueuesfeature \colon \statetype \rightarrow \processortype \nrightarrow \settype{\processortype} \nrightarrow \statetype}
	{\state.\lockrequestqueuesfeature(p, \overline{l})}
	{
		& \state.\processorregionsfeature.\processorsfeature.\containsfeature(p) \protect\\
		& \forall {x \in \overline{l}} \colon {\state.\processorregionsfeature.\processorsfeature.\containsfeature(x)} \protect\\
		& \forall {x \in \overline{l}} \colon {\state.\processorregionsfeature.\isrequestqueuelockedfeature(x) = \falsevalue}
	}
	{\state.\lockrequestqueuesfeature(p, \overline{l}) = \state.\setallfeature(\state.\processorregionsfeature.\lockrequestqueuesfeature(p, \overline{l}), \state.\heapfeature, \state.\storefeature)}

\feature
	{\popobtainedrequestqueuelocksfeature \colon \statetype \rightarrow \processortype \nrightarrow \statetype}
	{\state.\popobtainedrequestqueuelocksfeature(p)}
	{
		& \state.\processorregionsfeature.\processorsfeature.\containsfeature(p) \protect\\
		& \neg \state.\processorregionsfeature.\obtainedrequestqueuelocksfeature(p).\isemptyfeature \protect\\
		& \state.\processorregionsfeature.\arelockspassedfeature(p) = \falsevalue
	}
	{
		\protect\begin{split}
			& \state.\popobtainedrequestqueuelocksfeature(p) = \protect\\
			& \indentation \state.\setallfeature(\state.\processorregionsfeature.\popobtainedrequestqueuelocksfeature(p), \state.\heapfeature, \state.\storefeature)
		\protect\end{split}
	}

\feature
	{\unlockrequestqueuefeature \colon \statetype \rightarrow \processortype \nrightarrow \statetype}
	{\state.\unlockrequestqueuefeature(p)}
	{
		& \state.\processorregionsfeature.\processorsfeature.\containsfeature(p) \protect\\
		& \state.\processorregionsfeature.\isrequestqueuelockedfeature(p) = \truevalue \protect\\
		& \forall q \in \state.\processorregionsfeature.\processorsfeature \colon \neg \state.\processorregionsfeature.\obtainedrequestqueuelocksfeature(q).\flattenedfeature.\containsfeature(p)
	}
	{\state.\unlockrequestqueuefeature(p) = \state.\setallfeature(\state.\processorregionsfeature.\unlockrequestqueuefeature(p), \state.\heapfeature, \state.\storefeature)}

\feature
	{\delegateobtainedrequestqueuelocksfeature \colon \statetype \rightarrow \processortype \nrightarrow \settype{\processortype} \nrightarrow \statetype}
	{\state.\delegateobtainedrequestqueuelocksfeature(p, \overline{l})}
	{
		& \state.\processorregionsfeature.\processorsfeature.\containsfeature(p) \protect\\
		& \forall {x \in \overline{l}} \colon {\state.\processorregionsfeature.\processorsfeature.\containsfeature(x)} \protect\\
		& \forall {x \in \overline{l}} \colon {\neg \exists y \in \state.\processorregionsfeature.\processorsfeature \colon \state.\processorregionsfeature.\obtainedrequestqueuelocksfeature(y).\flattenedfeature.\containsfeature(x)} \protect\\
		& \forall {x \in \overline{l}} \colon {\state.\processorregionsfeature.\isrequestqueuelockedfeature(x) = \truevalue}
	}
	{
		\protect\begin{split}
			& \state.\delegateobtainedrequestqueuelocksfeature(p, \overline{l}) = \protect\\
			& \indentation \state.\setallfeature(\state.\processorregionsfeature.\delegateobtainedrequestqueuelocksfeature(p, \overline{l}), \state.\heapfeature, \state.\storefeature)
		\protect\end{split}
	}

\feature
	{\passlocksfeature \colon & \statetype \rightarrow \processortype \nrightarrow \processortype \nrightarrow \tupletype{\settype{\processortype}, \settype{\processortype}} \nrightarrow \statetype}
	{\state.\passlocksfeature(p, q, \tuple{\overline{l_{r}}, \overline{l_{c}}})}
	{
		& \state.\processorregionsfeature.\processorsfeature.\containsfeature(p) \wedge \state.\processorregionsfeature.\processorsfeature.\containsfeature(q) \protect\\
		& \forall {x \in \overline{l_{r}}} \colon  \state.\processorregionsfeature.\processorsfeature.\containsfeature(x) \wedge \forall {x \in \overline{l_{c}}} \colon  \state.\processorregionsfeature.\processorsfeature.\containsfeature(x) \protect\\
		& \protect\begin{split}
			& \forall x \in \overline{l_{r}} \colon \state.\processorregionsfeature.\obtainedrequestqueuelocksfeature(p).\flattenedfeature.\containsfeature(x) \vee \protect\\
			& \indentation  \state.\processorregionsfeature.\retrievedrequestqueuelocksfeature(p).\flattenedfeature.\containsfeature(x)
		\protect\end{split} \protect\\
		& \forall x \in \overline{l_{c}} \colon x = \state.\processorregionsfeature.\obtainedcallstacklockfeature(p) \vee \state.\processorregionsfeature.\retrievedcallstacklocksfeature(p).\flattenedfeature.\containsfeature(x) \protect\\
		& \state.\processorregionsfeature.\arelockspassedfeature(p) = \falsevalue
	}
	{\state.\passlocksfeature(p, q, \tuple{\overline{l_{r}}, \overline{l_{c}}}) = \state.\setallfeature(\state.\processorregionsfeature.\passlocksfeature(p, q, \tuple{\overline{l_{r}}, \overline{l_{c}}}), \state.\heapfeature, \state.\storefeature)}

\feature
	{\revokelocksfeature \colon \statetype \rightarrow \processortype \nrightarrow \processortype \nrightarrow \statetype}
	{\state.\revokelocksfeature(p, q)}
	{
		& \state.\processorregionsfeature.\processorsfeature.\containsfeature(p) \wedge \state.\processorregionsfeature.\processorsfeature.\containsfeature(q) \protect\\
		& \neg \state.\processorregionsfeature.\retrievedrequestqueuelocksfeature(q).\isemptyfeature \wedge \neg \state.\processorregionsfeature.\retrievedcallstacklocksfeature(q).\isemptyfeature \protect\\
		& \protect\begin{split}
			& \state.\processorregionsfeature.\retrievedrequestqueuelocksfeature(q).\topfeature \subseteq \protect\\
			& \indentation \state.\processorregionsfeature.\obtainedrequestqueuelocksfeature(p).\flattenedfeature \cup \state.\processorregionsfeature.\retrievedrequestqueuelocksfeature(p).\flattenedfeature
		\protect\end{split} \protect\\
		& \protect\begin{split}
			& \state.\processorregionsfeature.\retrievedcallstacklocksfeature(q).\topfeature \subseteq \protect\\
			& \indentation \set{\state.\processorregionsfeature.\obtainedcallstacklockfeature(p)} \cup \state.\processorregionsfeature.\retrievedcallstacklocksfeature(p).\flattenedfeature
		\protect\end{split} \protect\\
		& \protect\begin{split}
			& \state.\processorregionsfeature.\retrievedrequestqueuelocksfeature(q).\topfeature \cup \state.\processorregionsfeature.\retrievedcallstacklocksfeature(q).\topfeature \neq \set{} \rightarrow \protect\\
			& \indentation \state.\processorregionsfeature.\arelockspassedfeature(p) = \truevalue
		\protect\end{split} \protect\\
		& \state.\processorregionsfeature.\arelockspassedfeature(q) = \falsevalue
	}
	{\state.\revokelocksfeature(p, q) = \state.\setallfeature(\state.\processorregionsfeature.\revokelocksfeature(p, q), \state.\heapfeature, \state.\storefeature)}

\end{fortechnicalreport}

\begin{fortechnicalreport}
\subsection{Simplified state description}
Previous sections formalized the state as an instance of an ADT. Each instance is uniquely described by the values of all its queries. However, a description produced in this manner is not practical because it is too verbose. This section develops a \emph{simplified state description} that provides an abstract view on the queries of the state. The description is divided into four parts: the locks, the objects, the once status, and the environments. Each part is identified with a label.

The locks part shows for each processor the stack of obtained request queue locks, the stack of retrieved request queue locks, and the stack of retrieved call stack locks. It does not show the obtained call stack lock because these lock do not change. It uses two indicators to say when a processor's request queue is locked or unlocked. It also uses an indicator to state when a processor passed its locks. In absence of this indicator, the locks are not passed.

The objects part shows for each processor the handled objects with their references. It also shows the content of the objects. For objects other than arrays or objects of basic class type, the objects part shows a list of attribute values, omitting the ones that are void. For objects of basic type, the objects part shows the basic value. For arrays, it shows the cells of the array.

The once status part shows the status of each non-fresh routine. A once routine can be non-fresh either with respect to a subset of processors or with respect to all processors. In the first case, the once status part shows the once routine in connection with each processor in the subset. In the second case, it uses an indicator to denote all processors.

The environments part represents the store. It shows the environments for each processor.

\begin{example}[Simplified state description]
Assume a system with three processors $p_{1}$, $p_{2}$, and $p_{3}$. The following simplified state description shows a state in which all request queues are locked. Processor $p_{1}$ has a stack of obtained request queue locks with two items. On the bottom of the stack there is the set $\set{p_{2}}$ and on the top there is the set $\set{p_{3}}$. Processor $p_{1}$'s stack of retrieved request queue locks and retrieved call stack locks each consists of two empty sets. Processor $p_{1}$ passed its locks. From $p_{3}$'s entry, one can see that these locks have been passed from $p_{1}$ to $p_{3}$. Processor $p_{2}$ does not have any locks, except its own call stack lock.

In the objects part, one can see that processor $p_{1}$ handles an object $o_{1}$ that is referenced by $r_{1}$. Processors $p_{2}$ and $p_{3}$ each handle multiple objects. Object $o_{5}$ is an object with an attribute $\mathit{id}$ that references object $o_{6}$ through the reference $r_{6}$. Objects $o_{6}$ and $o_{4}$ are objects with the integer values $2$ and $1$. Object $o_{2}$ is a two dimensional array with $2 \times 2$ cells. Each of the cells references the object $o_{3}$ through the reference $r_{3}$.

The once status part shows two items for the once function $\mathit{id}$ and the once procedure $\mathit{initialize}$ of class \lstinline[language=SCOOP]!APPLICATION!. The once function $\mathit{id}$ is non-fresh with the value $r_{4}$ on processor $p_{2}$. The once procedure $\mathit{initialize}$ is non-fresh on all processors in the system.

The last part shows the environments for the processors. Processor $p_{1}$ has a stack with two environments. The environment on the left is at the bottom of the stack and the environment on the right is at the top of the stack. Processor $p_{2}$ has no environments and processor $p_{3}$ has one environment. In $p_{3}$'s environment there are three mappings. The entity $\mathit{root}$ has the value $r_{1}$, the current entity has the value $r_{5}$, and the result entity has the void value.

\isolatedsimplifiedstate
	{
		& \simplifiedstatelocksentry
			{p_{1}}
			{\set{p_{2}}, \set{p_{3}}}
			{\set{}, \set{}}
			{\set{}, \set{}}
			{\simplifiedstatelockedindicator}
			{\simplifiedstatepassedlocksindicator} \\
		& \simplifiedstatelocksentry
			{p_{2}}
			{}
			{}
			{}
			{\simplifiedstatelockedindicator}
			{\simplifiedstatenopassedlocksindicator} \\
		& \simplifiedstatelocksentry
			{p_{3}}
			{}
			{\set{p_{2}, p_{3}}}
			{\set{p_{1}}}
			{\simplifiedstatelockedindicator}
			{\simplifiedstatenopassedlocksindicator}
	}
	{
		& \simplifiedstateobjectsentry
			{p_{1}}
			{
				\simplifiedstatereferencedobject{r_{1}}{o_{1}}
			} \\
		& \simplifiedstateobjectsentry
			{p_{2}}
			{
				\simplifiedstatereferencedobject{r_{2}}{o_{2}[[r_{3}, r_{3}],[r_{3}, r_{3}]]},
				\simplifiedstatereferencedobject{r_{3}}{o_{3}},
				\simplifiedstatereferencedobject{r_{4}}{o_{4}(1)}
			} \\
		& \simplifiedstateobjectsentry
			{p_{3}}
			{
				\simplifiedstatereferencedobject{r_{5}}{o_{5}(\simplifiedstateentityvalue{id}{r_{6}})},
				\simplifiedstatereferencedobject{r_{6}}{o_{6}(2)}
			}
	}
	{
		& \simplifiedstateoncestatusentry
			{p_{2}}
			{
				\simplifiedstateoncefunctionstatus{APPLICATION}{id}{r_{4}}
			} \protect\\
		& \simplifiedstateoncestatusentry
			{\simplifiedstateallprocessorsindicator}
			{
				\simplifiedstateonceprocedurestatus{APPLICATION}{initialize}
			}
	}
	{
		& \simplifiedstateenvironmentsentry
			{p_{1}}
			{
				\simplifiedstateentityvalue{node}{r_{2}}, \simplifiedstatecurrententityvalue{r_{1}} \simplifiedstateenvironmentsentryseparator \simplifiedstateentityvalue{node}{r_{5}}, \simplifiedstatecurrententityvalue{r_{1}}  
			} \\
		& \simplifiedstateenvironmentsentry
			{p_{2}}
			{} \\
		& \simplifiedstateenvironmentsentry
			{p_{3}}
			{
				\simplifiedstateentityvalue{root}{r_{1}}, \simplifiedstatecurrententityvalue{r_{5}}, \simplifiedstateresultentityvalue{\voidvalue}
			}				
	}
\end{example}
\end{fortechnicalreport}

\begin{fortechnicalreport}
\section{Formalization of execution}
\end{fortechnicalreport}
\begin{forjournal}
\section{Formalization of Execution}
\end{forjournal}
\label{sec:execution-formalization}
This section formalizes the execution of a SCOOP program. It explains the general approach, defines the starting point of the execution, and explains the rules that drive the execution. The rules are divided into rules for mechanisms and rules for code elements.

\begin{fortechnicalreport}
\subsection{General approach}
\end{fortechnicalreport}
\begin{forjournal}
\subsection{General Approach}
\end{forjournal}
The formalization is based on \emph{structural operational semantics} \cite{plotkin:2004:structural_operational_semantics}, combined with parts of the terminology from Ostroff et al. \cite{ostroff-torshizi-huang-schoeller:2008:formal_semantics_for_SCOOP}. The idea behind a structural operational semantics is to define the behavior of a program in terms of its parts, i.e., the syntactical elements of the program. Such a semantics is intuitive because it talks directly about elements in the code. It is a very powerful semantics because it allows us to apply structural induction as a proof technique.

\subsubsection{Computations}
A \emph{computation} models the execution of a SCOOP program. It is a sequence of configurations, where each non-initial configuration is derived from a previous configuration through a transition. Each configuration defines a state and a list of statements for each processor. Each transition is described by an inference rule that maps one configuration to another. The transition from one configuration to the next models an atomic step of one processor. The concurrent execution of a SCOOP program is modeled by the interleaved transitions taken by different processors.
\begin{example}[Modeling of parallel execution]
Suppose there are two processors $p$ and $q$. Processor $p$ executes the following sequence of statements: $s_{p, 1} \statementseparator s_{p, 2}$. In parallel, processor $q$ executes the following sequence of statements: $s_{q, 1}$. This execution is modeled by any of the following simplified computations: $s_{p, 1} \statementseparator s_{p, 2} \statementseparator s_{q, 1}$ or $s_{p, 1} \statementseparator s_{q, 1} \statementseparator s_{p, 2}$ or $s_{q, 1} \statementseparator s_{p, 1} \statementseparator s_{p, 2}$.
\end{example}

\subsubsection{Configurations}
A \emph{configuration} models a snapshot in the execution of a SCOOP program. A configuration consists of a state and a set of processors, each with a queue of statements. The state is an instance of $\statetype$. A \emph{schedule} models the processors and the associated queues, called \emph{action queues}. Each processor must execute the statements in its action queue in a FIFO order. The beginning of the action queue contains the statements for the features that are being executed at the moment. The order of these statements models the way the call stack orders feature executions. The tail of the action queue is the request queue of the processor. A call stack lock is the right to add a feature request to the beginning of the action queue and a request queue lock is the right to add a feature request to the end of the action queue. The notation for a configuration with processors $p_{1}, \ldots, p_{n}$, respective action queues $s_{1}, \ldots, s_{n}$, and state $\state$ is:

\isolatedsinglelineconfiguration{p_{1} :: s_{1} \processorseparator \ldots \processorseparator p_{n} :: s_{n}}{\state}

The processor separator $\processorseparator$ is commutative and associative, i.e., $p_{1} :: s_{1} \processorseparator p_{2} :: s_{2} = p_{2} :: s_{2} \processorseparator p_{1} :: s_{1}$ and $p_{1} :: s_{1} \processorseparator (p_{2} :: s_{2} \processorseparator p_{3} :: s_{3}) = (p_{1} :: s_{1} \processorseparator p_{2} :: s_{2}) \processorseparator p_{3} :: s_{3}$. Within an action queue, $\statementseparator$ separates statements. The configuration is \emph{well-defined} if and only if $\neg \exists i, j \in \set{1, \ldots, n} \colon p_{i} = p_{j}$.

\subsubsection{Statements}
A \emph{statement} is an element of the action queue. A statement is either an instruction or an operation. An \emph{instruction} is user syntax, i.e.\ an action that occurs explicitly in the SCOOP program. An \emph{operation} is run-time syntax, i.e.\ an action that does not explicitly occur in a SCOOP program. For example, locking of request queues is not an action that is explicit in a SCOOP program. Instead, locking is based on the formal argument list. It is done implicitly before a feature gets executed.

\subsubsection{Transitions}
A \emph{transition} takes a system in a start configuration and leaves it in a result configuration. The following shows the general form of a transition definition that declares a start configuration $\singlelineconfiguration{P}{\state}$ with schedule $P \mathematicaldefinition p_{1} :: s_{1} \processorseparator \ldots \processorseparator p_{n} :: s_{n}$ and a result configuration $\singlelineconfiguration{P'}{\state'}$ with schedule $P' \mathematicaldefinition p'_{1} :: s'_{1} \processorseparator \ldots \processorseparator p'_{m} :: s'_{m}$:

\isolatedsinglelinetransition
	{\configuration{P}{\state}}
	{\configuration{P'}{\state'}}

The typing environment $\typingenvironment$ can be used in the transition definition to access static information about the SCOOP program.

\subsubsection{Inference rules}
An \emph{inference rule} describes the circumstances under which a transition can be used. The inference rule has a premise and a conclusion. The \emph{conclusion} is the transition and the \emph{premise} describes the circumstances under which the transition can be used. The premise consists of a number of transitions and a side condition. The premise is satisfied if all transitions in the premise can be taken and if the side condition is true.
\begin{fortechnicalreport}
The following shows a template for inference rules:

\singlelineinferencerule
	{General Inference Rule Template}
	{
		\mbox{\it{side condition}} \\
		\singlelinetransition{\configuration{P_{1}}{\state_{P_{1}}}}{\configuration{P_{1}'}{\state_{P_{1}}'}} \\
		\ldots \\
		\singlelinetransition{\configuration{P_{n}}{\state_{P_{n}}}}{\configuration{P_{n}'}{\state_{P_{n}}'}}
	}
	{\configuration{P_{n + 1}}{\state_{P_{n + 1}}}}
	{\configuration{P_{n + 1}'}{\state_{P_{n + 1}}'}}

\end{fortechnicalreport}
In this formalization, most of the rules have no transition in the premise. The following \emph{simplified inference rule template} takes this into account:

\singlelineinferencerule
	{Simplified Inference Rule Template}
	{
		\mbox{\it{condition}} \\
		\mbox{\it{new state}} \thinspace \state' \thinspace \mbox{\it{definition}}\\
		\mbox{\it{fresh channels definitions}}
	}
	{\configuration{P}{\state}}
	{\configuration{P'}{\state'}}

The side condition has three parts. The first part defines a \emph{condition} that is based on the typing environment and the start configuration. The second part is the \emph{new state definition} that defines the state of the result configuration. This new state is based on the state in the start configuration. The last part consists of the \emph{fresh channels definitions}. Auxiliary definitions can be used in the condition, the new state definition, and the fresh channels definitions. The inference rule can mention features of $\statetype$. The preconditions of these features serve as additional conditions in the side condition.

The following inference rule generalizes transitions by adding processors both to the start configuration and to the result configuration. These additional processors run in parallel but do not take any actions during the generalized transition.

\singlelineinferencerule
	{Parallelism}
	{
		\singlelinetransition
			{\configuration{P}{\state}}
			{\configuration{P'}{\state'}}
	}
	{\configuration{P \processorseparator Q}{\state}}
	{\configuration{P' \processorseparator Q}{\state'}}

\subsubsection{Scheduling}\label{sec:scheduling}
Before a processor can execute a feature it must obtain locks and it must wait until the wait condition is satisfied. A locking request encapsulates these two requirements; it consists of the requested locks and the wait condition. At every moment, multiple processors can have conflicting locking requests. The scheduler is the arbiter for these conflicts. The scheduler takes locking requests and stores them in a queue. It then approves locking requests according to a certain scheduling algorithm.

The model permits a number of possible scheduling algorithms. The algorithms differ in their level of fairness and their performance. This formalization does not focus on a particular scheduling algorithm. Instead, it uses the conditions of the inference rules to express locking requests. If more than one processor satisfies the conditions, then any of these processors can proceed.

\begin{fortechnicalreport}
\subsection{Initial configuration}
\end{fortechnicalreport}
\begin{forjournal}
\subsection{Initial Configuration}
\end{forjournal}
The initial configuration is defined by the SCOOP program. Each SCOOP program defines a root class type $c$ and a root procedure $f$. The root procedure is a creation procedure of the root class type that has no formal arguments and no precondition.

In the beginning, the runtime generates a bootstrap processor $p$ and root processor $q$ with a root object of the root class type. The request queue of the root processor is locked on behalf of the bootstrap processor. This defines our initial state $\state$:

\isolateddefinition
	{
		& \state_{x} \mathematicaldefinition \creation{\statetype}{\makefeature} \\
		& \state_{y} \mathematicaldefinition \state_{x}.\addprocessorfeature(\state_{x}.\newprocessorfeature) \\
		& p \mathematicaldefinition \state_{y}.\lastaddedprocessorfeature \\
		& \state_{z} \mathematicaldefinition \state_{y}.\addprocessorfeature(\state_{y}.\newprocessorfeature) \\
		& q \mathematicaldefinition \state_{z}.\lastaddedprocessorfeature \\
		& \state_{w} \mathematicaldefinition \state_{z}.\addobjectfeature(q, \state_{z}.\newobjectfeature(c)) \\
		& r \mathematicaldefinition \state_{w}.\referencefeature(\state_{w}.\lastaddedobjectfeature) \\
		& \state \mathematicaldefinition \state_{w}.\lockrequestqueuesfeature(p, \set{q})
	}

The bootstrap processor first asks the root processor to execute the root procedure on the root object and then asks the root processor to unlock its request queue as soon as it finished the execution. The bootstrap processor can do this because it has the request queue lock on the root processor. Finally, the bootstrap processor removes the request queue lock from its stack of obtained request queue locks. This is shown in the following initial configuration:

\isolatedconfiguration
	{
		p :: \ &\calloperation(r, f, \tuple{}, \tuple{}) \statementseparator \\
		& \issueoperation(q, \unlockrequestqueueoperation) \statementseparator \\
		& \popobtainedrequestqueuelocksoperation \processorseparator \\
		q :: &
	}
	{\state}

The statements $\calloperation$, $\issueoperation$, $\unlockrequestqueueoperation$, and $\popobtainedrequestqueuelocksoperation$ are operations. In a nutshell, the $\calloperation(r, f, \tuple{}, \tuple{})$ operation asks the handler of the target $r$ to make a call to the feature $f$ on target $r$. The $\unlockrequestqueueoperation$ operation unlocks the request queue of the processor that executes the operation. The $\issueoperation(q, \unlockrequestqueueoperation)$ operation adds the $\unlockrequestqueueoperation$ operation to $q$'s action queue. The $\popobtainedrequestqueuelocksoperation$ operation removes the top element from the stack of obtained request queue locks.

\begin{fortechnicalreport}
\begin{example}[Initial configuration]
This example defines the initial configuration of a share market application. The domain of the application consists of a number of markets, a number of investors, and a number of issuers. Each issuer can offer a number of shares on each market. Each investor can have an amount of cash available on each market. With this cash, the investor can buy the shares that are available on the market. Investors can sell a share on the market where they bought the share. Selling shares increases the investor's amount of cash on the market. Each market determines the price for each share. Financial regulations require the investors to keep track of the markets on which they operate. For simplicity, the price is constant and the application is restricted to one market, two investors, and one issuer with one share.

The class \lstinline[language=SCOOP]!MARKET! represents the market; the class \lstinline[language=SCOOP]!INVESTOR! represents the investor. The issuers are represented through identifiers of class \lstinline[language=SCOOP]!INTEGER!. The root class \lstinline[language=SCOOP]!APPLICATION! contains the root procedure \lstinline[language=SCOOP]!make!, where the actors get created and where the trade begins.

The execution starts with a bootstrap processor $p_{0}$, a root processor $p_{1}$ and a root object $o_{0}$ of root class type $\mathit{APPLICATION}$. The root object is referenced by $r_{0}$. The following initial configuration shows this:

\isolatedconfiguration
	{
		p_{0} :: \ & \calloperation(r_{0}, \mathit{make}, \tuple{}, \tuple{}) \statementseparator \\
		& \issueoperation(p_{1}, \unlockrequestqueueoperation) \statementseparator \\
		& \popobtainedrequestqueuelocksoperation \processorseparator \\
		p_{1} :: \ &
	}
	{
		\simplifiedstate
			{
				& \simplifiedstatelocksentry
					{p_{0}}
					{\set{p_{1}}}
					{}
					{}
					{\simplifiedstateunlockedindicator}
					{\simplifiedstatenopassedlocksindicator} \\
				& \simplifiedstatelocksentry
					{p_{1}}
					{}
					{}
					{}
					{\simplifiedstatelockedindicator}
					{\simplifiedstatenopassedlocksindicator}
			}
			{
				& \simplifiedstateobjectsentry
					{p_{0}}
					{} \\
				& \simplifiedstateobjectsentry
					{p_{1}}
					{
						\simplifiedstatereferencedobject{r_{0}}{o_{0}}
					}
			}
			{}
			{
				& \simplifiedstateenvironmentsentry
					{p_{0}}
					{} \\
				& \simplifiedstateenvironmentsentry
					{p_{1}}
					{}			
			}
	}

\end{example}
\end{fortechnicalreport}

\subsection{Mechanisms}\label{sec:mechanisms}
Mechanisms are the machinery for the execution of code elements. This section studies these mechanisms.

\subsubsection{Issuing mechanism}
With the issuing mechanism, a processor $p$ can add statements to the action queue of a processor $q$. It uses the $\issueoperation$ operation to get a result configuration in which a processor's action queue is extended with the new statements. There are two main cases: $p$ adds the statements to its own action queue, i.e., $p = q$, or $p$ adds the statements to the action queue of a different processor, i.e., $p \neq q$. The first case is the non-separate case and the second one is the separate case.

For the non-separate case $p$ puts the statements to the beginning of $q$'s action queue, which is the same as putting the statements on top of the call stack. This requires that $p$ is in possession of its own call stack lock.

\singlelineinferencerule
	{Issue Operation -- Non-Separate}
	{
		q = p \\
		\neg \state.\arelockspassedfeature(p) \\
		\state.\callstacklocksfeature(p).\containsfeature(q)
	}
	{\configuration{p :: \issueoperation(q, s_{w}) \statementseparator s_{p}}{\state}}
	{\configuration{p :: s_{w} \statementseparator s_{p}}{\state}}

For the separate case there is a difference between a normal and a callback case. In the normal case, $p$ adds the statements to the end of $q$'s action queue. This case requires that $p$ is in possession of $q$'s request queue lock. To distinguish the normal case from the callback case, this case also requires that $q$ does not have a lock on $p$.

\singlelineinferencerule
	{Issue Operation -- Separate}
	{
		q \neq p \\
		\neg \state.\arelockspassedfeature(p) \\
		\state.\requestqueuelocksfeature(p).\containsfeature(q) \\
		\neg (\state.\requestqueuelocksfeature(q).\containsfeature(p) \vee \state.\callstacklocksfeature(q).\containsfeature(p))
	}
	{\configuration{p :: \issueoperation(q, s_{w}) \statementseparator s_{p} \processorseparator q :: s_{q}}{\state}}
	{\configuration{p :: s_{p} \processorseparator q :: s_{q} \statementseparator s_{w}}{\state}}

The callback case occurs if $q$ has a lock on $p$. In this situation, $p$ could issue a statement $s_{w}$ on $q$ and then wait for $q$ to complete. On the other side, processor $q$ could already be waiting for $p$ to complete. Processor $q$ would be waiting for $p$ to finish and $p$ would be waiting for $q$ to finish. However, since $s_{w}$ would be at the end of $q$'s action queue and $q$ would be waiting there cannot be any progress. This type of deadlock can be prevented by adding $s_{w}$ not to the of $q$'s action queue but to the beginning. This will make sure that $q$ can execute the statement right away and hence $p$ can continue. This in return will enable $q$ to continue. As a prerequisite, $p$ must possess $q$'s call stack lock.

\singlelineinferencerule
	{Issue Operation -- Separate Callback}
	{
		q \neq p \\
		\neg \state.\arelockspassedfeature(p) \\
		\state.\callstacklocksfeature(p).\containsfeature(q) \\
		\state.\requestqueuelocksfeature(q).\containsfeature(p) \vee \state.\callstacklocksfeature(q).\containsfeature(p)
	}
	{\configuration{p :: \issueoperation(q, s_{w}) \statementseparator s_{p} \processorseparator q :: s_{q}}{\state}}
	{\configuration{p :: s_{p} \processorseparator q :: s_{w} \statementseparator s_{q}}{\state}}

\subsubsection{Delegated execution mechanism}
This section discusses how a processor $q$ can delegate the execution of statements to a different processor $p$. This mechanism is useful for the evaluation of asynchronous postconditions. Processor $q$ must make sure that the statements make sense in the context of processor $p$. The names that occur in these statements must be defined in the top environment of $p$ and $p$ must have the necessary locks to execute the statements. 
Statements that fulfill the following conditions can be delegated:
\begin{itemize}
	\item All names that occur in the statements are defined in $q$'s top environment.
	\item Their execution only requires the top set of $q$'s stack of obtained request queue locks.
\end{itemize}
These conditions exclude statements that involve non-separate calls or separate callbacks because such calls require a call stack lock. If these conditions are met, $q$ can transfer its top environment and the top of its obtained request queue locks to $p$. Given this context, $p$ can then execute the delegated statements instead of $q$.

The $\executedelegatedoperation(s_{w}, x, \set{q_{1}, \ldots, q_{m}})$ operation sets up a new context on $p$ with an environment $x$ and obtained request queue locks $\set{q_{1}, \ldots, q_{m}}$. To set up the new context, the operation uses a combination of the commands $\pushenvironmentfeature$ and $\delegateobtainedrequestqueuelocksfeature$. The command $\delegateobtainedrequestqueuelocksfeature$ requires that the request queue locks $\set{q_{1}, \ldots, q_{m}}$ are not in possession of another processor anymore. It also requires that the request queues of $\set{q_{1}, \ldots, q_{m}}$ are locked. Once the context is set up, processor $p$ executes the statements $s_{w}$ and then gets rid of the context, using the $\leavedelegatedexecutionoperation$ operation.

To delegate the execution of the statements $s_{w}$, processor $q$ must make sure that its top environment $x$ is set up correctly and it must make sure that the top set of its obtained request queue locks contains all locks $\set{q_{1}, \ldots, q_{m}}$ that are necessary for the execution of $s_{w}$. Processor $q$ must then issue a $\executedelegatedoperation(s_{w}, x, \set{q_{1}, \ldots, q_{m}})$ operation to processor $p$. Processor $q$ must then remove $\set{q_{1}, \ldots, q_{m}}$ from its stack of obtained request queue locks so that the $\delegateobtainedrequestqueuelocksfeature$ operation can take place.

\inferencerule
	{Execute Delegated Operation}
	{
		\forall x \in \set{q_{1}, \ldots, q_{m}} \colon \neg \exists y \in \state.\processorsfeature \colon \state.\requestqueuelocksfeature(y).\containsfeature(x) \\
		\forall x \in \set{q_{1}, \ldots, q_{m}} \colon \state.\isrequestqueuelockedfeature(x) \\
		\state' \mathematicaldefinition \state.\pushenvironmentfeature(p, x).\delegateobtainedrequestqueuelocksfeature(p, \set{q_{1}, \ldots, q_{m}})
	}
	{\configuration{p :: \executedelegatedoperation(s_{w}, x, \set{q_{1}, \ldots, q_{m}}) \statementseparator s_{p}}{\state}}
	{\configuration{p :: s_{w} \statementseparator \leavedelegatedexecutionoperation \statementseparator s_{p}}{\state'}}

\singlelineinferencerule
	{Leave Delegated Execution Operation}
	{
		\neg \state.\environmentsfeature(p).\isemptyfeature \\
		\neg \state.\obtainedrequestqueuelocksfeature(p).\isemptyfeature \\
		\state' \mathematicaldefinition \state.\popenvironmentfeature(p).\popobtainedrequestqueuelocksfeature(p)
	}
	{\configuration{p :: \leavedelegatedexecutionoperation \statementseparator s_{p}}{\state}}
	{\configuration{p :: s_{p}}{\state'}}

\subsubsection{Notification mechanism}\label{sec:notification mechanism}
Processors can notify each other. A notification can optionally include a value. The formalization uses channels to describe such communication. Channels are described in Milner's $\pi$-calculus \cite{milner:1999:Pi_calculus}. In the $\pi$-calculus, the expression $c(x).P$ denotes a process that is waiting for a notification sent on a channel $c$. Once the notification has been received, the value of the notification is bound to the variable $x$ and the process continues with the expression $P$. The notification comes from a process that executes $\overline{c}y.Q$ to emit the value $y$ on the channel $c$ before executing $Q$.

The formalization reuses the channel idea in two flavors: once as a notification mechanism with a value and once as a notification mechanism without a value. A processor sends a notification with a value $r$ over a channel $a$ as it executes the operation $\resultoperation(a, r)$. Similarly, the process sends a notification without a value over a channel $a$ by executing the operation $\notifyoperation(a)$. For both cases, any processor can wait for a notification by executing the operation $\waitoperation(a)$. In case a notification on a channel $a$ carries a value, the value can be accessed with $a.\datafeature$. This way of accessing the value of a channel is different from the way it is done in the $\pi$-calculus. In the $\pi$-calculus, each value is bound to a variable. This formalization does not define a new variable for the value. Instead, it uses $a.\datafeature$ to identify the value of a channel $a$.

A number of inference rules describe the interaction between a processor that sent a notification over a channel and a processor that is waiting for a notification over the same channel. Two main cases can be distinguished: either a processor sends a notification to itself or it sends a notification to a different processor. The first case is the non-separate case and the latter case is the separate case. In each of these two main cases, the channel carries a notification with or without a value. For each of these sub cases, there is one inference rule.

In the non-separate case, one processor has a $\resultoperation(a, r)$ operation or a $\notifyoperation(a)$ operation at the beginning of its action queue and a $\waitoperation(a)$ operation on the same channel later in the action queue. In this case, the $\waitoperation(a)$ operation can be removed along with the $\resultoperation(a, r)$ operation, respectively the $\notifyoperation(a)$ operation. If the channel carries a value, then the value must be installed on the processor, by substituting all occurrences of $a.\datafeature$ with the posted value in all the statements $s_{p}$ after the $\waitoperation(a)$ operation.

\singlelineinferencerule
	{Wait and Result Operation -- Non-Separate}
	{}
	{\configuration{p :: \resultoperation(a, r) \statementseparator s_{w} \statementseparator \waitoperation(a) \statementseparator s_{p}}{\state}}
	{\configuration{p :: s_{w} \statementseparator s_{p} \substitution{r}{a.\datafeature}}{\state}}

\singlelineinferencerule
	{Wait and Notify Operation -- Non-Separate}
	{}
	{\configuration{p :: \notifyoperation(a) \statementseparator s_{w} \statementseparator \waitoperation(a) \statementseparator s_{p}}{\state}}
	{\configuration{p :: s_{w} \statementseparator s_{p}}{\state}}

In the separate case, one processor has a $\resultoperation(a, r)$ or a $\notifyoperation(a)$ operation at the beginning of its action queue and a different processor has a $\waitoperation(a)$ somewhere in its action queue. In this situation, the $\waitoperation(a)$, $\resultoperation(a, r)$, and $\notifyoperation(a)$ can be removed from the action queues. In case the notification has a value, the value can be installed in the statements $s_{p}$, after the $\waitoperation(a)$ operation.

\singlelineinferencerule
	{Wait and Result Operation -- Separate}
	{}
	{\configuration{p :: s_{w} \statementseparator \waitoperation(a) \statementseparator s_{p} \processorseparator q :: \resultoperation(a, r) \statementseparator s_{q}}{\state}}
	{\configuration{p :: s_{w} \statementseparator s_{p} \substitution{r}{a.\datafeature} \processorseparator q :: s_{q}}{\state}}

\singlelineinferencerule
	{Wait and Notify Operation -- Separate}
	{}
	{\configuration{p :: s_{w} \statementseparator \waitoperation(a) \statementseparator s_{p} \processorseparator q :: \notifyoperation(a) \statementseparator s_{q}}{\state}}
	{\configuration{p :: s_{w} \statementseparator s_{p} \processorseparator q :: s_{q}}{\state}}

The operations presented here must be used so that each $\waitoperation$ operation can be resolved with exactly one $\resultoperation$ or $\notifyoperation$ operation. To define this condition more precisely, we define that one statement $s_{1}$ weakly precedes a statement $s_{2}$ if and only if $s_{1}$ occurs earlier than $s_{2}$ in the same action queue or $s_{1}$ and $s_{2}$ occur in different action queues. One statement $s_{1}$ strongly precedes a statement $s_{1}$ if and only if $s_{1}$ occurs earlier than $s_{2}$ in the same action queue. With these definitions, the condition says:
\begin{itemize}
	\item For each $\waitoperation(a)$ operation there must be either exactly one $\resultoperation(a, r)$ or exactly one $\notifyoperation(a)$ operation.
	\item For each $\resultoperation(a, r)$ or $\notifyoperation(a)$ operation there must be exactly one $\waitoperation(a)$ operation.
	\item Each $\resultoperation(a, r)$ or $\notifyoperation(a)$ operation weakly precedes the $\waitoperation(a)$ operation.
\end{itemize}

\subsubsection{Expression evaluation mechanism}\label{sec:expression evaluation mechanism}
An expression can either be a literal, an entity, or a query call. The query call can contain actual arguments that are expressions themselves. This section discusses the general mechanism to evaluate expressions. It focuses on the general approach and defers the evaluation of particular expressions to later sections.

The operation $\evaluateoperation(a, e)$ takes a channel $a$ and an expression $e$. Each $\evaluateoperation(a, e)$ operation determines the value $r$ of the expression $e$ and then sends a notification with value $r$ on channel $a$. This means that each $\evaluateoperation(a, e)$ operation creates a $\resultoperation(a, r)$ operation in the action queue. It is therefore important to follow each $\evaluateoperation(a, e)$ operation with exactly one $\waitoperation(a)$ to receive the notification with the value.
\subsubsection{Locking and unlocking mechanism}
A processor $p$ that wants to execute a feature must first obtain the request queue locks of a number of processors $\set{q_{1}, \ldots, q_{m}}$. Only then can $p$ issue statements to these processors. The $\lockrequestqueuesoperation(\set{q_{1}, \ldots, q_{m}})$ operation serves this purpose. It requires that none of the request queues is already locked.

\singlelineinferencerule
	{Lock Operation}
	{
		\neg \exists q_{i} \in \set{q_{1}, \ldots, q_{m}} \colon \state.\isrequestqueuelockedfeature(q_{i}) \\
		\state' \mathematicaldefinition \state.\lockrequestqueuesfeature(p, \set{q_{1}, \ldots, q_{m}})
	}
	{\configuration{p :: \lockrequestqueuesoperation(\set{q_{1}, \ldots, q_{m}}) \statementseparator s_{p}}{\state}}
	{\configuration{p :: s_{p}}{\state'}}

Once $p$ is done with the execution of the feature, it asks $\set{q_{1}, \ldots, q_{m}}$ to unlock their request queues once they are done with the issued statements. For this purpose, $p$ issues the $\unlockrequestqueueoperation$ operation to processors $\set{q_{1}, \ldots, q_{m}}$. This operation requires that the request queue is indeed locked and that no processor possesses the request queue lock.

\singlelineinferencerule
	{Unlock Operation}
	{
		\state.\isrequestqueuelockedfeature(p) \\
		\forall q \in \state.\processorsfeature \colon \neg \state.\requestqueuelocksfeature(q).\containsfeature(p) \\
		\state' \mathematicaldefinition \state.\unlockrequestqueuefeature(p)
	}
	{\configuration{p :: \unlockrequestqueueoperation \statementseparator s_{p}}{\state}}
	{\configuration{p :: s_{p}}{\state'}}

After $p$ issued the $\unlockrequestqueueoperation$ operations, it can remove $\set{q_{1}, \ldots, q_{m}}$ from its stack of obtained request queue locks using the $\popobtainedrequestqueuelocksoperation$ operation. This ensures that the $\unlockrequestqueueoperation$ operations can proceed.

\singlelineinferencerule
	{Pop Obtained Request Queue Locks}
	{
		\state' \mathematicaldefinition \state.\popobtainedrequestqueuelocksfeature(p)
	}
	{\configuration{p :: \popobtainedrequestqueuelocksoperation \statementseparator s_{p}}{\state}}
	{\configuration{p :: s_{p}}{\state'}}

Brooke, Paige, and Jacob \cite{brooke-paige-jacob:2007:formal_semantics_for_SCOOP} noticed that $\unlockrequestqueueoperation$ operations are not optimal. In essence, it could be possible to unlock the request queue of a processor $q_{i}$ directly after $p$ issued all statements. The request queue lock is important to guarantee exclusive access on $q_{i}$'s request queue. However, as soon as $p$ issued all statements on $q_{i}$, this lock is no longer needed. Unlocking the request queue right away could improve the performance in some situations because $q_{i}$'s request queue could be locked again earlier and hence another processor that is waiting for this lock could proceed earlier.

\subsubsection{Write and read mechanism}
A processor $p$ can use the $\writevalueoperation(x, v)$ operation to set a value $v$ of an entity with name $x$. This operation uses the $\setvaluefeature$ command. Hence, $p$ can both set attribute values of its current object and values of entities in its top environment.

\singlelineinferencerule
	{Write Value Operation}
	{\state' \mathematicaldefinition \state.\setvaluefeature(p, x, v)}
	{\configuration{p :: \writevalueoperation(x, v) \statementseparator s_{p}}{\state}}
	{\configuration{p :: s_{p}}{\state'}}

Similarly, processor $p$ can execute the $\readvalueoperation(x, a)$ operation to read a value of an entity with name $x$ and send the value over channel $a$. The $\readvalueoperation$ operation does not present its result in a $\resultoperation$ operation because, unlike an $\evaluateoperation$ operation, a $\readvalueoperation$ operation always produces a result for the surrounding action queue. It is easier to do the substitution of the channel access directly. A later section introduces the $\evaluateoperation$ operation for entity expressions. This variant of the $\evaluateoperation$ operation makes use of the $\readvalueoperation$ operation and presents the result in a $\resultoperation$ operation.

\singlelineinferencerule
	{Read Value Operation}
	{}
	{\configuration{p :: \readvalueoperation(x, a) \statementseparator s_{p}}{\state}}
	{\configuration{p :: s_{p} \substitution{\state.\valuefeature(p, x)}{a.\datafeature}}{\state}}

Finally, there is the $\setonceroutinenotfreshoperation$ operation in two variants for once functions and one variant for once procedures. This operation sets the once status of a once routine. The variant $\setonceroutinenotfreshoperation(f, r)$ sets the once status of a once function $f$ to non-fresh with value $r$. If $f$ is of separate type, then the once function becomes non-fresh on all processors in the system. If $f$ has a non-separate type, then $f$ becomes non-fresh only on processor $p$. The variant $\setonceroutinenotfreshoperation(f)$ sets the once status of a once procedure $f$ to non-fresh on processor $p$. The variant $\setonceroutinenotfreshwithresultoperation$ uses the value of the result entity to set the once status of a once function.

\singlelineinferencerule
	{Set Once Routine Not Fresh Operation -- Function}
	{
		f \in \functiontype \wedge f.\isonceroutinefeature \\
		\state' \mathematicaldefinition \state.\setoncefunctionnotfreshfeature(p, f, r)
	}
	{\configuration{p :: \setonceroutinenotfreshoperation(f, r) \statementseparator s_{p}}{\state}}
	{\configuration{p :: s_{p}}{\state'}}

\singlelineinferencerule
	{Set Once Routine Not Fresh Operation -- Procedure}
	{
		f \in \proceduretype \wedge f.\isonceroutinefeature \\
		\state' \mathematicaldefinition \state.\setonceprocedurenotfreshfeature(p, f)
	}
	{\configuration{p :: \setonceroutinenotfreshoperation(f) \statementseparator s_{p}}{\state}}
	{\configuration{p :: s_{p}}{\state'}}

\inferencerule
	{Set Once Routine Not Fresh Operation -- Function With Result}
	{
		f \in \functiontype \wedge f.\isonceroutinefeature \\
		\state.\environmentsfeature(p).\topfeature.\namesfeature.\containsfeature(\resultentityname) \\
		\freshchanneldefinition{a}
	}
	{\configuration{p :: \setonceroutinenotfreshwithresultoperation(f) \statementseparator s_{p}}{\state}}
	{\configuration{p :: \readvalueoperation(\resultentityname, a) \statementseparator \setonceroutinenotfreshoperation(f, a.\datafeature) \statementseparator s_{p}}{\state}}

\subsubsection{Flow control mechanism}
In addition to flow control instructions in the user code, there are flow control operations, which implement flow control in the inference rules. This way, fewer inference rules are required because multiple variants can be handled in one inference rule.

The $\singlelineconditionaloperation{x}{s_{t}}{s_{f}}$ operation takes the condition $x$ as an argument. The operation either executes $s_{t}$ if $x$ indicates that the condition is true or $s_{f}$ if $x$ indicates that the condition is false. For each possibility there is one inference rule. The condition $x$ can either be an instance of $\booleantype$ or it can be a reference that points to an object of class type $\booleanclasstype$. To decide which branch to take, the operation must evaluate $x$. If $x$ is an instance of $\booleantype$, then it can determine which instance $x$ is, i.e., $\truevalue$ or $\falsevalue$. If $x$ is a reference, then it must get the referenced object and see which boolean value it represents. For this purpose, it evaluates the attribute $\booleanclasstypeitemattributename$ of the referenced object.

\singlelineinferencerule
	{Conditional Operation -- True}
	{
		y \mathematicaldefinition 
			\left\{
				\begin{array}{ll}
					x & \condition{x \in \booleantype} \\
					\state.\attributevaluefeature(\state.\referencedobjectfeature(x), \booleanclasstypeitemattributename) & \condition{x \in \referencetype \wedge \state.\referencedobjectfeature(x).\classtypefeature = \booleanclasstype} \\
					\falsevalue & \otherwisecondition
				\end{array}
			\right. \\
		y = \truevalue
	}
	{
		\configuration
			{
				p :: \ \singlelineconditionaloperation
					{x}
					{s_{t}}
					{s_{f}} \statementseparator
				s_{p}
			}
			{\state}
	}
	{\configuration{p :: s_{t} \statementseparator s_{p}}{\state}}

\singlelineinferencerule
	{Conditional Operation -- False}
	{
		y \mathematicaldefinition 
			\left\{
				\begin{array}{ll}
					x & \condition{x \in \booleantype} \\
					\state.\attributevaluefeature(\state.\referencedobjectfeature(x), \booleanclasstypeitemattributename) & \condition{x \in \referencetype \wedge \state.\referencedobjectfeature(x).\classtypefeature = \booleanclasstype} \\
					\truevalue & \otherwisecondition
				\end{array}
			\right. \\
		y = \falsevalue
	}
	{
		\configuration
			{
				p :: \ \singlelineconditionaloperation
					{x}
					{s_{t}}
					{s_{f}} \statementseparator
				s_{p}
			}
			{\state}
	}
	{\configuration{p :: s_{f} \statementseparator s_{p}}{\state}}

The $\singlelineconditionaloperation{x}{s_{t}}{s_{f}}$ operation has two branches. Sometimes it is necessary to only have one branch. The $\nooperation$ operation can be executed without an effect. It can be used in the conditional operation to define an empty branch. The $\nooperation$ operation can also be used to indicate that an action queue is empty.

\singlelineinferencerule
	{No Operation}
	{}
	{\configuration{p :: \nooperation \statementseparator s_{p}}{\state}}
	{\configuration{p :: s_{p}}{\state}}

\begin{fortechnicalreport}
\subsection{Code elements}
\end{fortechnicalreport}
\begin{forjournal}
\subsection{Code Elements}
\end{forjournal}
This section explains the semantics of code elements: entity expressions, literal expressions, feature calls, feature applications, creation instructions, flow control instructions, and assignment instructions.

\subsubsection{Entity expressions}
A variant of the $\evaluateoperation(a, e)$ operation evaluates entity expressions. The operation uses the $\readvalueoperation$ operation to send a notification with the value of the entity over a new channel $a'$. It then uses the value of this channel to define the result of the $\evaluateoperation$ operation.

\singlelineinferencerule
	{Entity Expression}
	{
		e \in \entitytype \\
		\freshchanneldefinition{a'}
	}
	{\configuration{p :: \evaluateoperation(a, e) \statementseparator s_{p}}{\state}}
	{\configuration{p :: \readvalueoperation(e.\namefeature, a') \statementseparator \resultoperation(a, a'.\datafeature) \statementseparator s_{p}}{\state}}

\subsubsection{Literal expressions}
Another variant of the $\evaluateoperation(a, e)$ operation evaluates literal expressions. To evaluate a non-void literal expression, the operation creates a new object of the literal class type so that the new object represents the literal value. For this purpose, it uses the query $\objectfeature$ of $\literaltype$. Since the type of every literal is non-separate, it creates the new object on the processor that evaluates the literal expression. The reference $r$ to the new object is the result of the evaluation. To evaluate a void literal, the operation takes the void reference.

\singlelineinferencerule
	{Literal Expression}
	{
		e \in \literaltype \\
		\state' \mathematicaldefinition
			\left\{
				\begin{array}{ll}
					 \state & \condition{e = \voidliteral} \\
					 \state.\addobjectfeature(p, e.\objectfeature) & \otherwisecondition
				\end{array}
			\right. \\
		r \mathematicaldefinition
			\left\{
				\begin{array}{ll}
					 \voidvalue & \condition{e = \voidliteral} \\
					 \state'.\referencefeature(\state'.\lastaddedobjectfeature) & \otherwisecondition
				\end{array}
			\right.
	}
	{\configuration{p :: \evaluateoperation(a, e) \statementseparator s_{p}}{\state}}
	{\configuration{p :: \resultoperation(a, r) \statementseparator s_{p}}{\state'}}

\subsubsection{Feature calls}\label{sec:feature calls}
A feature call can occur in two ways. First, a feature call can be a call to a command in a command instruction. Second, a feature call can be a call to a query in an expression. This section studies both variants. A processor $p$ that executes a feature call $e_{0}.f(e_{1}, \ldots, e_{n})$ goes through the following steps:
\begin{enumerate}
	\item Target evaluation. Evaluate the target expression $e_{0}$ and let $q$ denote the handler of the target.
	\item Argument passing. Evaluate the actual arguments expressions $(e_{1}, \ldots, e_{n})$.
	\item Lock passing. Determine which locks to pass to $q$.
		\begin{itemize}
			\item Take all request queue locks and call stack locks if a controlled actual argument gets attached to an attached formal argument of reference type.
			\item Take all request queue locks and call stack locks if the feature call is a separate callback, i.e., $q$ has a lock on $p$.
			\item Otherwise, take no locks.
		\end{itemize}
	\item Feature request.
		\begin{itemize}
			\item Ask $q$ to apply $f$ to the target immediately and wait until the execution terminates if any of the following conditions holds:
				\begin{itemize}
					\item The feature call is non-separate, i.e., $p = q$.
					\item The feature call is a separate callback, i.e., $q$ has a lock on $p$.
				\end{itemize}
			\item Otherwise, ask $q$ to apply $f$ to the target after the previous feature requests.
		\end{itemize}
	\item Wait by necessity. If $f$ is a query, then wait for the result.
	\item Lock revocation. If lock passing happened, then wait for the locks to come back.
\end{enumerate}
A command instruction is a statement in the action queue. A query is an expression on the right hand side of an assignment, a condition in a flow control instruction, or an actual argument in a feature call. Whenever a query occurs in one of these constructs, the inference rule of the construct encloses the query in an $\evaluateoperation$ operation. To handle feature calls, there is an inference rule for command instructions and a variant of the $\evaluateoperation$ operation for query calls.

In each case, the statement first evaluates the target expression and all actual argument expressions. For each of these expressions $e_{i}$, it uses one $\evaluateoperation(a_{e_{i}}, e_{i})$ operation and a corresponding $\waitoperation(a_{e_{i}})$ operation with a fresh channel $a_{e_{i}}$. Each of the channel values gets used in the subsequent $\calloperation$ operation. With this, the statement handled the target evaluation and the argument passing step. It defers the attachment of the actual arguments to the formal arguments to the point where the called feature gets applied. The reason for this is simple: at this point the context for the feature application does not exist yet.

The $\calloperation$ operation takes care of the remaining steps. The operation exists in two variants, one for command instructions and one for queries. The variant for queries takes a channel $a'$ and uses it for the result of the query. Since a call to a command does not produce a result, such a channel is not required for command instructions. Both $\calloperation$ variants take the reference to the target $a_{e_{0}}$, the feature $f$ to be called, the references to the actual arguments $\tuple{a_{e_{1}}.\datafeature, \ldots, a_{e_{n}}.\datafeature}$, and the actual argument expressions $\tuple{e_{1}, \ldots, e_{n}})$. The actual argument expressions are used to check whether there is a controlled actual argument. This information determines whether the locks should be passed.

\inferencerule
	{Command Instruction}
	{\forall {i \in \set{0, \ldots, n}} \colon {\freshchanneldefinition{a_{e_{i}}}}}
	{\configuration{p :: e_{0}.f(e_{1}, \ldots, e_{n}) \statementseparator s_{p}}{\state}}
	{
		\configuration
			{
				p :: \ & \evaluateoperation(a_{e_{0}}, e_{0}) \statementseparator \evaluateoperation(a_{e_{1}}, e_{1}) \statementseparator \ldots \statementseparator \evaluateoperation(a_{e_{n}}, e_{n}) \statementseparator \\
				& \waitoperation(a_{e_{0}}) \statementseparator \waitoperation(a_{e_{1}}) \statementseparator \ldots \statementseparator \waitoperation(a_{e_{n}}) \statementseparator \\
				& \calloperation(a_{e_{0}}.\datafeature, f, \tuple{a_{e_{1}}.\datafeature, \ldots, a_{e_{n}}.\datafeature}, \tuple{e_{1}, \ldots, e_{n}}) \statementseparator \\
				& s_{p}
			}
			{\state}
	}

\inferencerule
	{Query Expression}
	{
		\forall {i \in \set{0, \ldots, n}} \colon {\freshchanneldefinition{a_{e_{i}}}} \\
		\freshchanneldefinition{a'}
	}
	{\configuration{p :: \evaluateoperation(a, e_{0}.f(e_{1}, \ldots, e_{n})) \statementseparator s_{p}}{\state}}
	{
		\configuration
			{
				p :: \ & \evaluateoperation(a_{e_{0}}, e_{0}) \statementseparator \evaluateoperation(a_{e_{1}}, e_{1}) \statementseparator \ldots \statementseparator \evaluateoperation(a_{e_{n}}, e_{n}) \statementseparator \\
				& \waitoperation(a_{e_{0}}) \statementseparator \waitoperation(a_{e_{1}}) \statementseparator \ldots \statementseparator \waitoperation(a_{e_{n}}) \statementseparator \\
				& \calloperation(a', a_{e_{0}}.\datafeature, f, \tuple{a_{e_{1}}.\datafeature, \ldots, a_{e_{n}}.\datafeature}, \tuple{e_{1}, \ldots, e_{n}}) \statementseparator \\
				& \resultoperation(a, a'.\datafeature) \statementseparator \\
				& s_{p}
			}
			{\state}
	}

Both variants of the $\calloperation$ operation take the reference to the target $r_{o}$, the feature $f$ to be called, the references to the actual arguments $\tuple{r_{1}, \ldots, r_{n}}$, and the actual argument expressions $\tuple{e_{1}, \ldots, e_{n}}$. The variant for queries takes an additional channel $a$ to be used for the result of the query. In a first step, the operation must evaluate the handler $q$ of the target. The handler is used in an $\issueoperation$ operation to issue a feature request on the responsible processor. The feature request comes in the form of an $\applyoperation$ operation. The $\applyoperation$ operation takes a channel $a$ for the communication between $p$ and $q$, the target reference $r_{0}$, the called feature $f$, the references to the actual arguments $\tuple{r_{1}, \ldots, r_{n}}$, the caller processor $p$, and the passed locks $\overline{l}$.

\begin{clarification}[Lock passing]
Processor $p$ passes all its request queue locks and all its call stack locks either if there is a controlled actual argument that will get attached to an attached formal argument of reference type or if the feature call is a separate callback. An attached formal argument of reference type means that the request queue lock or the call stack lock on the actual argument's handler is required during the application of $f$. A controlled actual argument means that $p$ has a request queue lock or a call stack lock on the handler of the actual argument. In short, $p$ has a lock that is required by $q$ and thus $p$ has to pass the locks. A separate callback occurs if $q$ has a lock on $p$. In this situation, $p$ can issue a statement to $q$ and then wait for $q$ to complete. However, processor $q$ could already be waiting for $p$ to complete. To handle this case, the $\issueoperation$ operation in the $\calloperation$ operation triggers an immediate execution by adding the $\applyoperation$ to the beginning of $q$'s action queue. The $\issueoperation$ operation requires that $p$ has the call stack lock of $q$. To enable $q$ to perform an immediate execution, $p$ has to give back $q$'s call stack lock.

In both cases, $p$ has to wait for the locks to come back. Thus it does not hurt to pass all the locks in both cases. In contrast to Nienaltowski's \cite{nienaltowski:2007:SCOOP} description of SCOOP, $p$ only passes the locks that it really has. In particular, $p$ does not pass its own request queue lock in situations where $p$ does not possess this lock, such as when the processor that called $p$ possesses $p$'s request queue lock.
\end{clarification}

In the cases where the operation passes the locks, $\overline{l}$ is $\tuple{\state.\requestqueuelocksfeature(p), \state.\callstacklocksfeature(p)}$. In all other cases there is no lock passing and thus $\overline{l} = \tuple{\set{}, \set{}}$. The operation just determines which locks to pass. The actual lock passing action will be executed by $q$. Similarly, the actual lock revocation action will be executed by $q$.

For command calls, lock passing is the only reason to wait. In this case, the operation creates a fresh channel $a$ to wait for a notification from $q$. The notification arrives when $q$ is ready to return the locks. For query calls, the operation has to wait for the result. The operation uses the given channel $a$ to wait for the result. This has the advantage that once the result arrives, it will be substituted after the $\calloperation$ operation, i.e.\ in the $\resultoperation$ operation of the $\evaluateoperation$ operation.

\inferencerule
	{Call Operation -- Command}
	{
		q \mathematicaldefinition \state.\handlerfeature(r_{0}) \\
		\overline{l} \mathematicaldefinition 
			\left\{
				\begin{split}
					& \multilinecondition{
						& q \neq p \wedge 
						\exists i \in \set{1, \ldots, n} \colon \typingenvironmentderivation{e_{i}: t \wedge \iscontrolledfeature(t)} \wedge 
						\typingenvironmentderivation{f.\formalargumentsfeature(i): (!, g, c) \wedge c.\isreferenceclasstypefeature}
					} \\
					& \indentation \tuple{\state.\requestqueuelocksfeature(p), \state.\callstacklocksfeature(p)} \\
					& \multilinecondition{
						& q \neq p \wedge
						(\state.\requestqueuelocksfeature(q).\containsfeature(p) \vee \state.\callstacklocksfeature(q).\containsfeature(p))
					} \\
					& \indentation \tuple{\state.\requestqueuelocksfeature(p), \state.\callstacklocksfeature(p)} \\
					& \otherwisecondition \\
					& \indentation \tuple{\set{}, \set{}}
				\end{split}
			\right. \\
		\freshchanneldefinition{a}
	}
	{\configuration{p :: \calloperation(r_{0}, f, \tuple{r_{1}, \ldots, r_{n}}, \tuple{e_{1}, \ldots, e_{n}}) \statementseparator s_{p}}{\state}}
	{
		\configuration
			{
				p :: \ & \issueoperation(q, \applyoperation(a, r_{0}, f, \tuple{r_{1}, \ldots, r_{n}}, p, \overline{l})) \statementseparator \\
				& \singlelineconditionaloperation
					{\overline{l} \neq \tuple{\set{}, \set{}}}
					{\waitoperation(a)}
					{\nooperation} \statementseparator \\
				& s_{p}
			}
			{\state}
	}

\inferencerule
	{Call Operation -- Query}
	{
		q \mathematicaldefinition \state.\handlerfeature(r_{0}) \\
		\overline{l} \mathematicaldefinition 
			\left\{
				\begin{split}
					& \multilinecondition{
						& q \neq p \wedge
						\exists i \in \set{1, \ldots, n} \colon \typingenvironmentderivation{e_{i}: t \wedge \iscontrolledfeature(t)} \wedge
						\typingenvironmentderivation{f.\formalargumentsfeature(i): (!, g, c) \wedge c.\isreferenceclasstypefeature}
					} \\
					& \indentation \tuple{\state.\requestqueuelocksfeature(p), \state.\callstacklocksfeature(p)} \\
					& \multilinecondition{
						& q \neq p \wedge
						(\state.\requestqueuelocksfeature(q).\containsfeature(p) \vee \state.\callstacklocksfeature(q).\containsfeature(p))
					} \\
					& \indentation \tuple{\state.\requestqueuelocksfeature(p), \state.\callstacklocksfeature(p)} \\
					& \otherwisecondition \\
					& \indentation \tuple{\set{}, \set{}}
				\end{split}
			\right.
	}
	{\configuration{p :: \calloperation(a, r_{0}, f, \tuple{r_{1}, \ldots, r_{n}}, \tuple{e_{1}, \ldots, e_{n}}) \statementseparator s_{p}}{\state}}
	{
		\configuration
			{p ::
				\issueoperation(q, \applyoperation(a, r_{0}, f, \tuple{r_{1}, \ldots, r_{n}}, p, \overline{l})) \statementseparator
				\waitoperation(a) \statementseparator
				s_{p}
			}
			{\state}
	}

\begin{fortechnicalreport}
\begin{example}[Feature call]
This example demonstrates the inference rules for feature calls. For this purpose, consider again the share market example. Suppose the root processor $p_{1}$ started with the execution of the procedure $do\_transaction$ on the root object $o_{0}$. This procedure is shown in \listingreference{lst:application class with implementation}.

\begin{lstlisting}[caption=Application class with implementation, label=lst:application class with implementation, language=SCOOP, escapechar=\%]
class APPLICATION

create
  make

feature -- Initialization
  make
    do
      ...
    end

feature {APPLICATION} -- Implementation
  market: separate MARKET
      -- The market.
			
  do_transaction (
    first_investor: separate INVESTOR;
    second_investor: separate INVESTOR;
    issuer_id: INTEGER
  )
      -- Make each of the two investors buy and then sell a share of the issuer.
    do
      first_investor.buy (Current.market, issuer_id)
      second_investor.buy (Current.market, issuer_id)
      first_investor.sell (Current.market, issuer_id)
      second_investor.sell (Current.market, issuer_id)
    end
end
\end{lstlisting}

The following configuration is our starting point. Processor $p_{1}$ has one environment for the callee procedure $make$ and one for the called procedure $do\_transaction$. Processor $p_{2}$ handles the market object $o_{35}$, processor $p_{3}$ handles the first investor object $o_{44}$, and processor $p_{4}$ handles the second investor object $o_{46}$. The market object references three arrays. The cash array $o_{25}$ shows that each investor has the same amount of cash. The available shares array $o_{33}$ shows that there is one issuer with one available share. The owned shares array $o_{38}$ shows that none of the investors owns a share. The action queue of $p_{1}$ indicates that $p_{1}$ is about to make the feature calls on the two investors. For this purpose, $p_{1}$ obtained the request queue locks of their handlers.

\isolatedconfiguration
	{
		p_{1} :: \ & first\_investor.buy(\currententity.market, issuer\_id) \statementseparator \\
		& second\_investor.buy(\currententity.market, issuer\_id) \statementseparator \\
		& first\_investor.sell(\currententity.market, issuer\_id) \statementseparator \\
		& second\_investor.sell(\currententity.market, issuer\_id) \statementseparator \\
		& \ldots \processorseparator \\
		p_{2} :: & \processorseparator
		p_{3} :: \processorseparator
		p_{4} ::
	}
	{
		\simplifiedstate
			{
				& \simplifiedstatelocksentry
					{p_{1}}
					{\set{}, \set{p_{3}, p_{4}}}
					{\set{}, \set{}}
					{\set{}, \set{}}
					{\simplifiedstatelockedindicator}
					{\simplifiedstatenopassedlocksindicator} \\
				& \simplifiedstatelocksentry
					{p_{2}}
					{}
					{}
					{}
					{\simplifiedstateunlockedindicator}
					{\simplifiedstatenopassedlocksindicator} \\
				& \simplifiedstatelocksentry
					{p_{3}}
					{}
					{}
					{}
					{\simplifiedstatelockedindicator}
					{\simplifiedstatenopassedlocksindicator} \\
				& \simplifiedstatelocksentry
					{p_{4}}
					{}
					{}
					{}
					{\simplifiedstatelockedindicator}
					{\simplifiedstatenopassedlocksindicator}
			}
			{
				& \simplifiedstateobjectsentry
					{p_{1}}
					{
						\simplifiedstatereferencedobject{r_{0}}{o_{0}(\simplifiedstateentityvalue{market}{r_{1}})}, \simplifiedstatereferencedobject{r_{39}}{o_{48}(1)}
					} \\
				& \simplifiedstateobjectsentry
					{p_{2}}
					{
						& \simplifiedstatereferencedobject{r_{1}}{o_{35}(\simplifiedstateentityvalue{cash}{r_{16}}, \simplifiedstateentityvalue{available\_shares}{r_{23}}, \simplifiedstateentityvalue{owned\_shares}{r_{29}})}, \\
						& \simplifiedstatereferencedobject{r_{16}}{o_{25}[r_{21}, r_{22}]}, \simplifiedstatereferencedobject{r_{21}}{o_{23}(100)}, \simplifiedstatereferencedobject{r_{22}}{o_{24}(100)}, \\
						& \simplifiedstatereferencedobject{r_{23}}{o_{33}[r_{28}]}, \simplifiedstatereferencedobject{r_{28}}{o_{32}(1)}, \\
						& \simplifiedstatereferencedobject{r_{29}}{o_{38}[[r_{34}], [r_{35}]]}, \simplifiedstatereferencedobject{r_{34}}{o_{41}(0)}, \simplifiedstatereferencedobject{r_{35}}{o_{42}(0)}
					} \\
				& \simplifiedstateobjectsentry
					{p_{3}}
					{
						\simplifiedstatereferencedobject{r_{6}}{o_{44}(\simplifiedstateentityvalue{id}{r_{36}})},
						\simplifiedstatereferencedobject{r_{36}}{o_{43}(1)}
					} \\
				& \simplifiedstateobjectsentry
					{p_{4}}
					{
						\simplifiedstatereferencedobject{r_{8}}{o_{46}(\simplifiedstateentityvalue{id}{r_{37}})},
						\simplifiedstatereferencedobject{r_{37}}{o_{45}(2)}
					}
			}
			{}
			{
				& \simplifiedstateenvironmentsentry
					{p_{1}}
					{
						& \simplifiedstateentityvalue{first\_investor}{r_{6}}, \simplifiedstateentityvalue{second\_investor}{r_{8}}, \simplifiedstatecurrententityvalue{r_{0}} \simplifiedstateenvironmentsentryseparator \\
						& \simplifiedstateentityvalue{first\_investor}{r_{6}}, \simplifiedstateentityvalue{second\_investor}{r_{8}}, \simplifiedstateentityvalue{issuer\_id}{r_{39}}, \simplifiedstatecurrententityvalue{r_{0}}
					} \\
				& \simplifiedstateenvironmentsentry
					{p_{2}}
					{} \\
				& \simplifiedstateenvironmentsentry
					{p_{3}}
					{} \\
				& \simplifiedstateenvironmentsentry
					{p_{4}}
					{}
			}
	}

The inference rule for command instructions leads to the following configuration, with fresh channels $a_{59}$, $a_{60}$, and $a_{61}$:

\isolatedconfiguration
	{
		p_{1} :: \ & \evaluateoperation(a_{59}, first\_investor) \statementseparator \\
		& \evaluateoperation(a_{60}, \currententity.market) \statementseparator \\
		& \evaluateoperation(a_{61}, issuer\_id) \statementseparator \\
		& \waitoperation(a_{59}) \statementseparator \\
		& \waitoperation(a_{60}) \statementseparator \\
		& \waitoperation(a_{61}) \statementseparator \\
		& \calloperation(a_{59}.\datafeature, buy, \tuple{a_{60}.\datafeature, a_{61}.\datafeature}, \tuple{\currententity.market, issuer\_id}) \statementseparator \\
		& second\_investor.buy(\currententity.market, issuer\_id) \statementseparator \\
		& first\_investor.sell(\currententity.market, issuer\_id) \statementseparator \\
		& second\_investor.sell(\currententity.market, issuer\_id) \statementseparator \\
		& \ldots \processorseparator \\
		p_{2} :: & \processorseparator
		p_{3} :: \processorseparator
		p_{4} ::
	}
	{\ldots}

First $p_{1}$ has to evaluate the target expression and the actual argument expressions. This leads to the following configuration:

\isolatedconfiguration
	{
		p_{1} :: \ & \calloperation(r_{6}, buy, \tuple{r_{1}, r_{39}}, \tuple{\currententity.market, issuer\_id}) \statementseparator \\
		& second\_investor.buy(\currententity.market, issuer\_id) \statementseparator \\
		& first\_investor.sell(\currententity.market, issuer\_id) \statementseparator \\
		& second\_investor.sell(\currententity.market, issuer\_id) \statementseparator \\
		& \ldots \processorseparator \\
		p_{2} :: & \processorseparator
		p_{3} :: \processorseparator
		p_{4} ::
	}
	{\ldots}

During the execution of the $\calloperation$ operation, $p_{1}$ determines that no locks need to be passed. Then $p_{1}$ executes an $\issueoperation$ operation to enqueue an $\applyoperation$ operation to the action queue of the first investor's handler.
\end{example}
\end{fortechnicalreport}

\subsubsection{Feature applications}\label{sec:feature applications}
A feature call by a client processor $q$ results in a feature request for a supplier processor $p$. A \emph{feature application} is the serving of the feature request. This section discusses how $p$ applies a feature $f$ on a target referenced by $r_{0}$. Processor $p$ takes the following steps:
\begin{enumerate}
	\item Once status update. If $f$ is a once routine, then set its status to non-fresh.
	\item Lock passing. Pass the locks from $q$ to $p$.
	\item Argument passing. Bind the actual arguments to the formal arguments. Arguments of expanded type that are handled by a different processor than $p$ must be deep imported by $p$.
	\item Synchronization. Involve the scheduler to wait until the following synchronization conditions are satisfied atomically:
		\begin{itemize}
			\item Processor $p$ owns the request queue lock of each processor $q$ such that:
				\begin{itemize}
					\item Processor $q$ handles an actual argument of $f$ and the corresponding formal argument has an attached reference type.
					\item Processor $p$ and processor $q$ are different.
					\item Processor $p$ does not have $q$'s request queue lock.
					\item Processor $q$ does not have $p$'s request queue lock.
				\end{itemize}
			\item The precondition of $f$ holds.
		\end{itemize}
	\item Execution.
		\begin{itemize}
			\item If $f$ is a non-once routine or a fresh once routine, then run its body.
			\item If $f$ is a non-fresh procedure, then do nothing. If $f$ is a non-fresh function, then take its once value as the result.
			\item If $f$ is an attribute, then evaluate it.
		\end{itemize}
	\item Postcondition evaluation. Evaluate the postcondition if any of the following conditions is satisfied:
		\begin{itemize}
			\item A feature call in the postcondition requires a lock that was not obtained in the synchronization step.
			\item The evaluation of the postcondition involves lock passing.
		\end{itemize}
		Otherwise ask any processor whose request queue lock was obtained in the synchronization step to evaluate the postcondition.
	\item Lock releasing. Ask each processor whose request queue has been locked in the synchronization step to unlock its request queue after it is done with the feature requests issued by $p$.
	\item Invariant evaluation. If $f$ is a routine, then evaluate the invariant.
	\item Result returning. If $f$ is a query, then return the result to $q$. If the result is of expanded type and $p \neq q$, then the result must be deep imported by $q$.
	\item Lock revocation. Return the passed locks from $p$ to $q$.
\end{enumerate}
Each feature application starts with an operation $\applyoperation(a, r_{0}, f, \tuple{r_{1}, \ldots, r_{n}}, q, \overline{l})$ in the action queue of processor $p$. The channel $a$ is used to communicate with the client processor $q$. If the called feature $f$ is a procedure and the caller processor $q$ passed some locks, then $a$ is used to signal that the locks returned. If $f$ is query, then $a$ is used to return the value. The reference $r_{0}$ points to the target of the call. The references $\tuple{r_{1}, \ldots, r_{n}}$ point to the actual arguments. The tuple $\overline{l}$ contains the locks to be passed from $q$ to $p$.

If one takes a look at the execution step, one can differentiate three cases:
\begin{itemize}
	\item The feature $f$ is a non-once routine or a fresh once routine.
	\item The feature $f$ is a non-fresh once routine.
	\item The feature $f$ is an attribute.
\end{itemize}
For each of these cases, there is one inference rule. Each inference rule covers one variant of the $\applyoperation$ operation. The discussion continues with the most involved case: the feature $f$ is a non-once routine or a fresh once routine.

\inferencerule
	{Application Operation -- Non-Once Routine or Fresh Once Routine}
	{
		f \in \routinetype \wedge f.\isonceroutinefeature \rightarrow \state.\isonceroutinefreshfeature(p, f) \\
		\state.\handlerfeature(r_{0}) = p \\
		\neg \state.\passlocksfeature(q, p, \overline{l}).\arelockspassedfeature(p) \\
		\state' \mathematicaldefinition
			\left\{
				\begin{array}{ll}
					\state.\setoncefunctionnotfreshfeature(p, f, \voidvalue) & \condition{f \in \functiontype \wedge f.\isonceroutinefeature} \\
					\state.\setonceprocedurenotfreshfeature(p, f) & \condition{f \in \proceduretype \wedge f.\isonceroutinefeature} \\
					\state & \otherwisecondition
				\end{array}
			\right. \\
		\state'' \mathematicaldefinition \state'.\passlocksfeature(q, p, \overline{l}).\pushenvironmentwithfeaturefeature(p, f, r_{0}, \tuple{r_{1}, \ldots, r_{n}}) \\
		\begin{split}
			& \overline{g}_{required\_locks} \mathematicaldefinition \set{p} \cup \\
			& \indentation \setinference{x \in \processortype}{\exists i \in \set{1, \ldots, n}, g, c \colon \typingenvironmentderivation{f.\formalargumentsfeature(i): (!, g, c) \wedge c.\isreferenceclasstypefeature} \wedge x = \state''.\handlerfeature(r_{i})}
		\end{split} \\
		\begin{split}
			& \overline{g}_{required\_cs\_locks} \mathematicaldefinition \\
			& \indentation \setinference{x \in \overline{g}_{required\_locks}}{x = p \vee (x \neq p \wedge (\state''.\requestqueuelocksfeature(x).\containsfeature(p) \vee \state''.\callstacklocksfeature(x).\containsfeature(p)))}
		\end{split} \\
		\overline{g}_{required\_rq\_locks} \mathematicaldefinition \overline{g}_{required\_locks} \setminus \overline{g}_{required\_cs\_locks} \\
		\overline{g}_{missing\_rq\_ locks} \mathematicaldefinition \setinference{x \in \overline{g}_{required\_rq\_locks}}{\neg \state''.\requestqueuelocksfeature(p).\containsfeature(x)} \\
		\forall x \in \overline{g}_{required\_cs\_locks} \colon \state''.\callstacklocksfeature(p).\containsfeature(x) \\
		\freshchanneldefinition{a_{inv}} \wedge \freshchanneldefinition{a'}
	}
	{\configuration{p :: \applyoperation(a, r_{0}, f, \tuple{r_{1}, \ldots, r_{n}}, q, \overline{l}) \statementseparator s_{p}}{\state}}
	{
		\configuration
			{
				p :: \ & \checkpreconditionandlockrequestqueuesoperation(\overline{g}_{missing\_rq\_ locks}, f) \statementseparator \\
				\multilineconditionaloperation
					{f \in \functiontype \wedge f.\isonceroutinefeature}
					{
						\begin{split}
							& f.\bodyfeature \\
							& \indentation \begin{split}
								& \begin{split}
									\substitution
										{
											\resultentity \eassignment y \statementseparator
											\setonceroutinenotfreshwithresultoperation(f)
										}
										{
											\resultentity \eassignment y
										}
								\end{split} \\
								& \begin{split}
									\substitution
										{
											\ecreate \resultentity.y \statementseparator
											\setonceroutinenotfreshwithresultoperation(f)
										}
										{
											\ecreate \resultentity.y
										}
								\end{split}
							\end{split}
						\end{split}
					}
					{f.\bodyfeature} \statementseparator \\
				& \checkpostconditionandunlockrequestqueuesoperation(\overline{g}_{missing\_rq\_ locks}, f) \statementseparator \\
				\multilineconditionaloperation
					{f.\classtypefeature.\invariantexistsfeature \wedge f.\isexportedfeature}
					{\evaluateoperation(a_{inv}, f.\classtypefeature.\invariantfeature) \statementseparator \waitoperation(a_{inv})}
					{\nooperation} \statementseparator \\
				\multilineconditionaloperation
					{f \in \functiontype}
					{\readvalueoperation(\resultentityname, a') \statementseparator \returnoperation(a, a'.\datafeature, q)}
					{\returnoperation(a, q)} \statementseparator \\
				& s_{p}}
			{\state''}
	}

The condition of the inference rule states that each processor can only apply a feature on one of its own objects. The condition also states the $p$ must not have passed its locks. This part of the condition is always given because $p$ waits whenever it passes its locks. In a first step, the operation defines an updated state $\state'$ to set $f$'s once status to non-fresh, in case $f$ is a once routine. The operation does this before deep importing the actual arguments to avoid the following contradiction.

\begin{clarification}[When to change the status of a fresh once routine]
Assume $f$ is either a once procedure or a non-separate once routine. The feature $f$ was fresh at the beginning of the $\applyoperation$ operation. Assume that the caller passed an expanded actual argument that is handled by a processor $g \neq p$. Therefore $p$ has to deep import the actual argument. Assume furthermore that the class type of the actual argument has the once routine $f$ and that $f$ is non-fresh on $g$. If the operation would deep import before setting $f$ as non-fresh on $p$, then the deep import operation would take over the once status of $f$ from processor $g$ to processor $p$. But then the $\applyoperation$ operation on $p$ would not make much sense anymore because $f$ would now be non-fresh on $p$. If the operation sets $f$ as non-fresh at the beginning of the $\applyoperation$ operation, then the deep import operation does not take over the once status from $g$ because $f$ is already non-fresh on $p$.
\end{clarification}

The operation defines an updated state $\state''$ in which the locks are passed from $q$ to $p$ and in which there is a new environment with the actual arguments $\tuple{r_{1}, \ldots, r_{n}}$. The call to the $\pushenvironmentwithfeaturefeature$ feature takes care of copying and deep importing actual arguments of expanded type. The caller processor $q$ can also pass an empty tuple $\tuple{\set{}, \set{}}$ which simply means that $q$ did not pass any locks.

In the next step, the operation synchronizes. For each target expressions in the body of $f$, the operation can get the controlling entity. Each of these controlling entities is mapped to an object and each of these objects is handled by a processor. For each of these processors the operation must either get a request queue lock or a call stack lock. There are three types of calls: non-separate calls, separate calls, and separate callbacks. Non-separate calls and separate callbacks require a call stack lock. Separate calls require a request queue lock. This leads to two sets of required locks: one set with required request queue locks and another set with required call stack locks. The set of required call stack locks is composed of $p$ that will lead to a non-separate call and all the processors that will lead to separate callbacks. The set of required request queue locks is composed of the processors that will lead to separate calls. The operation defines two sets for these two categories: $\overline{g}_{required\_cs\_locks}$ and $\overline{g}_{required\_rq\_locks}$.

Each processor initially has its own call stack lock as its obtained call stack lock. This call stack never gets unlocked. This means that other call stack locks cannot be obtained; they must be retrieved through lock passing. The condition of the inference rule expresses this: $\forall x \in \overline{g}_{required\_cs\_locks} \colon \state''.\callstacklocksfeature(p).\containsfeature(x)$. The operation can be assured that $p$ did not pass its own call stack lock because otherwise $p$ would be waiting. The remaining required call stack locks are the ones for the processors that will lead to separate callbacks. Note that the lock passing conditions are not sufficient to guarantee that the call stack locks for separate callbacks are always available.

As for the request queue locks, the operation calculates $\overline{g}_{missing\_rq\_ locks}$ as the required request queue locks minus the already owned request queue locks. The already owned request queue locks are the previously obtained request queue locks and the retrieved request queue locks. In the synchronization step, the operation must obtain the difference. If this is not possible because some of the missing request queue locks are not available, then the operation must wait. The $\checkpreconditionandlockrequestqueuesoperation$ operation takes care of this; it takes $\overline{g}_{missing\_rq\_ locks}$ and the feature $f$. Once the execution succeeds, $p$ has the request queue locks of $\overline{g}_{missing\_rq\_ locks}$ and the precondition of $f$ holds.

The $\applyoperation$ operation can be assured that each processor $g$, whose obtained request queue lock the operation got in the synchronization step, must be in possession of its call stack lock. If $g$ was not in possession of its call stack lock, it must have passed its locks. This means that $g$ is executing a feature call and still waiting for the locks to return. In order to execute the feature call, there must have been a lock on $g$'s request queue lock so that its action queue can contain the feature call. The request queue must still be locked because $g$ is still executing the feature call. Hence, it would not have been possible to obtain $g$'s request queue lock. The only exception is the bootstrap processor. However this processor only plays a role in the system setup and it never passes its own call stack lock.

Once the operation got all the required locks, it can execute the body. For once functions it must update the once status whenever it writes to the result entity as part of an assignment instruction or as part of a creation instruction. For this purpose, it adds a $\setonceroutinenotfreshwithresultoperation$ operation after each assignment instruction or creation instruction.

After the execution of the body, the operation has to evaluate the postcondition and it has to make sure that the locked request queues get unlocked at the right time. These two steps are performed by another operation $\checkpostconditionandunlockrequestqueuesoperation$ that takes the missing request queue locks $\overline{g}_{missing\_rq\_locks}$ and the feature $f$. This operation evaluates the postcondition either synchronously or asynchronously. After the evaluation of the postcondition, the operation enqueues an $\unlockrequestqueueoperation$ operation to each request queue in $\overline{g}_{missing\_rq\_locks}$.

SCOOP relies on the Eiffel invariant mechanism. This mechanism is described in \externalsectionreference{7.5} and \externalsectionreference{8.9.16} of the Eiffel ECMA standard \cite{ecma:2006:Eiffel}. On one hand, \externalsectionreference{7.5} describes the semantics of invariants: invariants must be satisfied after the execution of every exported routine and after the execution of every creation procedure. On the other hand, \externalsectionreference{8.9.16} describes the runtime monitoring of invariants: invariants get evaluated on both start and termination of a qualified call to a routine and after every call to a creation procedure. We had to decide whether to rely on the semantics of invariants or on the runtime monitoring of invariants. We decided to rely on the semantics of invariants for two reasons. First, the runtime invariant monitoring mechanism is only one possible implementation of the invariant semantics. Second, the runtime invariant monitoring mechanism relies on the notion of unqualified calls. However, for simplicity this work assumes feature calls to be in the canonical qualified form. The $\applyoperation$ operation reflects this decision: the operation evaluates the invariant whenever $f$ is exported. Note that the invariant can only contain non-separate target expressions. Hence, each call in the invariant will only require $p$'s call stack lock.

Finally, the operation has to return the locks and it has to return the result if $f$ is a function. The $\returnoperation$ operation takes care of this. It comes in a variant for queries and in a variant for commands. Both variants take the channel $a$ and the caller processor $q$ in order to communicate with $q$. The variant for queries additionally takes the value to be returned to $q$.

Before explaining the variants of the $\applyoperation$ operation for non-fresh once routines and attributes, the discussion continues with the operations that have not been discussed in detail so far, namely $\checkpreconditionandlockrequestqueuesoperation$, $\checkpostconditionandunlockrequestqueuesoperation$, and $\returnoperation$.

\begin{fortechnicalreport}
\inferencerule
	{Check Precondition and Lock Request Queues Operation}
	{\freshchanneldefinition{a}}
	{\configuration{p :: \checkpreconditionandlockrequestqueuesoperation(\set{q_{1}, \ldots, q_{m}}, f) \statementseparator s_{p}}{\state}}
	{
		\configuration
			{
				p :: \ & \lockrequestqueuesoperation(\set{q_{1}, \ldots, q_{m}}) \statementseparator \\
				\multilineconditionaloperation
					{f.\preconditionexistsfeature}
					{
						\begin{split}
							& \evaluateoperation(a, f.\preconditionfeature) \statementseparator \\
							& \waitoperation(a) \statementseparator \\
							\multilineconditionaloperation
								{a.\datafeature}
								{\nooperation}
								{
									\begin{split}
										& \issueoperation(q_{1}, \unlockrequestqueueoperation) \statementseparator \\
										& \ldots \\
										& \issueoperation(q_{m}, \unlockrequestqueueoperation) \statementseparator \\
										& \popobtainedrequestqueuelocksoperation \statementseparator \\
										& \checkpreconditionandlockrequestqueuesoperation(\set{q_{1}, \ldots, q_{m}}, f)
									\end{split}
								}
						\end{split}
					}
					{\nooperation} \statementseparator \\
				& s_{p}}
			{\state}
	}

\end{fortechnicalreport}

\begin{forjournal}
\inferencerule
	{Check Precondition and Lock Request Queues Operation}
	{\freshchanneldefinition{a}}
	{\configuration{p :: \checkpreconditionandlockrequestqueuesoperation(\set{q_{1}, \ldots, q_{m}}, f) \statementseparator s_{p}}{\state}}
	{
		\configuration
			{
				p :: \ & \lockrequestqueuesoperation(\set{q_{1}, \ldots, q_{m}}) \statementseparator \\
				\multilineconditionaloperation
					{f.\preconditionexistsfeature}
					{
						\begin{split}
							& \evaluateoperation(a, f.\preconditionfeature) \statementseparator \\
							& \waitoperation(a) \statementseparator \\
							\multilineconditionaloperation
								{a.\datafeature}
								{\nooperation}
								{
									\begin{split}
										& \issueoperation(q_{1}, \unlockrequestqueueoperation) \statementseparator \\
										& \ldots \\
										& \issueoperation(q_{m}, \unlockrequestqueueoperation) \statementseparator \\
										& \popobtainedrequestqueuelocksoperation \statementseparator \\
										& \checkpreconditionandlockrequestqueuesoperation(\set{q_{1}, \ldots, q_{m}}, f)
									\end{split}
								}
						\end{split}
					}
					{\nooperation} \statementseparator \\
				& s_{p}}
			{\state}
	}

\end{forjournal}

The $\checkpreconditionandlockrequestqueuesoperation(\set{q_{1}, \ldots, q_{m}}, f)$ operation, executed by processor $p$, takes a processor set $\set{q_{1}, \ldots, q_{m}}$ whose request queues must be locked on behalf of $p$ and it takes a feature $f$ whose precondition must be satisfied. The operation treats the precondition as a wait condition. It goes through a number of iterations. Each iteration obtains the request queue locks and then evaluates the precondition. If the precondition is satisfied, then the $\checkpreconditionandlockrequestqueuesoperation$ operation finishes. Otherwise it unlocks the request queues and then starts a new iteration.

\begin{fortechnicalreport}
\inferencerule
	{Check Postcondition and Unlock Request Queues Operation}
	{
		\overline{q} \mathematicaldefinition \set{q_{1}, \ldots, q_{m}} \\
		p \notin \overline{q} \\
		\targetsfeature(e) \mathematicaldefinition
		\left\{
			\begin{array}{ll}
				\set{e_{0}} \cup \bigcup_{i = 0 \ldots n}^{}{\targetsfeature(e_{i})} & \condition{e = e_{0}.w(e_{1}, \ldots, e_{n})} \\
				\set{} & \otherwisecondition
			\end{array}
		\right. \\
		\actualargumentsfeature(e) \mathematicaldefinition
		\left\{
			\begin{array}{ll}
				\bigcup_{i = 1 \ldots n}^{}{\set{\tuple{e_{i}, w, i}} \cup \actualargumentsfeature(e_{i})} & \condition{e = e_{0}.w(e_{1}, \ldots, e_{n})} \\
				\set{} & \otherwisecondition
			\end{array}
		\right. \\
		g_{0} \mathematicalelementdefinition
			\left\{
				\begin{split}
					& \multilinecondition{
						& \overline{q} \neq \set{} \wedge \\
						& \forall {x \in \targetsfeature(f.\postconditionfeature)} \colon (\typingenvironmentderivation{\state.\handlerfeature(\state.\valuefeature(p, \controllingentityfeature(x).\namefeature)) \in \overline{q}}) \wedge \\
						& \begin{split}
							& \neg \exists \tuple{x, y, z} \in \actualargumentsfeature(f.\postconditionfeature), t, h, c \colon \\
							& \indentation (\typingenvironmentderivation{x: t \wedge \iscontrolledfeature(t) \wedge y.\formalargumentsfeature(z): (!, h, c) \wedge c.\isreferenceclasstypefeature})
						\end{split}
					} \\
					& \indentation \overline{q} \\
					& \otherwisecondition \\
					& \indentation \set{p}
				\end{split}
			\right. \\
		\set{g_{1}, \ldots, g_{j}} \mathematicaldefinition \overline{q} \setminus g_{0} \\
		\freshchanneldefinition{a}
	}
	{\configuration{p :: \checkpostconditionandunlockrequestqueuesoperation(\set{q_{1}, \ldots, q_{m}}, f) \statementseparator s_{p}}{\state}}
	{
		\configuration
			{
				p :: \ \multilineconditionaloperation
					{f.\postconditionexistsfeature \wedge g_{0} \neq p}
					{
						\begin{split}
							& \begin{split}
								& \issueoperation( \\
								& \indentation \begin{split}
									& g_{0}, \\
									& \executedelegatedoperation( \\
									& \indentation \begin{split}
										& \indentation \begin{split}
											& \evaluateoperation(a, f.\postconditionfeature) \statementseparator \waitoperation(a) \statementseparator \\
											& \issueoperation(g_{1}, \unlockrequestqueueoperation) \statementseparator \ldots \statementseparator \issueoperation(g_{j}, \unlockrequestqueueoperation)
										\end{split} \\
										& , \\
										& \indentation \state.\environmentsfeature(p).\topfeature, \set{q_{1}, \ldots, q_{m}}
									\end{split} \\
									& ) \statementseparator \\
									& \unlockrequestqueueoperation
								\end{split}	\\
								& ) \statementseparator
							\end{split} \\
							& \popobtainedrequestqueuelocksoperation
						\end{split}
					}
					{
						\begin{split}
							\multilineconditionaloperation
								{f.\postconditionexistsfeature}
								{\evaluateoperation(a, f.\postconditionfeature) \statementseparator \waitoperation(a)}
								{\nooperation} \statementseparator \\
							& \issueoperation(q_{1}, \unlockrequestqueueoperation) \statementseparator \ldots \statementseparator \issueoperation(q_{m}, \unlockrequestqueueoperation) \statementseparator \\
							& \popobtainedrequestqueuelocksoperation
						\end{split}
					} \statementseparator \\
				& s_{p}
			}
			{\state}
	}

\end{fortechnicalreport}

The $\checkpostconditionandunlockrequestqueuesoperation$ operation also takes a processor set $\set{q_{1}, \ldots, q_{m}}$ and a feature $f$. The processor set is the same as for the  $\checkpreconditionandlockrequestqueuesoperation$ operation, i.e., the set of processors whose request queues got locked in the synchronization step. The operation first determines whether the postcondition should be evaluated synchronously or asynchronously. Then the operation starts the evaluation. Finally, the operation enqueues an $\unlockrequestqueueoperation$ operation to each request queue in $\set{q_{1}, \ldots, q_{m}}$.

\begin{forjournal}

\end{forjournal}

\begin{clarification}[Asynchronous postcondition evaluation]
The postcondition can be evaluated asynchronously if every feature call in the postcondition only requires a request queue lock that was obtained in the synchronization step and if the postcondition does not involve lock passing. If the postcondition has a feature call that requires a lock different from the obtained request queue locks, then $p$ cannot delegate its obtained request queue lock and then continue because the required lock would be required in another feature execution context as well. Hence the postcondition must be evaluated synchronously in this case. If the postcondition involves lock passing, then one of $p$'s lock might be necessary for the evaluation of the postcondition. Hence, $p$ must pass its locks and cannot proceed until the postcondition is evaluated and the passed locks returned. Once again, the postcondition must be evaluated synchronously. In Nienaltowski's description of SCOOP \cite{nienaltowski:2007:SCOOP} a postcondition can be evaluated asynchronously if the current processor is not involved in the postcondition evaluation. This rule permits configurations in which the evaluating processor does not have the necessary locks for the evaluation.
\end{clarification}

If the postcondition can be evaluated asynchronously, then the operation can take one of the processors in $\set{q_{1}, \ldots, q_{m}}$. This set does not contain processor $p$ because processor $p$ never obtains its own request queue lock. Each processor in this set is exclusively available in the current execution context and can thus be used to evaluate the postcondition asynchronously. The $\checkpostconditionandunlockrequestqueuesoperation$ operation defines $g_{0}$ to be the evaluating processor according to the rule just presented. It also defines $\set{g_{1}, \ldots, g_{j}}$ to be the set $\set{q_{1}, \ldots, q_{m}}$ minus the request queue lock of $g_{0}$. If $p$ is the evaluating processor, then this set is the same as $\set{q_{1}, \ldots, q_{m}}$. As a result of these definitions, the postcondition can be evaluated asynchronously if $g_{0} \neq p$. Otherwise, the postcondition must be evaluated synchronously.

In the synchronous case, processor $p$ evaluates the postcondition, enqueues $\unlockrequestqueueoperation$ operations to each request queue in $\set{q_{1}, \ldots, q_{m}}$, and then removes the corresponding locks from its stack of obtained request queue locks. The $\unlockrequestqueueoperation$ operations will not proceed until the locks have been removed from $p$'s stack of obtained request queue locks. In the asynchronous case, processor $p$ must delegate the postcondition evaluation to processor $g_{0}$. For this purpose, $p$ enqueues an $\executedelegatedoperation$ operation to $g_{0}$. The workload involves the postcondition evaluation along with the subsequent issuing of $\unlockrequestqueueoperation$ operations to all processor in $\set{g_{1}, \ldots, g_{j}}$. Processor $g_{0}$ unlocks its own request queue after the delegated execution. The evaluation of the postcondition on $g_{0}$ requires the environment that defines the values of the entities in the postcondition. Furthermore, the evaluation requires the request queue locks $\set{q_{1}, \ldots, q_{m}}$. These locks are sufficient because the postcondition only gets evaluated asynchronously if the evaluation only requires these locks. To satisfy these two requirements, $p$ gives its top environment and $\set{q_{1}, \ldots, q_{m}}$ to $g_{0}$. After $g_{0}$ performed the delegated execution, it can unlock its own request queue. In the meantime, processor $p$ removes $\set{q_{1}, \ldots, q_{m}}$ from its obtained request queue locks to enable $g_{0}$ to proceed with the delegated execution.

The $\returnoperation$ operation comes in two variants: one for queries and one for commands.

\singlelineinferencerule
	{Return Operation -- Query}
	{
		\begin{split}
			& \tuple{\state', r'} \mathematicaldefinition
				\left\{
					\begin{split}
						& \condition{r \neq \voidvalue \wedge \state.\referencedobjectfeature(r).\classtypefeature.\isexpandedclasstypefeature \wedge \state.\handlerfeature(r) \neq q} \\
						& \indentation \where
							{\tuple{\state_{x}, \state_{x}.\lastimportedreferencefeature}}
							{\state_{x} \mathematicaldefinition \state.\deepimportfeature(q, r)} \\
							& \otherwisecondition \\
							& \indentation \tuple{\state, r}
					\end{split}
				\right.
		\end{split} \\
		\state'' \mathematicaldefinition \state'.\popenvironmentfeature(p).\revokelocksfeature(q, p)
	}
	{\configuration{p :: \returnoperation(a, r, q) \statementseparator s_{p}}{\state}}
	{
		\configuration
			{p :: \resultoperation(a, r') \statementseparator s_{p}}
			{\state''}
	}

\inferencerule
	{Return Operation -- Command}
	{\state' \mathematicaldefinition \state.\popenvironmentfeature(p).\revokelocksfeature(q, p)}
	{\configuration{p :: \returnoperation(a, q) \statementseparator s_{p}}{\state}}
	{
		\configuration
			{
				p :: \ & \singlelineconditionaloperation
					{\state.\arelockspassedfeature(q)}
					{\notifyoperation(a)}
					{\nooperation} \statementseparator
				s_{p}
			}
			{\state'}
	}

The variant for queries returns the result and the locks. The variant for commands only returns the locks. Both variants take a channel $a$ and the caller processor $q$. For queries, the channel is used to return the result. For this purpose, the operation takes a reference $r$ that points to the result. Processor $q$ is waiting for this result on channel $a$. This can be seen in the $\calloperation$ operation, which issues an $\applyoperation$ operation and a subsequent $\waitoperation(a)$ operation. The $\applyoperation$ operation calls the $\returnoperation$ operation with the same channel $a$. To return the result to $q$, processor $p$ executes a $\resultoperation$ on $a$. The value to be returned is not always $r$ directly. If $r$ points to an object of expanded class type and $q \neq p$, then $q$ must deep import the object. In all other cases, $q$ can take $r$ as the return value. An explanation why the deep import operation is necessary can be found in \sectionreference{sec:setting values}. For commands, the channel is used to signal to $q$ that the locks have been returned in case $q$ passed its locks. This can be determined by looking at the state: $\state.\arelockspassedfeature(q)$. In both variants of the $\returnoperation$ operation, $p$ removes the passed locks from the stacks of retrieved locks. In case $q$ did not pass any locks, the removed entries might be the empty set. Processor $p$ also removes its top environment because this environment is no longer needed. In case of an asynchronous postcondition evaluation, this environment temporarily gets delegated to the evaluating processor.

Until now, the discussion left out the non-fresh once routines and the attributes. Non-fresh once functions already have a result. The $\applyoperation$ operation just needs to get this result from the state and return it. For non-fresh once procedures it does not even have to do this. The only obligation is the evaluation of the invariant. The evaluation of the invariant requires the call stack lock of $p$. This lock is given if the condition $\neg \state.\arelockspassedfeature(p)$ holds. For attributes, note that an instance of $\attributetype$ is also an instance of $\expressiontype$. Hence, the operation evaluates the attribute expression and returns the result of the evaluation. The invariant does not have to be evaluated in this case.

\inferencerule
	{Application Operation -- Non-Fresh Once Routine}
	{
		f \in \routinetype \wedge f.\isonceroutinefeature \wedge \neg \state.\isonceroutinefreshfeature(p, f) \\
		\state.\handlerfeature(r_{0}) = p \\
		\neg \state.\passlocksfeature(q, p, \overline{l}).\arelockspassedfeature(p) \\
		\state' \mathematicaldefinition \state.\passlocksfeature(q, p, \overline{l}).\pushenvironmentwithfeaturefeature(p, f, r_{0}, \tuple{r_{1},\ldots, r_{n}}) \\
		\freshchanneldefinition{a_{inv}}
	}
	{\configuration{p :: \applyoperation(a, r_{0}, f, \tuple{r_{1}, \ldots, r_{n}}, q, \overline{l}) \statementseparator s_{p}}{\state}}
	{
		\configuration
			{
				p :: \ \multilineconditionaloperation
					{f.\classtypefeature.\invariantexistsfeature \wedge f.\isexportedfeature}
					{\evaluateoperation(a_{inv}, f.\classtypefeature.\invariantfeature) \statementseparator \waitoperation(a_{inv})}
					{\nooperation} \statementseparator \\
				\multilineconditionaloperation
					{f \in \functiontype}
					{\returnoperation(a, \state'.\oncefunctionresultfeature(p, f), q)}
					{\returnoperation(a, q)} \statementseparator \\
				& s_{p}}
			{\state'}
	}

\inferencerule
	{Application Operation -- Attribute}
	{
		f \in \attributetype \\
		\state.\handlerfeature(r_{0}) = p \\
		\neg \state.\passlocksfeature(q, p, \overline{l}).\arelockspassedfeature(p) \\
		\state' \mathematicaldefinition \state.\passlocksfeature(q, p, \overline{l}).\pushenvironmentwithfeaturefeature(p, f, r_{0}, \tuple{}) \\
		\freshchanneldefinition{a'}
	}
	{\configuration{p :: \applyoperation(a, r_{0}, f, \tuple{}, q, \overline{l}) \statementseparator s_{p}}{\state}}
	{
		\configuration
			{
				p :: \ & \evaluateoperation(a', f) \statementseparator \\
				& \waitoperation(a') \statementseparator \\
				& \returnoperation(a, a'.\datafeature, q) \statementseparator \\
				& s_{p}}
			{\state'}
	}

\begin{fortechnicalreport}
\newpage
\begin{example}[Feature application]
This example looks at the application of the feature $buy$ on the first investor. This feature is shown in \listingreference{lst:investor class}.

\begin{lstlisting}[caption=Investor class, label=lst:investor class, language=SCOOP]
class INVESTOR

create
  make
	
feature -- Initialization
  make (new_id: INTEGER)
      -- Create an investor with a new identifier.
    do
      id := new_id
    end
		
feature -- Access
  id: INTEGER
      -- The identifier.
	
  log: separate UUID
      -- The identifier of the last market.

  buy (market: separate MARKET; issuer_id: INTEGER)
      -- Buy a share of the issuer on the market.
    require
      market.can_buy (Current.id, issuer_id)
    do
      market.buy (Current, issuer_id)
      log := market.id
    ensure
      market.can_sell (Current.id, issuer_id)
    end

    ...		
end
\end{lstlisting}

The execution starts with a configuration where processor $p_{1}$ finished executing the feature calls in feature $do\_transaction$. These feature calls led to one $\applyoperation$ operation for the $buy$ feature and one for the $sell$ feature in each investor's action queue.

\isolatedconfiguration
	{
		p_{1} :: \ & \ldots \processorseparator \\
		p_{2} :: \ & \processorseparator \\
		p_{3} :: \ & \applyoperation(a_{62}, r_{6}, buy, \tuple{r_{1}, r_{39}}, p_{1}, \tuple{\set{}, \set{}}) \statementseparator \\
		& \applyoperation(a_{70}, r_{6}, sell, \tuple{r_{1}, r_{39}}, p_{1}, \tuple{\set{}, \set{}}) \processorseparator \\
		p_{4} :: \ & \applyoperation(a_{66}, r_{8}, buy, \tuple{r_{1}, r_{39}}, p_{1}, \tuple{\set{}, \set{}}) \statementseparator \\
		& \applyoperation(a_{74}, r_{8}, sell, \tuple{r_{1}, r_{39}}, p_{1}, \tuple{\set{}, \set{}})
	}
	{
		\simplifiedstate
			{
				& \simplifiedstatelocksentry
					{p_{1}}
					{\set{}, \set{p_{3}, p_{4}}}
					{\set{}, \set{}}
					{\set{}, \set{}}
					{\simplifiedstatelockedindicator}
					{\simplifiedstatenopassedlocksindicator} \\
				& \simplifiedstatelocksentry
					{p_{2}}
					{}
					{}
					{}
					{\simplifiedstateunlockedindicator}
					{\simplifiedstatenopassedlocksindicator} \\
				& \simplifiedstatelocksentry
					{p_{3}}
					{}
					{}
					{}
					{\simplifiedstatelockedindicator}
					{\simplifiedstatenopassedlocksindicator} \\
				& \simplifiedstatelocksentry
					{p_{4}}
					{}
					{}
					{}
					{\simplifiedstatelockedindicator}
					{\simplifiedstatenopassedlocksindicator}
			}
			{
				& \simplifiedstateobjectsentry
					{p_{1}}
					{
						\simplifiedstatereferencedobject{r_{0}}{o_{0}(\simplifiedstateentityvalue{market}{r_{1}})}, \simplifiedstatereferencedobject{r_{39}}{o_{48}(1)}
					} \\
				& \simplifiedstateobjectsentry
					{p_{2}}
					{
						& \simplifiedstatereferencedobject{r_{1}}{o_{35}(\simplifiedstateentityvalue{cash}{r_{16}}, \simplifiedstateentityvalue{available\_shares}{r_{23}}, \simplifiedstateentityvalue{owned\_shares}{r_{29}})}, \\
						& \simplifiedstatereferencedobject{r_{16}}{o_{25}[r_{21}, r_{22}]}, \simplifiedstatereferencedobject{r_{21}}{o_{23}(100)}, \simplifiedstatereferencedobject{r_{22}}{o_{24}(100)}, \\
						& \simplifiedstatereferencedobject{r_{23}}{o_{33}[r_{28}]}, \simplifiedstatereferencedobject{r_{28}}{o_{32}(1)}, \\
						& \simplifiedstatereferencedobject{r_{29}}{o_{38}[[r_{34}], [r_{35}]]}, \simplifiedstatereferencedobject{r_{34}}{o_{41}(0)}, \simplifiedstatereferencedobject{r_{35}}{o_{42}(0)}
					} \\
				& \simplifiedstateobjectsentry
					{p_{3}}
					{
						\simplifiedstatereferencedobject{r_{6}}{o_{44}(\simplifiedstateentityvalue{id}{r_{36}})},
						\simplifiedstatereferencedobject{r_{36}}{o_{43}(1)}
					} \\
				& \simplifiedstateobjectsentry
					{p_{4}}
					{
						\simplifiedstatereferencedobject{r_{8}}{o_{46}(\simplifiedstateentityvalue{id}{r_{37}})},
						\simplifiedstatereferencedobject{r_{37}}{o_{45}(2)}
					}
			}
			{}
			{
				& \simplifiedstateenvironmentsentry
					{p_{1}}
					{
						& \simplifiedstateentityvalue{first\_investor}{r_{6}}, \simplifiedstateentityvalue{second\_investor}{r_{8}}, \simplifiedstatecurrententityvalue{r_{0}} \simplifiedstateenvironmentsentryseparator \\
						& \simplifiedstateentityvalue{first\_investor}{r_{6}}, \simplifiedstateentityvalue{second\_investor}{r_{8}}, \simplifiedstateentityvalue{issuer\_id}{r_{39}}, \simplifiedstatecurrententityvalue{r_{0}}
					} \\
				& \simplifiedstateenvironmentsentry
					{p_{2}}
					{} \\
				& \simplifiedstateenvironmentsentry
					{p_{3}}
					{} \\
				& \simplifiedstateenvironmentsentry
					{p_{4}}
					{}
			}
	}

At this point, processors $p_{3}$ and $p_{4}$ can each take the transition that is described by the inference rule for the $\applyoperation$ operation for non-once routines. Each processor can then take an additional transition by executing the $\checkpreconditionandlockrequestqueuesoperation$ operation. The result configuration is shown below. The channel $a_{75}$ is a fresh channel. Both processors added a new environment that maps the expanded formal argument to a copy of the expanded actual argument. In case of processor $p_{3}$, the copied object is referenced by $r_{40}$. On processor $p_{4}$, the copied object is referenced by $r_{41}$. Since processor $p_{1}$ did not pass its locks, both $p_{3}$ and $p_{4}$ added empty lock sets to their stack of retrieved locks.

\isolatedconfiguration
	{
		p_{1} :: \ & \ldots \processorseparator \\
		p_{2} :: \ & \processorseparator \\
		p_{3} :: \ & \lockrequestqueuesoperation(\set{p_{2}}) \statementseparator \\
		& \evaluateoperation(a_{75}, market.can\_buy(\currententity.id, issuer\_id)) \statementseparator \\
		& \waitoperation(a_{75}) \statementseparator \\
		\multilineconditionaloperation
				{a_{75}.\datafeature}
				{\nooperation}
				{
					\begin{split}
						& \issueoperation(p_{2}, \unlockrequestqueueoperation) \statementseparator \\
						& \popobtainedrequestqueuelocksoperation \statementseparator \\
						& \checkpreconditionandlockrequestqueuesoperation(\set{p_{2}}, buy)
					\end{split}
				} \statementseparator \\
		& market.buy (\currententity, issuer\_id) \statementseparator \\
		& log \eassignment market.id \statementseparator \\
		& \checkpostconditionandunlockrequestqueuesoperation(\set{p_{2}}, buy) \statementseparator \\
		& \returnoperation(a_{62}, p_{1}) \statementseparator \\
		& \applyoperation(a_{70}, r_{6}, sell, \tuple{r_{1}, r_{39}}, p_{1}, \tuple{\set{}, \set{}}) \processorseparator \\
		p_{4} :: \ & \lockrequestqueuesoperation(\set{p_{2}}) \statementseparator \\
		& \ldots
	}
	{
		\simplifiedstate
			{
				& \simplifiedstatelocksentry
					{p_{1}}
					{\set{}, \set{p_{3}, p_{4}}}
					{\set{}, \set{}}
					{\set{}, \set{}}
					{\simplifiedstatelockedindicator}
					{\simplifiedstatenopassedlocksindicator} \\
				& \simplifiedstatelocksentry
					{p_{2}}
					{}
					{}
					{}
					{\simplifiedstateunlockedindicator}
					{\simplifiedstatenopassedlocksindicator} \\
				& \simplifiedstatelocksentry
					{p_{3}}
					{}
					{\set{}}
					{\set{}}
					{\simplifiedstatelockedindicator}
					{\simplifiedstatenopassedlocksindicator} \\
				& \simplifiedstatelocksentry
					{p_{4}}
					{}
					{\set{}}
					{\set{}}
					{\simplifiedstatelockedindicator}
					{\simplifiedstatenopassedlocksindicator}
			}
			{
				& \simplifiedstateobjectsentry
					{p_{1}}
					{
						\simplifiedstatereferencedobject{r_{0}}{o_{0}(\simplifiedstateentityvalue{market}{r_{1}})}, \simplifiedstatereferencedobject{r_{39}}{o_{48}(1)}
					} \\
				& \simplifiedstateobjectsentry
					{p_{2}}
					{
						& \simplifiedstatereferencedobject{r_{1}}{o_{35}(\simplifiedstateentityvalue{cash}{r_{16}}, \simplifiedstateentityvalue{available\_shares}{r_{23}}, \simplifiedstateentityvalue{owned\_shares}{r_{29}})}, \\
						& \simplifiedstatereferencedobject{r_{16}}{o_{25}[r_{21}, r_{22}]}, \simplifiedstatereferencedobject{r_{21}}{o_{23}(100)}, \simplifiedstatereferencedobject{r_{22}}{o_{24}(100)}, \\
						& \simplifiedstatereferencedobject{r_{23}}{o_{33}[r_{28}]}, \simplifiedstatereferencedobject{r_{28}}{o_{32}(1)}, \\
						& \simplifiedstatereferencedobject{r_{29}}{o_{38}[[r_{34}], [r_{35}]]}, \simplifiedstatereferencedobject{r_{34}}{o_{41}(0)}, \simplifiedstatereferencedobject{r_{35}}{o_{42}(0)}
					} \\
				& \simplifiedstateobjectsentry
					{p_{3}}
					{
						\simplifiedstatereferencedobject{r_{6}}{o_{44}(\simplifiedstateentityvalue{id}{r_{36}})},
						\simplifiedstatereferencedobject{r_{36}}{o_{43}(1)},
						\simplifiedstatereferencedobject{r_{40}}{o_{49}(1)}
					} \\
				& \simplifiedstateobjectsentry
					{p_{4}}
					{
						\simplifiedstatereferencedobject{r_{8}}{o_{46}(\simplifiedstateentityvalue{id}{r_{37}})},
						\simplifiedstatereferencedobject{r_{37}}{o_{45}(2)},
						\simplifiedstatereferencedobject{r_{41}}{o_{50}(1)}
					}
			}
			{}
			{
				& \simplifiedstateenvironmentsentry
					{p_{1}}
					{
						& \simplifiedstateentityvalue{first\_investor}{r_{6}}, \simplifiedstateentityvalue{second\_investor}{r_{8}}, \simplifiedstatecurrententityvalue{r_{0}} \simplifiedstateenvironmentsentryseparator \\
						& \simplifiedstateentityvalue{first\_investor}{r_{6}}, \simplifiedstateentityvalue{second\_investor}{r_{8}}, \simplifiedstateentityvalue{issuer\_id}{r_{39}}, \simplifiedstatecurrententityvalue{r_{0}}
					} \\
				& \simplifiedstateenvironmentsentry
					{p_{2}}
					{} \\
				& \simplifiedstateenvironmentsentry
					{p_{3}}
					{
						& \simplifiedstateentityvalue{market}{r_{1}}, \simplifiedstateentityvalue{issuer\_id}{r_{40}}, \simplifiedstatecurrententityvalue{r_{6}}
					} \\
				& \simplifiedstateenvironmentsentry
					{p_{4}}
					{
						& \simplifiedstateentityvalue{market}{r_{1}}, \simplifiedstateentityvalue{issuer\_id}{r_{41}}, \simplifiedstatecurrententityvalue{r_{8}}
					}
			}
	}

The inference rule for the $\lockrequestqueuesoperation$ operation shows that both $p_{3}$ and $p_{4}$ require the request queue lock of $p_{2}$. We decide to give priority to $p_{3}$. This leads to the following configuration, where $p_{2}$'s request queue is locked on behalf of $p_{3}$:

\isolatedconfiguration
	{
		p_{1} :: \ & \ldots \processorseparator \\
		p_{2} :: \ & \processorseparator \\
		p_{3} :: \ & \evaluateoperation(a_{75}, market.can\_buy(\currententity.id, issuer\_id)) \statementseparator \\
		& \waitoperation(a_{75}) \statementseparator \\
		\multilineconditionaloperation
			{a_{75}.\datafeature}
			{\nooperation}
			{
				\begin{split}
					& \issueoperation(p_{2}, \unlockrequestqueueoperation) \statementseparator \\
					& \popobtainedrequestqueuelocksoperation \statementseparator \\
					& \checkpreconditionandlockrequestqueuesoperation(\set{p_{2}}, buy)
				\end{split}
			} \statementseparator \\
		& market.buy (\currententity, issuer\_id) \statementseparator \\
		& log \eassignment market.id \statementseparator \\
		& \checkpostconditionandunlockrequestqueuesoperation(\set{p_{2}}, buy) \statementseparator \\
		& \returnoperation(a_{62}, p_{1}) \statementseparator \\
		& \applyoperation(a_{70}, r_{6}, sell, \tuple{r_{1}, r_{39}}, p_{1}, \tuple{\set{}, \set{}}) \processorseparator \\
		p_{4} :: & \ldots
	}
	{
		\simplifiedstate
			{
				& \simplifiedstatelocksentry
					{p_{1}}
					{\set{}, \set{p_{3}, p_{4}}}
					{\set{}, \set{}}
					{\set{}, \set{}}
					{\simplifiedstatelockedindicator}
					{\simplifiedstatenopassedlocksindicator} \\
				& \simplifiedstatelocksentry
					{p_{2}}
					{}
					{}
					{}
					{\simplifiedstatelockedindicator}
					{\simplifiedstatenopassedlocksindicator} \\
				& \simplifiedstatelocksentry
					{p_{3}}
					{\set{p_{2}}}
					{\set{}}
					{\set{}}
					{\simplifiedstatelockedindicator}
					{\simplifiedstatenopassedlocksindicator} \\
				& \simplifiedstatelocksentry
					{p_{4}}
					{}
					{\set{}}
					{\set{}}
					{\simplifiedstatelockedindicator}
					{\simplifiedstatenopassedlocksindicator}
			}
			{\ldots}
			{}
			{\ldots}
	}

The evaluation of the precondition produces a result on $p_{2}$, which is awaited by $p_{3}$. The result is a deep imported object of class type $\booleanclasstype$, referenced by $r_{58}$. The boolean value of this object indicates that the precondition is satisfied and hence $p_{3}$ can continue with the execution of the body. In the following configuration, the arrays referenced by $r_{16}$, $r_{23}$, and $r_{29}$ have been updated; the first investor bought a share of the issuer. Consequently, the first investor has a lower amount of cash and there is one fewer share available. Furthermore, the $log$ attribute of the first investor object has been updated with the identifier of the market. This identifier object, referenced by $r_{65}$, has been created by $p_{2}$ in a once function of separate type. Hence the object is available as a once result on all processors in the system.

\isolatedconfiguration
	{
		p_{1} :: \ & \ldots \processorseparator
		p_{2} :: \processorseparator \\
		p_{3} :: \ & \checkpostconditionandunlockrequestqueuesoperation(\set{p_{2}}, buy) \statementseparator \\
		& \returnoperation(a_{62}, p_{1}) \statementseparator \\
		& \applyoperation(a_{70}, r_{6}, sell, \tuple{r_{1}, r_{39}}, p_{1}, \tuple{\set{}, \set{}}) \processorseparator
		p_{4} :: \ldots
	}
	{
		\simplifiedstate
			{\ldots}
			{
				& \simplifiedstateobjectsentry
					{p_{1}}
					{
						\simplifiedstatereferencedobject{r_{0}}{o_{0}(\simplifiedstateentityvalue{market}{r_{1}})}, \simplifiedstatereferencedobject{r_{39}}{o_{48}(1)}
					} \\
				& \simplifiedstateobjectsentry
					{p_{2}}
					{
						& \simplifiedstatereferencedobject{r_{1}}{o_{35}(\simplifiedstateentityvalue{cash}{r_{16}}, \simplifiedstateentityvalue{available\_shares}{r_{23}}, \simplifiedstateentityvalue{owned\_shares}{r_{29}})}, \\
						& \simplifiedstatereferencedobject{r_{16}}{o_{71}[r_{61}, r_{22}]}, \simplifiedstatereferencedobject{r_{61}}{o_{70}(90)}, \simplifiedstatereferencedobject{r_{22}}{o_{24}(100)}, \\
						& \simplifiedstatereferencedobject{r_{23}}{o_{73}[r_{62}]}, \simplifiedstatereferencedobject{r_{62}}{o_{72}(0)}, \\
						& \simplifiedstatereferencedobject{r_{29}}{o_{75}[[r_{63}], [r_{35}]]}, \simplifiedstatereferencedobject{r_{63}}{o_{74}(1)}, \simplifiedstatereferencedobject{r_{35}}{o_{42}(0)}, \\
						& \simplifiedstatereferencedobject{r_{65}}{o_{77}}
					} \\
				& \simplifiedstateobjectsentry
					{p_{3}}
					{
						\simplifiedstatereferencedobject{r_{6}}{o_{78}(\simplifiedstateentityvalue{id}{r_{36}}, \simplifiedstateentityvalue{log}{r_{65}})},
						\simplifiedstatereferencedobject{r_{36}}{o_{43}(1)},
						\simplifiedstatereferencedobject{r_{40}}{o_{49}(1)},
						\simplifiedstatereferencedobject{r_{58}}{o_{67}(true)}
					} \\
				& \simplifiedstateobjectsentry
					{p_{4}}
					{
						\simplifiedstatereferencedobject{r_{8}}{o_{46}(\simplifiedstateentityvalue{id}{r_{37}})},
						\simplifiedstatereferencedobject{r_{37}}{o_{45}(2)},
						\simplifiedstatereferencedobject{r_{41}}{o_{50}(1)}
					}
			}
			{
				& \simplifiedstateoncestatusentry
					{\simplifiedstateallprocessorsindicator}
					{
						\simplifiedstateoncefunctionstatus{MARKET}{id}{r_{65}}
					}
			}
			{\ldots}
	}

The postcondition of $buy$ contains the expression $current.id$ with target $current$. The handler of the controlling entity is $p_{3}$. Because $p_{3}$ did not obtain its own request queue lock, it must evaluate the postcondition synchronously. After the postcondition evaluation, $p_{3}$ issues the $\unlockrequestqueueoperation$ operation to $p_{2}$ and removes $p_{2}$'s request queue lock. This enables $p_{2}$ to unlock its request queue lock. Processor $p_{3}$ then executes the $\returnoperation$ operation to get rid of its top environment and its retrieved locks:

\isolatedconfiguration
	{
		p_{1} :: \ & \ldots \processorseparator
		p_{2} :: \processorseparator 
		p_{3} :: \applyoperation(a_{70}, r_{6}, sell, \tuple{r_{1}, r_{39}}, p_{1}, \tuple{\set{}, \set{}}) \processorseparator
		p_{4} :: \ldots
	}
	{
		\simplifiedstate
			{
				& \simplifiedstatelocksentry
					{p_{1}}
					{\set{}, \set{p_{3}, p_{4}}}
					{\set{}, \set{}}
					{\set{}, \set{}}
					{\simplifiedstatelockedindicator}
					{\simplifiedstatenopassedlocksindicator} \\
				& \simplifiedstatelocksentry
					{p_{2}}
					{}
					{}
					{}
					{\simplifiedstateunlockedindicator}
					{\simplifiedstatenopassedlocksindicator} \\
				& \simplifiedstatelocksentry
					{p_{3}}
					{}
					{}
					{}
					{\simplifiedstatelockedindicator}
					{\simplifiedstatenopassedlocksindicator} \\
				& \simplifiedstatelocksentry
					{p_{4}}
					{}
					{\set{}}
					{\set{}}
					{\simplifiedstatelockedindicator}
					{\simplifiedstatenopassedlocksindicator}
			}
			{\ldots}
			{\ldots}
			{
				& \simplifiedstateenvironmentsentry
					{p_{1}}
					{
						& \simplifiedstateentityvalue{first\_investor}{r_{6}}, \simplifiedstateentityvalue{second\_investor}{r_{8}}, \simplifiedstatecurrententityvalue{r_{0}} \simplifiedstateenvironmentsentryseparator \\
						& \simplifiedstateentityvalue{first\_investor}{r_{6}}, \simplifiedstateentityvalue{second\_investor}{r_{8}}, \simplifiedstateentityvalue{issuer\_id}{r_{39}}, \simplifiedstatecurrententityvalue{r_{0}}
					} \\
				& \simplifiedstateenvironmentsentry
					{p_{2}}
					{} \\
				& \simplifiedstateenvironmentsentry
					{p_{3}}
					{} \\
				& \simplifiedstateenvironmentsentry
					{p_{4}}
					{
						& \simplifiedstateentityvalue{market}{r_{1}}, \simplifiedstateentityvalue{issuer\_id}{r_{41}}, \simplifiedstatecurrententityvalue{r_{8}}
					}
			}
	}
\end{example}
\end{fortechnicalreport}

\subsubsection{Creation instructions}
A creation instruction has the form $\ecreate b.f(e_{1}, \ldots, e_{n})$ where $b$ is the target entity, $f$ is the creation procedure, and $e_{1}, \ldots, e_{n}$ are the actual arguments. Assume that $b$ is of type $(d, g, c)$. A processor $p$ that executes this instruction takes the following steps:
\begin{enumerate}
	\item Processor $q$ creation.
		\begin{itemize}
			\item If $b$ is separate, i.e., $g = \top$, then create a new processor.
			\item If $b$ has an explicit processor specification, i.e., $g = \alpha$, then
				\begin{itemize}
					\item take the processor denoted by $\alpha$ if it already exists.
					\item create a new processor if the processor denoted by $\alpha$ does not exist yet.
				\end{itemize}
			\item If $b$ is non-separate, i.e., $g = \bullet$, then take $p$.
		\end{itemize}
	\item Locking. Lock the request queue of $q$ if the following conditions hold:
		\begin{itemize}
			\item Processor $p$ and processor $q$ are different.
			\item Processor $p$ does not have $q$'s request queue lock.
			\item Processor $q$ does not have $p$'s request queue lock.
		\end{itemize}
	\item Object creation. Ask $q$ to create a new instance with class type $c$ using the creation procedure $f$. Attach the newly created object to $b$.
	\item Invariant evaluation. If $f$ is not exported, then ask $q$ to evaluate the invariant.
	\item Lock releasing. If $q$'s request queue has been locked in the locking step, then ask $q$ to unlock its request queue after it is done with the feature request.
\end{enumerate}
There are four cases in the processor creation step:
\begin{itemize}
	\item The entity $b$ has a separate type.
	\item The entity $b$ has an explicit processor specification and the denoted processor already exists.
	\item The entity $b$ has an explicit processor specification and the denoted processor does not yet exist.
	\item The entity $b$ has a non-separate type.
\end{itemize}
For each of these cases, there is one inference rule. The discussion starts with the variant where $b$ has a separate type. In this case, the instruction defines $q$ as a new processor and $o$ as a new object of class type $c$. The reference $r$ points to this object. First the instruction obtains a request queue lock on the new processor $q$ so that it can issue statements on $q$. Next, it writes the value $r$ into the entity $b$. To make a call to the creation procedure, it executes a command instruction. Once this is done, it checks whether there is an invariant to evaluate. If $f$ is exported, then the invariant will be evaluated as part of $f$'s feature application. In this case the instruction does nothing. However, if $f$ is not exported, then it must issue the invariant evaluation to $q$. After this step, it can issue an $\unlockrequestqueueoperation$ operation to $q$ and remove the request queue lock from $p$'s obtained request queue locks.

\inferencerule
	{Create Instruction -- Top}
	{
		(d, h, c) \mathematicaldefinition \typefromtypingenvironment(\typingenvironment, b) \\
		h = \top \\
		q \mathematicaldefinition \state.\newprocessorfeature \\
		o \mathematicaldefinition \state.\newobjectfeature(c) \\
		\state' \mathematicaldefinition \state.\addprocessorfeature(q).\addobjectfeature(q, o) \\
		r \mathematicaldefinition \state'.\referencefeature(o) \\
		\freshchanneldefinition{a}
	}
	{\configuration{p :: \ecreate b.f(e_{1}, \ldots, e_{n}) \statementseparator s_{p}}{\state}}
	{
		\configuration
			{
				p :: \ & \lockrequestqueuesoperation(\set{q}) \statementseparator \\
				& \writevalueoperation(b.\namefeature, r) \statementseparator \\
				& b.f(e_{1}, \ldots, e_{n}) \statementseparator \\
				\multilineconditionaloperation
					{\neg f.\classtypefeature.\invariantexistsfeature \vee f.\isexportedfeature}
					{\nooperation}
					{\issueoperation(q, \evaluateoperation(a, f.\classtypefeature.\invariantfeature) \statementseparator \waitoperation(a))} \statementseparator \\
				& \issueoperation(q, \unlockrequestqueueoperation) \statementseparator \\
				& \popobtainedrequestqueuelocksoperation \statementseparator \\
				& s_{p} \processorseparator q :: \nooperation
			}
			{\state'}
	}

The following discussion looks at the two variants for the cases where $b$ has an explicit processor specification. There are two forms of explicit processor specifications: unqualified and qualified. An unqualified explicit processor specification, i.e., $<x>$, is based on a processor attribute $x$ with an attached type. The processor denoted by this explicit processor specification is the processor stored in $x$. A qualified explicit processor specification, i.e., $<y.\handlerfeature>$, is based on a non-writable entity $y$ of attached type. The processor denoted by this explicit processor specification is the same processor as the one handling the object referenced by $y$. A qualified explicit processor specification always denotes an existing processor because this specification is based on an attached entity. This means that there is already an object attached to this entity and thus its handler must exist. This insight helps to write the conditions for the two inference rule variants.

\begin{forjournal}
\inferencerule
	{Create Instruction -- Existing Explicit Processor}
	{
		(d, h, c) \mathematicaldefinition \typefromtypingenvironment(\typingenvironment, b) \\
		h = <x> \vee h = <y.\handlerfeature> \\
		q \mathematicaldefinition 
			\left\{
				\begin{array}{ll}
					\state.\valuefeature(p, x) & \condition{t = (d, <x>, c)} \\
					\state.\handlerfeature(\state.\valuefeature(p, y)) & \condition{t = (d, <y.\handlerfeature>, c)}
				\end{array}
			\right. \\
		\state.\processorsfeature.\containsfeature(q) \\
		\overline{g}_{required\_cs\_locks} \mathematicaldefinition 
			\left\{
				\begin{array}{ll}
					\set{q} & \condition{q \neq p \wedge (\state.\requestqueuelocksfeature(q).\containsfeature(p) \vee \state.\callstacklocksfeature(q).\containsfeature(p))} \\
					\set{} & \otherwisecondition
				\end{array}
			\right. \\
		\forall x \in \overline{g}_{required\_cs\_locks} \colon \neg \state.\arelockspassedfeature(p) \wedge \state.\callstacklocksfeature(p).\containsfeature(x) \\
		o \mathematicaldefinition  \state.\newobjectfeature(c) \\
		\state' \mathematicaldefinition \state.\addobjectfeature(q, o) \\
		r \mathematicaldefinition  \state'.\referencefeature(o) \\
		\freshchanneldefinition{a}
	}
	{\configuration{p :: \ecreate b.f(e_{1}, \ldots, e_{n}) \statementseparator s_{p}}{\state}}
	{
		\configuration
			{
				p :: \ \multilineconditionaloperation
					{q \neq p \wedge \neg \state'.\requestqueuelocksfeature(p).\containsfeature(q) \wedge \neg \state'.\requestqueuelocksfeature(q).\containsfeature(p)}
					{\lockrequestqueuesoperation(\set{q})}
					{\nooperation} \statementseparator \\
				& \writevalueoperation(b.\namefeature, r) \statementseparator \\
				& b.f(e_{1}, \ldots, e_{n}) \statementseparator \\
				\multilineconditionaloperation
					{\neg f.\classtypefeature.\invariantexistsfeature \vee f.\isexportedfeature}
					{\nooperation}
					{\issueoperation(q, \evaluateoperation(a, f.\classtypefeature.\invariantfeature) \statementseparator \waitoperation(a))} \statementseparator \\
				\multilineconditionaloperation
					{q \neq p \wedge \neg \state'.\requestqueuelocksfeature(p).\containsfeature(q) \wedge \neg \state'.\requestqueuelocksfeature(q).\containsfeature(p)}
					{
						\begin{split}
							& \issueoperation(q, \unlockrequestqueueoperation) \statementseparator \\
							& \popobtainedrequestqueuelocksoperation 
						\end{split}
					}
					{\nooperation} \statementseparator \\
				& s_{p}
			}
			{\state'}
	}

\end{forjournal}

The variant that handles existing processors states that the specified processor must exist. To check this, one must consider both the qualified and the unqualified possibility. For the qualified option, one can simply lookup the value of the attribute $x$. For the unqualified option, one first looks up the value of the entity $y$ and then determines the handler of the referenced object. In either case, the result $q$ is either the denoted processor or the void value. One then checks whether $q$ is in the set of processors of our system. The overall idea of this inference rule is the same as in the case where $b$ has a separate type. The difference is in the processor creation, locking, and lock releasing steps. Instead of creating a new processor, the instruction takes the existing processor $q$. If $q = p$, then the call to the creation procedure will be a non-separate call. In this case, the instruction requires $p$'s call stack lock. This lock is given because otherwise $p$ would be waiting. If $p \neq q$ and $q$ has a lock on $p$, then the call to the creation procedure will be a separate callback. In this case, the instruction requires $q$'s call stack lock. This is expressed in the condition with the help of the set $\overline{g}_{required\_cs\_locks}$. If $p \neq q$ and $q$ does not have $p$'s request queue lock, then the call to the creation procedure will be a separate call. In this case, the instruction must obtain $q$'s request queue lock, provided it does not already have this lock. Only when it obtained $q$'s request queue lock, does the instruction have to issue an $\unlockrequestqueueoperation$ operation and remove $q$ from $p$'s stack of obtained request queue locks.

\begin{forjournal}
\inferencerule
	{Create Instruction -- Non-Existing Explicit Processor}
	{
		(d, h, c) \mathematicaldefinition \typefromtypingenvironment(\typingenvironment, b) \\
		h = <x> \\
		\neg \state.\processorsfeature.\containsfeature(\state.\valuefeature(p, x)) \\
		q \mathematicaldefinition \state.\newprocessorfeature \\
		o \mathematicaldefinition \state.\newobjectfeature(c) \\
		\state' \mathematicaldefinition \state.\addprocessorfeature(q).\addobjectfeature(q, o) \\
		r \mathematicaldefinition \state'.\referencefeature(o) \\
		\freshchanneldefinition{a}
	}
	{\configuration{p :: \ecreate b.f(e_{1}, \ldots, e_{n}) \statementseparator s_{p}}{\state}}
	{
		\configuration
			{
				p :: \ & \writevalueoperation(x.\namefeature, q) \statementseparator \\
				& \lockrequestqueuesoperation(\set{q}) \statementseparator \\
				& \writevalueoperation(b.\namefeature, r) \statementseparator \\
				& b.f(e_{1}, \ldots, e_{n}) \statementseparator \\
				\multilineconditionaloperation
					{\neg f.\classtypefeature.\invariantexistsfeature \vee f.\isexportedfeature}
					{\nooperation}
					{\issueoperation(q, \evaluateoperation(a, f.\classtypefeature.\invariantfeature) \statementseparator \waitoperation(a))} \statementseparator \\
				& \issueoperation(q, \unlockrequestqueueoperation) \statementseparator \\
				& \popobtainedrequestqueuelocksoperation  \statementseparator \\
				& s_{p} \processorseparator q :: \nooperation
			}
			{\state'}
	}

\end{forjournal}

For the variant that handles non-existing processors, one has to verify that the specified processor does not exist. To do so, one considers only unqualified processor specifications. In this case, the instruction creates a new processor $q$ with a new object $o$ and reference $r$. The steps in this variant are similar to those in the variant where $b$ has a separate type. However, the instruction has to set the value of the processor attribute $x$ to the newly created processor. This ensures that the denoted processor will be found to exist in the future.

\begin{fortechnicalreport}

\inferencerule
	{Create Instruction -- Non-Existing Explicit Processor}
	{
		(d, h, c) \mathematicaldefinition \typefromtypingenvironment(\typingenvironment, b) \\
		h = <x> \\
		\neg \state.\processorsfeature.\containsfeature(\state.\valuefeature(p, x)) \\
		q \mathematicaldefinition \state.\newprocessorfeature \\
		o \mathematicaldefinition \state.\newobjectfeature(c) \\
		\state' \mathematicaldefinition \state.\addprocessorfeature(q).\addobjectfeature(q, o) \\
		r \mathematicaldefinition \state'.\referencefeature(o) \\
		\freshchanneldefinition{a}
	}
	{\configuration{p :: \ecreate b.f(e_{1}, \ldots, e_{n}) \statementseparator s_{p}}{\state}}
	{
		\configuration
			{
				p :: \ & \writevalueoperation(x.\namefeature, q) \statementseparator \\
				& \lockrequestqueuesoperation(\set{q}) \statementseparator \\
				& \writevalueoperation(b.\namefeature, r) \statementseparator \\
				& b.f(e_{1}, \ldots, e_{n}) \statementseparator \\
				\multilineconditionaloperation
					{\neg f.\classtypefeature.\invariantexistsfeature \vee f.\isexportedfeature}
					{\nooperation}
					{\issueoperation(q, \evaluateoperation(a, f.\classtypefeature.\invariantfeature) \statementseparator \waitoperation(a))} \statementseparator \\
				& \issueoperation(q, \unlockrequestqueueoperation) \statementseparator \\
				& \popobtainedrequestqueuelocksoperation  \statementseparator \\
				& s_{p} \processorseparator q :: \nooperation
			}
			{\state'}
	}

\end{fortechnicalreport}

Lastly, there is a variant for the case where $b$ has a non-separate type. In this case, the instruction creates the object on $p$. Processor creation, locking, and lock releasing is not necessary. The required call stack lock on $p$ is given because otherwise $p$ would be waiting.

\inferencerule
	{Create Instruction -- Non-Separate}
	{
		(d, h, c) \mathematicaldefinition \typefromtypingenvironment(\typingenvironment, b) \\
		h = \bullet \\
		o \mathematicaldefinition \state.\newobjectfeature(c) \\
		\state' \mathematicaldefinition \state.\addobjectfeature(p, o) \\
		r \mathematicaldefinition \state'.\referencefeature(o) \\
		\freshchanneldefinition{a}
	}
	{\configuration{p :: \ecreate b.f(e_{1}, \ldots, e_{n}) \statementseparator s_{p}}{\state}}
	{
		\configuration
			{
				p :: \ & \writevalueoperation(b.\namefeature, r) \statementseparator \\
				& b.f(e_{1}, \ldots, e_{n}) \statementseparator \\
				\multilineconditionaloperation
						{\neg f.\classtypefeature.\invariantexistsfeature \vee f.\isexportedfeature}
						{\nooperation}
						{\evaluateoperation(a, f.\classtypefeature.\invariantfeature) \statementseparator \waitoperation(a)} \statementseparator \\
				& s_{p}
			}
			{\state'}
	}

\begin{fortechnicalreport}
\begin{example}[Object creation]
To illustrate object creation, consider a configuration where the root processor $p_{1}$ is executing the root procedure $make$ on the root object $o_{0}$. This procedure is shown in \listingreference{lst:application class with initialization}.

\begin{lstlisting}[caption=Application class with initialization, label=lst:application class with initialization, language=SCOOP]
class APPLICATION

create
  make

feature -- Initialization
  make
      -- Create a market with investors and issuers. Then do some transactions.
    local
      first_investor: separate INVESTOR
      second_investor: separate INVESTOR
    do
      -- Create the market with two investors and one issuer. Each investor has 100 units of cash. The issuer has one share.
      create market.make (2, 100, 1, 1)
      create first_investor.make (1)
      create second_investor.make (2)
			
      -- Do a transaction.
      Current.do_transaction (first_investor, second_investor, 1)
    end

feature {APPLICATION} -- Implementation
  market: separate MARKET
      -- The market.
	
  do_transaction
    ...	
end
\end{lstlisting}

The following configuration is our starting point:

\isolatedconfiguration
	{
		p_{1} :: \ & \ecreate market.make(2, 100, 1, 1) \statementseparator \\
		& \ldots
	}
	{
		\simplifiedstate
			{
				& \simplifiedstatelocksentry
					{p_{1}}
					{\set{}}
					{\set{}}
					{\set{}}
					{\simplifiedstatelockedindicator}
					{\simplifiedstatenopassedlocksindicator}
			}
			{
				& \simplifiedstateobjectsentry
					{p_{1}}
					{
						\simplifiedstatereferencedobject{r_{0}}{o_{0}(\simplifiedstateentityvalue{market}{\voidvalue})}
					}
			}
			{}
			{
				& \simplifiedstateenvironmentsentry
					{p_{1}}
					{
						& \simplifiedstateentityvalue{first\_investor}{\voidvalue}, \simplifiedstateentityvalue{second\_investor}{\voidvalue}, \simplifiedstatecurrententityvalue{r_{0}}
					}
			}
	}

Processor $p_{1}$ starts executing the creation instruction. The result is a new configuration, where $o_{1}$ is the new market object handled by a new processor $p_{2}$:

\isolatedconfiguration
	{
		p_{1} :: \ & \lockrequestqueuesoperation(\set{p_{2}}) \statementseparator \\
		& \writevalueoperation(market.\namefeature, r_{1}) \statementseparator \\
		& market.make(2, 100, 1, 1) \statementseparator \\
		& \issueoperation(p_{2}, \unlockrequestqueueoperation) \statementseparator \\
		& \popobtainedrequestqueuelocksoperation \statementseparator \\
		& \ldots \processorseparator p_{2} :: \nooperation
	}
	{
		\simplifiedstate
			{
				& \simplifiedstatelocksentry
					{p_{1}}
					{\set{}}
					{\set{}}
					{\set{}}
					{\simplifiedstatelockedindicator}
					{\simplifiedstatenopassedlocksindicator} \\
				& \simplifiedstatelocksentry
					{p_{2}}
					{}
					{}
					{}
					{\simplifiedstateunlockedindicator}
					{\simplifiedstatenopassedlocksindicator}
			}
			{
				& \simplifiedstateobjectsentry
					{p_{1}}
					{
						\simplifiedstatereferencedobject{r_{0}}{o_{0}(\simplifiedstateentityvalue{market}{\voidvalue})}
					} \\
				& \simplifiedstateobjectsentry
					{p_{2}}
					{
						\simplifiedstatereferencedobject{r_{1}}{o_{1}}
					}
			}
			{}
			{
				& \simplifiedstateenvironmentsentry
					{p_{1}}
					{
						& \simplifiedstateentityvalue{first\_investor}{\voidvalue}, \simplifiedstateentityvalue{second\_investor}{\voidvalue}, \simplifiedstatecurrententityvalue{r_{0}}
					} \\
				& \simplifiedstateenvironmentsentry
					{p_{2}}
					{}
			}
	}

Now processor $p_{1}$ locks the request queue of processor $p_{2}$. It then stores the reference $r_{1}$ into the entity $market$. With these two steps, processor $p_{1}$ set up the context to execute a feature call to the creation procedure $make$. The resulting feature request will be executed by processor $p_{2}$. Processor $p_{1}$ then asks processor $p_{2}$ to unlock its request queue after it is done with the feature request. Then processor $p_{1}$ removes the obtained request queue lock from its stack.
\end{example}
\end{fortechnicalreport}

\subsubsection{Flow control instructions}
The $\eif{e}{s_{t}}{s_{f}}$ instruction executes $s_{t}$ if the expression $e$ evaluates to true. Otherwise the instruction executes $s_{f}$. There is one inference rule for this instruction. In a first step, the instruction evaluates the expression $e$ using a fresh channel $a$ and then waits for a notification on $a$. In a second step, it uses the $\conditionaloperationname$ operation to either execute $s_{t}$ or $s_{f}$, depending on the value of the expression.

\inferencerule
	{If Instruction}
	{\freshchanneldefinition{a}}
	{\configuration{p :: \eif{e}{s_{t}}{s_{f}} \statementseparator s_{p}}{\state}}
	{
		\configuration
			{
				p :: \ & \evaluateoperation(a, e) \statementseparator \\
				& \waitoperation(a) \statementseparator \\
				\multilineconditionaloperation
					{a.\datafeature}
					{s_{t}}
					{s_{f}} \statementseparator \\
				& s_{p}
			}
			{\state}
	}

The $\euntil{e}{s_{l}}$ instruction executes a sequence of $s_{l}$ instructions until the expression $e$ evaluates to true. If $e$ is true initially, then $s_{l}$ never gets executed. There is one inference rule for this instruction. First, the instruction evaluates $e$ using a fresh channel $a$. Then it waits for a notification on $a$. Next, it uses the $\conditionaloperationname$ operation to check whether $e$ evaluates to true or false. If $e$ is true, then it is done. Otherwise, it executes $s_{l}$ followed by another $\euntil{e}{s_{t}}$ operation.

\inferencerule
	{Loop Instruction}
	{\freshchanneldefinition{a}}
	{\configuration{p :: \euntil{e}{s_{l}} \statementseparator s_{p}}{\state}}
	{
		\configuration
			{
				p :: \ & \evaluateoperation(a, e) \statementseparator \\
				& \waitoperation(a) \statementseparator \\
				\multilineconditionaloperation
					{a.\datafeature}
					{\nooperation}
					{s_{l} \statementseparator \euntil{e}{s_{l}}} \statementseparator \\
				& s_{p}
			}
			{\state}
	}

\subsubsection{Assignment instructions}
An assignment instruction $b \eassignment e$ assigns the value of the expression $e$ to the entity $b$. The instruction first evaluates the expression $e$ and then waits for a notification on a fresh channel $a$. Once it gets this notification, it uses the $\writevalueoperation$ operation to set the value to the entity $b$.

\singlelineinferencerule
	{Assignment}
	{\freshchanneldefinition{a}}
	{\configuration{p :: b \eassignment e; s_{p}}{\state}}
	{
		\configuration
			{p :: \evaluateoperation(a, e) \statementseparator \waitoperation(a) \statementseparator \writevalueoperation(b.\namefeature, a.\datafeature) \statementseparator s_{p}}
			{\state}
	}

\subsection{Termination}
The system terminates when it reaches a configuration where all action queues are empty, i.e., when there is no more work to do.

\section{Conclusion}\label{sec:conclusion}
In this paper we have presented a formal specification of the SCOOP
model, based on operational semantics. We have demonstrated that this
level of rigor is necessary if the specification is to be used as a
guideline for an implementation. In particular, we were able to
clarify a number of omissions and ambiguities in the available
informal specification, which had gone undetected in other
formalizations:
\begin{itemize}
	\item Are processor locks fine-grained enough? We require request queue locks and call stack locks.
	\item Which locks must be passed? Which locks can be passed? We pass all the locks we actually have. We pass these locks both for normal lock passing and for separate callbacks.
	\item How do we move object structures from one processor to another processor without violating the invariant? The deep import operation must be used.
	\item When do we set the status of a fresh once routine to non-fresh? The status of the once routine must be set to non-fresh before deep importing.
	\item When can a postcondition be evaluated asynchronously? The postcondition can be evaluated asynchronously if every feature call in the postcondition only requires a lock that was obtained in the synchronization step and if the postcondition does not involve lock passing.
\end{itemize}
Because of the complexity of the SCOOP model, our resulting
specification is large, the management of which is a challenge for a
fully formal development. To address this problem, we used abstract
data types and a notation with an object-oriented flavor, which made
the specification more readable and more easily extendable, without
sacrificing any of the rigors of operational semantics. Furthermore, we
introduced a distinction between two kinds of statements, namely instructions (user
syntax) and operations (run-time syntax).  This made it possible to treat within one
inference system both the actual language elements and the
implementation details of the runtime system, and to distinguish
clearly between them.

The main application of this work is to
guide the implementation of the SCOOP model. This has led to a
successful implementation of SCOOP on top of the Eiffel language,
which supersedes the previous prototype implementation and is publicly
available~\cite{eth:2011:SCOOP}. The SCOOP model can however be
implemented on top of any object-oriented language (support for
contracts, as offered by Java or Spec\#, is beneficial), and our work
also facilitates such future implementation efforts. In the case of
Java, first steps towards such an implementation have been
taken~\cite{torshizi-ostroff-paige-chechik:2009:JSCOOP}, which could certainly be supported
by our work.

A number of other applications of our semantics can be
envisioned. First, the semantics can be used to prove correct various
properties of the model which have so far only been postulated, such
as absence of object-level data races and type safety (absence of
traitors). In light of the complexity of the full model, these
properties are no longer obvious. For example, as processor locks
serve as an abstraction only, it must be shown that locks are not
misused in situations such as separate callbacks, which involve call stack
locks. Second, our operational semantics can also be used to prove
correct an axiomatic semantics for the SCOOP model, which is planned
for future work. In the case of sequential Eiffel, a similar
development is documented in~\cite{nordio-calcagno-mueller-meyer:2009:formal_specification_for_Eiffel}.
Third, we feel our semantics is detailed enough that its rules can
directly be implemented as an interpreter for SCOOP programs. Such an
interpreter could serve as a true reference implementation, which
could in turn be used for conformance checking of real
implementations.

\paragraph{Acknowledgments}
We thank Stephan van Staden for interesting discussions on the lock model. This work is part of the SCOOP project at
ETH Zurich, which has benefited from grants from the Hasler
Foundation, the Swiss National Foundation, Microsoft (Multicore award),
and ETH (ETHIIRA).


\begin{forjournal}
\bibliographystyle{style/splncs03}
\end{forjournal}
\bibliography{bibliography}

\end{document}